\RequirePackage{fix-cm}

\documentclass[]{pasj02} 

\usepackage[switch,mathlines]{lineno} 
\usepackage[percent]{overpic}
\usepackage{comment}
\usepackage{ulem}
\jyear{2026}
\Received{}
\Accepted{}

\newcommand{\msun}{\rm M_\odot}

\newcommand{\rfig}[1]{Fig. \ref{#1}}

\newcommand{\citealt}[1]{\cite{#1}}

\begin{document} 
\title{SIRIUS: The relation between the diversity of dwarf galaxies and their formation histories}

\author{
 Chihong \textsc{Lin},\altaffilmark{1}\orcid{0000-0002-9885-587X} 
 Michiko S \textsc{Fujii},\altaffilmark{1}\altemailmark\orcid{0000-0002-6465-2978} \email{fujii@astron.s.u-tokyo.ac.jp}
 Takayuki R \textsc{Saitoh},\altaffilmark{2,3}\orcid{0000-0001-8226-4592}
 and Yutaka \textsc{Hirai},\altaffilmark{4}\orcid{0000-0002-5661-033X}
}
\altaffiltext{1}{Department of Astronomy, Graduate School of Science, The University of Tokyo, 7-3-1 Hongo, Bunkyo-ku, Tokyo 113-0033, Japan}
\altaffiltext{2}{Department of Planetology, Graduate School of Science, Kobe University, 1-1 Rokkodai-cho, Nada-ku, Kobe, Hyogo 657-8501, Japan}
\altaffiltext{3}{Center for Planetary Science (CPS), Graduate School of Science, Kobe University, 1-1 Rokkodai, Nada-ku, Kobe, Hyogo 657-8501, Japan}
\altaffiltext{4}{Department of Community Service and Science, Tohoku University of Community Service and Science, 3-5-1 Iimoriyama, Sakata, Yamagata 998-8580, Japan}




\maketitle

\begin{abstract}
Low-mass dwarf galaxies ($M_{\rm vir} \lesssim 10^9\rm\ M_\odot$) are fundamental cosmological building blocks, yet the physical processes driving their structural diversity remain poorly understood. 
Recent numerical simulations have suggested a diversity in the stellar-to-halo mass ratio in this halo mass range, but either the number of samples obtained from the same simulation setup or the numerical resolution was limited. 
We performed high-resolution cosmological zoom-in simulations for eight galaxies with a dark matter halo mass of $\sim 10^9\rm\ M_{\odot}$ up to $t=1.2$\,Gyr at which most gas in the galaxies has been expelled. 
Our samples have a scatter of an order of magnitude in the halo mass at the reionization epoch. 
The stellar-to-halo mass ratio expected at $z=0$ scatters nearly two orders of magnitude with $5\times10^{-5}$ to $2\times10^{-3}$.
We also observed variation in the compactness of their stellar distributions.
Some of our simulated galaxies exhibit a stellar half-mass radius of $\sim30$\,pc, which is as small as that of ultra-compact dwarfs. 
The formation condition for such a compact stellar distribution is understood as an analog of the condition for the formation of dense, massive star clusters. We found that when the central gas surface density exceeds a critical threshold ($\Sigma_{\rm gas} \gtrsim 30\rm\ M_\odot \rm\ {pc}^{-2}$), the star formation becomes highly efficient and results in dense stellar systems. These results suggest that UCDs can form in situ even in isolated dark matter halos. 
\end{abstract}


\section{Introduction}

{Dwarf galaxies are generally defined as galaxies with stellar masses of $M_\ast \sim 10^3-10^9 \rm\ M_\odot$ and sizes ranging from 10 pc to 2 kpc (e.g., \citealt{bullock2017small}). 
They are frequently categorized by their stellar mass into three distinct groups: bright dwarfs ($M_{\ast} \approx 10^{7}$--$10^{9}\,\rm\ M_{\odot}$), classical dwarfs ($M_{\ast} \approx 10^{5}$--$10^{7}\,\rm\ M_{\odot}$), and ultra-faint dwarfs (UFDs; $M_{\ast} \approx 10^{2}$--$10^{5}\,\rm\ M_{\odot}$). These groups exhibit stellar-to-halo mass ratios of $10^{-2}$--$10^{-3}$ for bright dwarfs, $10^{-3}$--$10^{-4}$ for classical dwarfs, and $10^{-4}$--$10^{-5}$ for UFDs. This scaling relation between stellar mass and dark matter halo mass is supported by both theoretical numerical simulations and observational abundance matching techniques.}

\citet{behroozi2013average} and \citet{2013MNRAS.428.3121M} have shown the relationship between the stellar masses of galaxies and their host dark matter halo masses using abundance matching. Because of observational limitations, the minimum halo mass in their study is $\sim 10^{10} \rm\ M_{\odot}$. Simple extrapolations for $< 10^9$ M$_{\odot}$ halos are discrepant between these studies.

Galaxy formation simulations present difficulties in estimating the stellar-to-halo mass ratios for halos with $< 10^9$ M$_{\odot}$. In large-scale cosmological galaxy formation simulations, the mass resolution is $10^4\rm\ M_{\odot}$ even for the highest one, such as TNG50 \citep{2019MNRAS.490.3234N}. The number of star particles for a dwarf galaxy with $10^6 \rm\ M_{\odot}$ is only 100. For higher resolution, cosmological zoom-in simulations have been performed. The FIRE-2 simulation \citep{hopkins2018fire} has performed $\sim 30$ cosmological zoom-in simulations, which include a galaxy with $\sim 10^9 \rm\ M_{\odot}$ at $z=0$ and a few more in \citet{Wheeler_2019}. The mass resolution for gas particles is typically a few thousand solar masses, but the highest one was $30 \rm\ M_{\odot}$ \citep{Wheeler_2019}. The stellar-to-halo mass of their simulation was $< 10^{-5}$, which is consistent with the extrapolation of the relation obtained by \citet{2013MNRAS.428.3121M}.

\citet{2018A&A...616A..96R} performed cosmological zoom-in simulations of dwarf galaxies (GEAR simulation) with a mass resolution similar to that of FIRE-2. They focused on galaxies with halo masses of $\sim 10^9\rm\ M_{\odot}$, and the stellar mass resolution was $1024\ \rm\ M_{\odot}$. In contrast to FIRE-2 simulations, their dwarf galaxies showed stellar-to-halo mass ratios of $10^{-3}$--$10^{-4}$, which is close to the extrapolation of the relation obtained by \citet{behroozi2013average}. These ratios are two orders of magnitude higher than those of FIRE-2 (see also \cite{sales2022baryonic}). 
The discrepancies between the two simulations are likely driven by their respective baryonic physics implementations. While both models include essential processes such as star formation and UV background heating, their feedback prescriptions differ in complexity and energy coupling. The GEAR model primarily accounts for thermal energy injection from supernovae and chemical enrichment; in contrast, FIRE-2 employs a more comprehensive, multi-channel feedback framework. This includes not only Type I and II supernovae but also stellar winds (OB/AGB mass loss), photo-ionization heating, and radiation pressure. Such differences in how energy and momentum are deposited into the interstellar medium may drive divergence in dwarf properties.

MARVEL simulations include samples of dwarf galaxies with a minimum halo mass of $\sim 10^8\rm\ M_{\odot}$ \citep{munshi2021quantifying}. Their simulations were performed with an $N$-body/SPH code, \textsc{CHANGA} \citep{2015ComAC...2....1M}, with a mass resolution of $1400 \rm\ M_{\odot}$. Their simulated stellar-to-halo mass ratios are $10^{-5}$--$10^{-3}$ for $10^8$--$10^9\rm\ M_{\odot}$ halos, although the stellar mass depended on their feedback models. Their stellar-to-halo mass ratios distribute between those of FIRE-2 and GEAR.

\citet{Wheeler_2019} performed cosmological zoom-in simulations for dwarf galaxies with halo mass of $\sim 10^8$--$\sim 10^{10} \rm\ M_{\odot}$ at $z=0$ using the FIRE-2 setup with the resolution of $30\rm\ M_{\odot}$. 
Their stellar-to-halo mass ratios were $10^{-5}$--$10^{-4}$, so that the stellar mass were $10^4$--$10^6\rm\ M_{\odot}$. The half-mass radius of their stellar distribution was an order of a hundred parsecs; therefore, their samples are close to ultra-diffuse dwarf galaxies (UFDs) in the Milky Way. 

EDGE is also a series of dwarf-galaxy simulations with a resolution similar to that of \citet{Wheeler_2019}. Their galaxies with the halo mass of $\sim 10^9\rm\ M_{\odot}$ came to the stellar-to-halo mass ratio between those of FIRE-2 and GEAR samples \citep{rey2019edge,2020MNRAS.497.1508R} (see also \cite{sales2022baryonic}). EDGE series was performed using the \textsc{RAMSES} code \citep{teyssier2002cosmological}, in which the hydrodynamics is solved using adaptive mesh refinement. Their spatial resolution was $3\,$pc. Their minimum halo mass was slightly larger than $10^9 \rm\ M_{\odot}$, and the stellar-to-halo mass ratio was $10^{-5}$--$10^{-4}$, which is slightly lower than that typical for the GEAR sample but much higher than that of FIRE-2. The half-mass radii of their simulated galaxies were typically over 100\,pc \citep{2020MNRAS.491.1656A}. 

\citet{jeon2017connecting} investigated the formation of isolated ultra-faint dwarf galaxies (UFDs) and gas-rich dwarfs (Leo P analogs) using cosmological hydrodynamic zoom-in simulations with the \textsc{Gadget} code \citep{Springel_2005b}. Their mass resolution was $m_{\mathrm{gas}}\sim 500\,\rm\ M_{\odot}$ for gas and $m_{\mathrm{DM}}\sim 2000\,\rm\ M_{\odot}$ for dark matter. Their samples focused on the lower mass regime, with halo masses of $M_{\mathrm{vir}} < 2 \times 10^9\,\rm\ M_{\odot}$ for UFD analogs and $\sim 4 \times 10^9\,\rm\ M_{\odot}$ for gas-rich dwarfs. The resulting stellar masses were in the range of $10^4$--$10^5\,\rm\ M_{\odot}$. Therefore, their stellar-to-halo mass ratios were about $10^{-5}$--$10^{-4}$, consistent with the range for UFDs.

LYRA is the most recent series of dwarf galaxy formation simulations with the highest resolution, $4\rm\ M_{\odot}$ for gas particles \citep{Gutcke_2021}. Their simulations were performed with \textsc{Arepo} code \citep{2020ApJS..248...32W}, which adopts a moving-mesh method. Their resolution of $4\rm\ M_{\odot}$ for gas particles reaches the range where we do not need to use subgrid models for supernova feedback \citep{2019MNRAS.483.3363H}. Therefore, they did not adopt any subgrid model for supernova feedback. 
\citet{sureda2025co} have performed cosmological zoom-in simulations for six halos with halo masses of 0.7--$4.8\times10^9\rm\ M_{\odot}$ at $z=0$. Their stellar-to-halo mass ratios were consistent with those of GEAR. The LYRA samples also showed radii larger than 100\,pc, but one had a radius smaller than the others. Such compact galaxies can be found only in high-resolution simulations.

In \citet{2026arXiv260113765K}, we performed cosmological zoom-in simulations of two dwarf galaxies with a mass resolution of $\sim 20\rm\ M_{\odot}$ for baryon particles and star-by-star for star particles. In these simulations, we found that an ultra-compact dwarf (UCD)-like galaxy, which has a stellar mass of $\sim 10^6 \rm\ M_{\odot}$ and a half-mass radius of $\sim 30$\,pc. This galaxy is much more compact than the sample in LYRA.

Dwarf galaxies with stellar masses of $10^6$--$10^8\rm\ M_{\odot}$ and half-light radius of $10 \leq r \leq 100~\rm pc$ are known as UCDs \citep{mieske2008nature,2020ApJS..250...17L,saifollahi2021ultra}. They are as dense as globular clusters in the Milky Way \citep{baumgardt2018catalogue} but more massive. 
The formation of UCDs is considered to be (1) tidally stripped nuclei of nucleated dwarf galaxies, produced mainly in dense environments where stripping is efficient \citep{bassino1994globular,bekki2003galaxy,goerdt2008formation}, (2) massive super star clusters during gas‑rich, violent galaxy mergers; merged cluster complexes then relaxed into compact systems \citep{fellhauer2002formation}, or (3) compact dwarf galaxies that formed in situ from high‑peak fluctuations of the small‑scale dark‑matter density field \citep{drinkwater2004ultra}.

In \citet{2026arXiv260113765K}, we found one UCD-like simple in high-resolution zoom-in simulations. 
In the present paper, we perform galaxy formation simulations for additional samples and investigate the diversity of dwarf galaxies in terms of mass and the size of their stellar distributions. 
We further investigate the conditions necessary to form UCD-like galaxies in an isolated environment.

\begin{table*}
  \caption{Summary of Previous Simulations}
  \label{tab:simulations}
  \centering
  \renewcommand{\arraystretch}{1.2}
  \begin{tabular}{ccccccc}
    \hline
    Simulation &  $M_{\rm Halo}$ [$\rm M_{\odot}$]& $m_{\mathrm{gas}}\,[\rm M_{\odot}]$ & $\epsilon_{\mathrm{gas}}\,[\mathrm{pc}]$ & $m_{\mathrm{DM, min}}\,[\mathrm{M_{\odot}}]$ & $\epsilon_{\mathrm{DM, min}}\,[\mathrm{pc}]$ & $M_\ast/M_{\rm Halo}$ for $M_{\rm Halo}<2\times10^9 \rm\ M_\odot$ \\
    
    \hline
    Jeon+17  &  $\lesssim 2\times 10^9$ & 495   & 2.8  & 2,000  & 40 &  $2.5$--$5.1 \times 10^{-5}$  \\
    GEAR     & $5.0\times 10^{8}$ -- $ 10^{10}$  & 1,024 & 10   & 22,462 & 50  & $2.7\times 10^{-4}$ to $2.0 \times 10^{-2}$ \\
    FIRE-2   & $\lesssim10^{10}$--$\lesssim10^{12}$ & 7,070 & 2.0  & 43,000 & 40  & $2.6$-$8.5 \times 10^{-6}$\\
    EDGE     & $ \lesssim3\times  10^9$ & 300   & 3.0  & 960    & 3.0 & $3.0 \times 10^{-5} $ to $3.6 \times 10^{-4}$ \\
    MARVEL   &  $10^{7}$--$ 10^{11}$  & 420   & 60   & 6,650  & 60  & $5.1 \times 10^{-5}$ to $3.4\times 10^{-4}$\\
    LYRA     &  $\sim  10^9$ & 4.0    & 0.5  & 80     & 8.0 & $5.6$-$9.4 \times 10^{-4}$    \\
    \hline
  \end{tabular}

  \vspace{1ex}
  \raggedright
  \footnotesize
Comparison of different simulations: Jeon+17 (GADGET; \citealt{jeon2017connecting}), GEAR (\citealt{2018A&A...616A..96R}), FIRE-2 (\citealt{hopkins2018fire}), EDGE (\citealt{2020MNRAS.497.1508R}), MARVEL (\citealt{munshi2021quantifying}), and LYRA (\citealt{sureda2025co}). From left to right: (1) Simulation name; (2) Target halo mass; (3) Gas mass resolution ($m_{\rm gas}$); (4) Gas softening length ($\epsilon_{\rm gas}$); (5) Minimum dark matter particle mass ($m_{\rm DM, min}$); (6) Minimum dark matter softening length ($\epsilon_{\rm DM, min}$); (7) Stellar-to-halo mass ratio range for simulation samples at $M_{\rm Halo}$ $<2 \times 10^{9}\,\rm M_{\odot}$.
\end{table*}

\section{Methods}\label{sec_method}

We performed cosmological zoom-in simulations of dwarf galaxy formation. Hereafter, we describe the initial condition and simulation setups. 

\subsection{Initial Condition}
In order to find halos for the following zoom-in simulations, we carried out a dark‑matter‑only cosmological simulation in a comoving box of (\(4\ \mathrm{Mpc}\,h^{-1}\) )$^3$. The initial condition of this simulation was generated with \textsc{MUSIC} \citep{Hahn_2011}. We adopted a flat \(\Lambda\)CDM cosmology with parameters from \citet{Planck_2020}: \(H_0=67.32\ \mathrm{km\,s^{-1}\,Mpc^{-1}}\), \(\Omega_{m,0}=0.3158\), \(\Omega_{\Lambda,0}=0.6842\), \(\Omega_{b,0}=0.04939\), and \
$\sigma_8 = 0.812$. The mass resolution of this low‑resolution run is \(m_{\mathrm{DM}}=5\times10^{5}\,\rm\ M_{\odot}\). Starting from \(z=100\), the simulation was performed to \(z=0\) with \textsc{GADGET-2} \citep{Springel_2005b}.
We identified halos at \(z=0\) using Amiga Halo Finder (AHF; \cite{knollmann2009ahf}). From the halos with $\sim 10^9\rm\ M_{\odot}$ at $z=0$, we selected eight isolated halos for the following zoom-in simulations. Here, we chose halos with varying mass at the reionization epoch.
Their properties at $z=0$ and the virial mass at the reionization epoch are listed in Table~\ref{tab:HaloPropertyGadget}. The properties are calculated in AHF. We note that Halos 230 and 284 are identical to those calculated in \citet{2026arXiv260113765K}, but we re-simulated them in this study to have more frequent snapshots. 

For each selected halo, we generated refined (zoom‑in) initial conditions with \textsc{MUSIC}. The zoomed region follows \citet{griffen2016caterpillar}: (1) select particles within \(4\,R_{\rm vir}\) of the halo center at \(z=0\); (2) compute the minimal enclosing ellipsoid of these particles at \(z=100\); and (3) enlarge the ellipsoid by a factor of 1.05 to mitigate contamination by low‑resolution boundary particles.

\begin{table}
 \caption{Properties of the halos selected from the dark-matter-only simulations.}
 \label{tab:HaloPropertyGadget}
 \centering  
  \renewcommand{\arraystretch}{1.2} 
  \begin{tabular}{cccccc}
   \hline
    Halo ID & $M_{\mathrm{vir},\ z=0}$ & $R_{\mathrm{vir},\ z=0}$ &  $V_{\mathrm{max},\ z=0}$ & $M_{\mathrm{vir,\ EoR}}$ \\
    & $[\rm\ M_{\odot}]$ & [kpc] &  $[\mathrm{km}\ \mathrm{s^{-1}}]$ & $[\rm\ M_{\odot}]$ \\
   \hline 

   198 &         $ 1.33 \times 10^9$   &  23.2        &    19.41  &  $6.3\times 10^8$   \\ 
   215 &         $ 1.14 \times 10^9$    &  22.1       &   19.33   &  $4.0\times 10^7$  \\ 
   219 &         $ 1.12 \times 10^9$     & 21.9       &  17.55    &  $5.0\times 10^7$ \\ 
   230 &         $ 1.08 \times 10^9$    & 21.6       & 18.54    &  $2.0\times 10^7$ \\
   236 &         $ 1.02 \times 10^9$  &  21.2        &   20.10   & $3.2\times 10^8$   \\ 
   281 &         $ 8.91\times 10^8$    &  20.3       &  20.14    &  $3.2\times 10^7$    \\ 
   284 &        $ 8.77 \times 10^8$     & 20.2           & 17.70  & $3.6\times 10^8$   \\
   299 &        $ 8.13 \times 10^8$   & 19.7         &     18.27  & $1.6\times 10^8$  \\ 
   \hline 
  \end{tabular}
  
  \vspace{1ex} 
  \raggedright
  \footnotesize 
  From left to right: (1) Halo ID in AHF, (2) virial mass at $z=0$, (3) virial radius at $z=0$, (4) circular velocity at $z=0$, (5) virial mass at the epoch of the reionization.
\end{table}

\subsection{Simulation Code and Numerical Setup}\label{sub_code}

The following cosmological zoom-in simulations were performed using the \textsc{ASURA+BRIDGE} code, which couples $N$-body gravity and smoothed particle hydrodynamics (SPH) (\cite{Saitoh_2009,Hirai_2021,Fujii_2021}) as a part of the SIRIUS project. Gravity was computed using the tree method \citep{Barnes_1986} with an opening angle $\theta=0.5$, and hydrodynamics was solved with the density‑independent SPH (DISPH) formulation \citep{Saitoh_2013}. 
To model reionization, we adopt a uniform ultraviolet (UV) background switched on at $z=8.5$ \citep{Haardt_2012}, corresponding to $t\simeq0.59\ \mathrm{Gyr}$ in our cosmology. Radiative cooling and heating were treated with metallicity‑dependent tables covering $10$--$10^{9}\ \mathrm{K}$, generated with \texttt{Cloudy} v13.05 (\cite{Ferland_1998, Ferland_2013, Ferland_2017}). 

Chemical evolution of gas and stars is followed with the \textsc{CELib} library \citep{Hirai_2021,Saitoh_2017}.

We set the gravitational softening for dark matter (DM) following \citet{hopkins2018fire}:
\begin{equation}
    \epsilon_{\mathrm{DM}} = 30\ \mathrm{pc}\left(\frac{m_{\mathrm{DM}}}{1000\ \rm\ M_{\odot}}\right)^{1/2}
    \left(\frac{M_{\mathrm{vir}}}{10^{12}\ \rm\ M_{\odot}}\right)^{-0.2},
\label{equation:SofteningDM}
\end{equation}
where $m_{\mathrm{DM}}$ is the DM particle mass and $M_{\mathrm{vir}}$ is the virial mass; for our runs we adopt $M_{\mathrm{vir}}=10^{9}\ M_{\odot}$.
Gas softening is set following \citet{Dutton_2020} with a modified normalization:
\begin{equation}
    \epsilon_{\mathrm{gas}} = 2.13\ \mathrm{pc}\left(\frac{m_{\mathrm{gas}}}{1\ M_{\odot}}\right)^{1/3}
    \left(\frac{\rho_{\mathrm{th}}}{100\ \mathrm{cm}^{-3}}\right)^{-1/3},
\label{equation:SofteningGas}
\end{equation}
where $m_{\mathrm{gas}}$ is the initial gas particle mass and $\rho_{\mathrm{th}}$ the star‑formation threshold density. Newly formed star particles inherit the gas softening, i.e., $\epsilon_{\rm star}=\epsilon_{\rm gas}$.
The softening length used in the gravity calculation, $\epsilon_{\mathrm{soft}}^{\mathrm{calc}}$, varies with redshift as
\begin{equation}
\epsilon_{\mathrm{soft}}^{\mathrm{calc}}(z)=
\begin{cases}
\epsilon_{\mathrm{soft}}, & z < z_{\mathrm{f}},\\[6pt]
\epsilon_{\mathrm{soft}}\dfrac{1+z_{\mathrm{f}}}{1+z}, & z \ge z_{\mathrm{f}},
\end{cases}
\label{equation:SofteningRedshift}
\end{equation}
where $\epsilon_{\mathrm{soft}}$ denotes either $\epsilon_{\mathrm{DM}}$ or $\epsilon_{\mathrm{gas}}$ (or $\epsilon_{\rm star}$), and we adopt $z_{\mathrm{f}}=9.0$.\\

We adopt a star‑by‑star formation scheme \citep{Hirai_2021,Hirai_2025}, in which individual star particles are assigned masses sampled from a prescribed initial mass function (IMF), the Salpeter IMF \citep{Salpeter_1955} in this study.
The star formation procedure starts when a gas particle satisfies all of the following criteria: (i) hydrogen nuclei number density \(n_{\rm H}>100\ \mathrm{cm}^{-3}\); (ii) temperature \(T<1000\ \mathrm{K}\); (iii) the local flow is converging, \(\nabla\cdot\mathbf{v}<0\); and (iv) the particle has not been recently heated by supernova feedback. 

The star formation occurs probabilistically, with the probability that the star formation efficiency per free-fall (dynamical) time becomes a constant value. To capture unresolved, rapid gravitational collapse, particles exceeding \(100\times\) the density threshold are converted instantaneously.
Once a gas particle has satisfied conditions to form a star, its mass and the forming stellar mass, which is randomly drawn from the given IMF, are compared. When the metallicity of the star-forming gas particle was less than $10^{-5} \rm\ Z_{\odot}$, where $\rm\ Z_{\odot}$ is the Solar metallicity, we adopted the Susa IMF for Population III stars \citep{susa2014mass}. 

To conserve mass locally, the sampled stellar mass is constrained by the available gas. If the sampled mass does not exceed the parent gas‑particle mass, the star is formed from that particle. If the sampled mass exceeds the parent mass, neighbouring gas within \(r_{\rm max}=3\ \mathrm{pc}\) is searched. If the total gas mass inside \(r_{\rm max}\) is at least twice the sampled stellar mass, the star particle is assembled by removing mass from nearby gas particles (preferentially the nearest) until the target mass is reached; otherwise, the stellar mass is re‑sampled. The position and velocity of the new star particle are set to the center of mass of the contributing gas particles.

Stellar evolution and feedback were handled with \texttt{CELib}. Stellar lifetimes are taken from the metallicity‑dependent tables of \citet{Portinari_1998}. We include three principal feedback channels: H\textsc{ii} regions from massive stars and core‑collapse supernovae (CCSNe).

For radiative feedback, we do not perform full radiative transfer; instead, each massive star (\(>15\,\rm\ M_{\odot}\)) is assumed to carve a spherical H\textsc{ii} region. The Str\"{o}mgren radius is computed from the local gas density, and the gas inside this radius is maintained at \(T=10^{4}\ \mathrm{K}\) \citep{Fujii_2021}. To avoid unrealistically large H\textsc{ii} regions in nearly gas‑free environments, we impose a maximum radius of \(30\ \mathrm{pc}\) \citep{Fujii_2021}.

We adopted a simple supernova feedback model in which all supernova energy is injected as thermal energy into surrounding gas particles. We assume that CCSNe arise from progenitors in the mass range \(8\text{--}100\,\msun\), with each event releasing \(10^{51}\ \mathrm{erg}\). Additionally, we assume that \(5\%\) of stars in the \(20\text{--}100\,\msun\) range explode as broad-line Type Ic events (hypernovae), releasing \(10^{52}\ \mathrm{erg}\). The nucleosynthetic yields for CCSNe are adopted from \citet{nomoto2013nucleosynthesis}.

\begin{table}
 \caption{Resolution of our simulations}
 \label{table:resolution}
  \renewcommand{\arraystretch}{1.2} 
 \begin{center}
  \begin{tabular}{cccc}
   \hline
    $m_{\mathrm{gas}}\,[\rm M_{\odot}]$ & $\epsilon_{\mathrm{gas}}\,[\mathrm{pc}]$ &  $m_{\mathrm{DM, min}}\,[\mathrm{M_{\odot}}]$ & $\epsilon_{\mathrm{DM, min}}\,[\mathrm{pc}]$\\
    
   \hline 
   19.0 & 5.68 & 102 & 12.7 \\
   \hline
  \end{tabular}
  \end{center}
  From left to right: (1) gas-mass resolution ($m_{\rm gas}$), (2) gas softening length ($\epsilon_{\rm gas}$) (3) minimum dark-matter mass ($m_{\rm DM, min}$) (4) minimum dark-matter softening length ($\epsilon_{\rm DM, min}$). 
\end{table}

\section{Results}\label{sec:results}

\subsection{{Overview of the formation processes}}

\begin{table*}
 \caption{Summary of Dwarf Basic Properties.}
 \label{tab:halo_properties}
 \centering
 \renewcommand{\arraystretch}{1.2} 
  \begin{tabular}{cccccc}
   \hline
    Halo ID  & $M_{\mathrm{vir,\ end}}$ & $M_{\ast,\ \rm end}$ & $M_{\rm \ast,\ end}/M_{{\rm Halo},\ z=0}$ & $R_{\mathrm{vir,\ end}}$ & $r_{h,\mathrm{2D}}$ \\
            & $[M_{\odot}]$  & $[M_{\odot}]$ &  &  [kpc]            & [pc] \\
   \hline
    198    &$ 1.3\times 10^9$&    $3.2\times 10^6$ & $2.4\times10^{-3}$ & 2.43 & 28.9   \\
    215    &$6.3\times 10^7$&     $6.7\times 10^4$ & $5.9\times10^{-5}$ & 0.98 & 55.9   \\
    219    &$7.9\times 10^7$&     $1.3\times 10^5$ & $1.2\times10^{-4}$ & 1.05 & 85.1   \\
    230    &$1.3\times 10^8$ &     $5.6\times 10^4$ & $5.2\times10^{-5}$ & 1.20 & 76.9   \\
    236    &$1.0\times 10^9$&     $2.4\times 10^6$ & $2.3\times10^{-3}$ & 2.35 & 109  \\
    281    & $7.9\times 10^7$&     $9.6\times 10^4$ & $1.1\times10^{-4}$ & 1.00 & 74.1   \\
    284    &$8.0\times 10^8$&     $1.8\times 10^6$ & $2.1\times10^{-3}$ & 2.17 & 26.3  \\
    299    &$5.0\times 10^8$&      $1.0\times 10^6$ & $1.3\times10^{-3}$ & 1.81 & 56.1   \\
   \hline
  \end{tabular}

  \vspace{1ex} 
  \raggedright
  \footnotesize 
From left to right: (1) Halo ID in AHF, (2) Stellar mass at the end of our simulation ($t=1.2$ Gyr), (3) Stellar-to-halo mass ratio at $z=0$, (4) Virial radius at $t=1.2$ Gyr, (5) Half-mass radius in parsecs (pc), derived from our density radial profiles. 
\end{table*}

\begin{figure*}
    \begin{center}
    \includegraphics[width=0.33\textwidth]{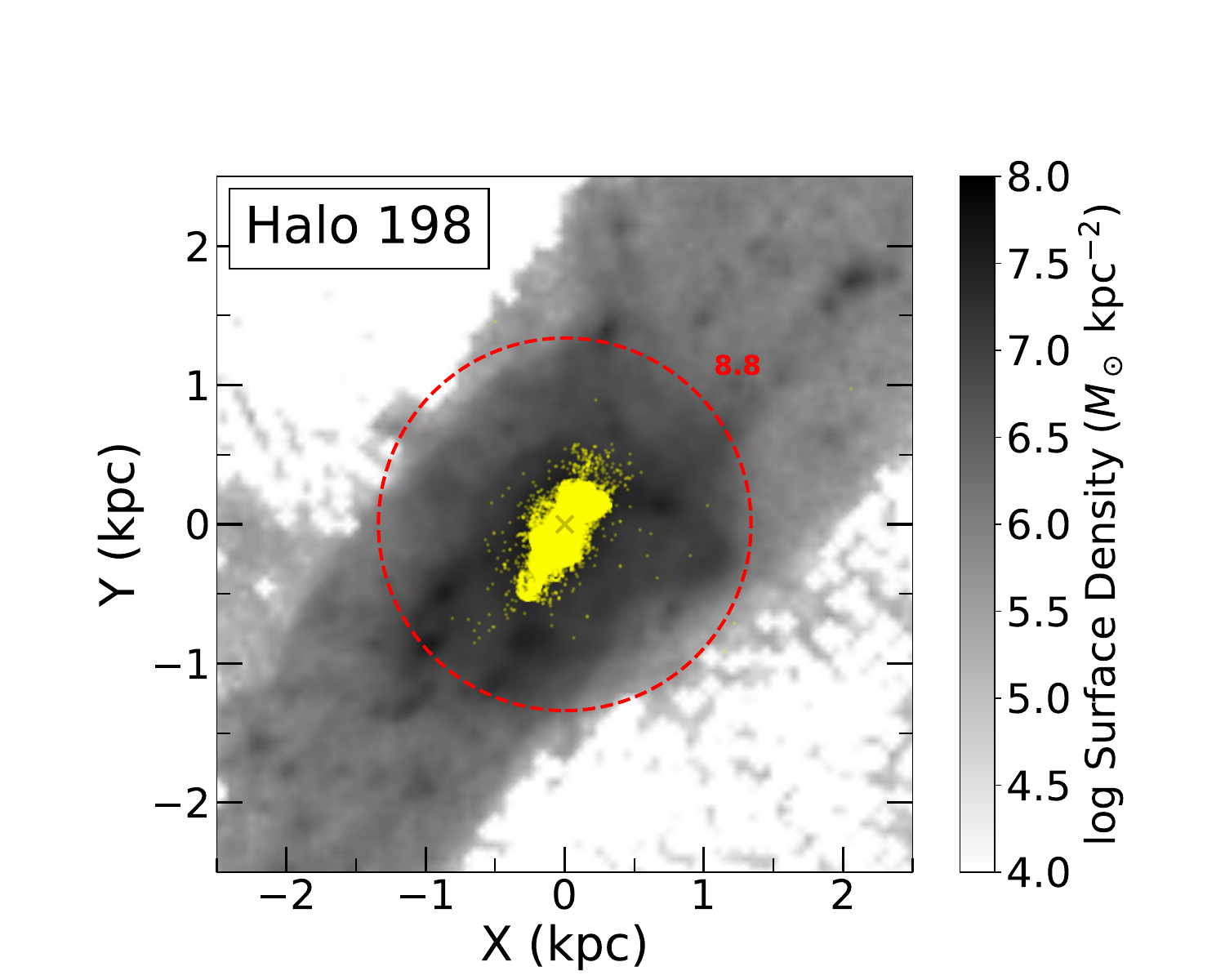}
    \includegraphics[width=0.33\textwidth]{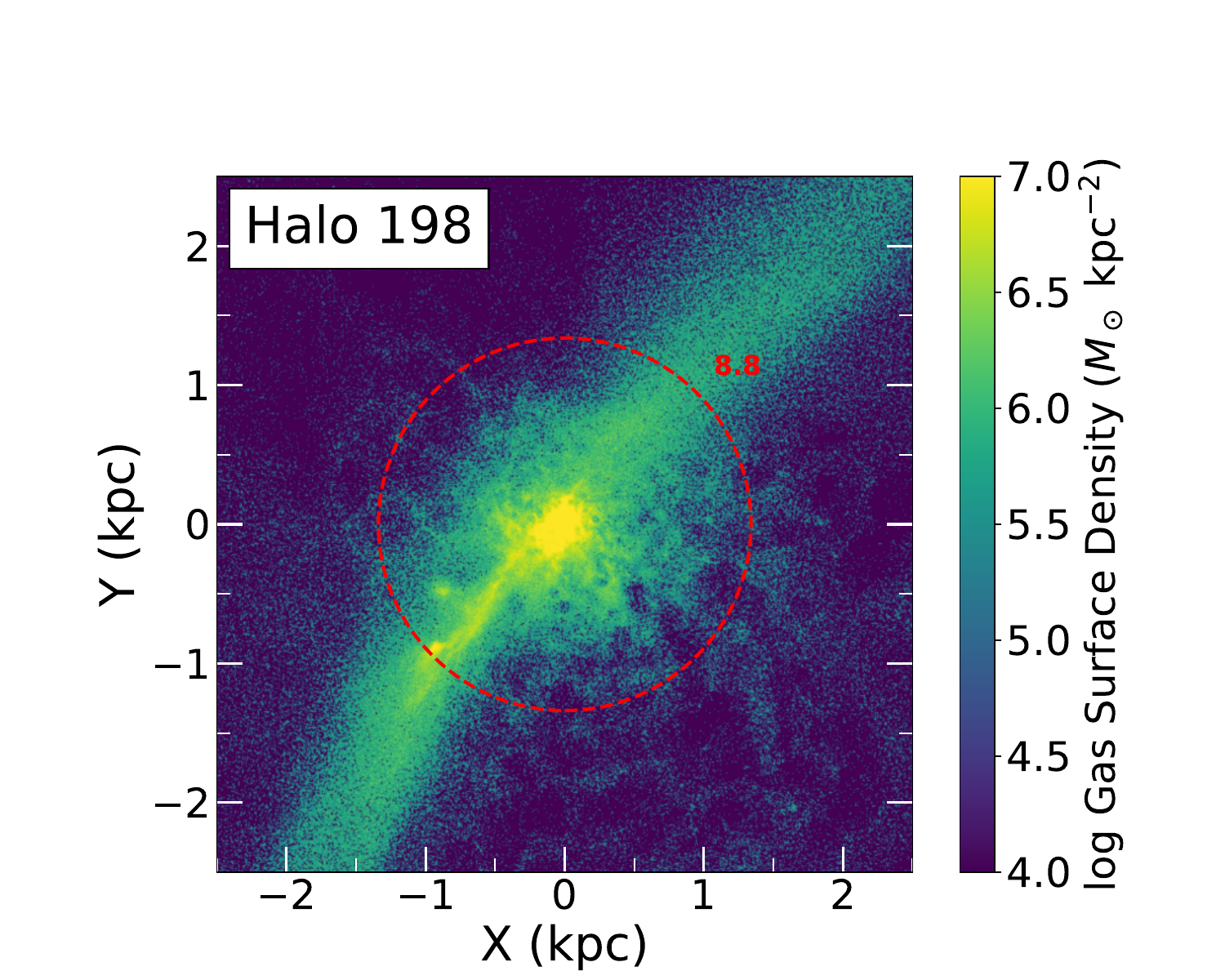}
    \includegraphics[width=0.33\textwidth]{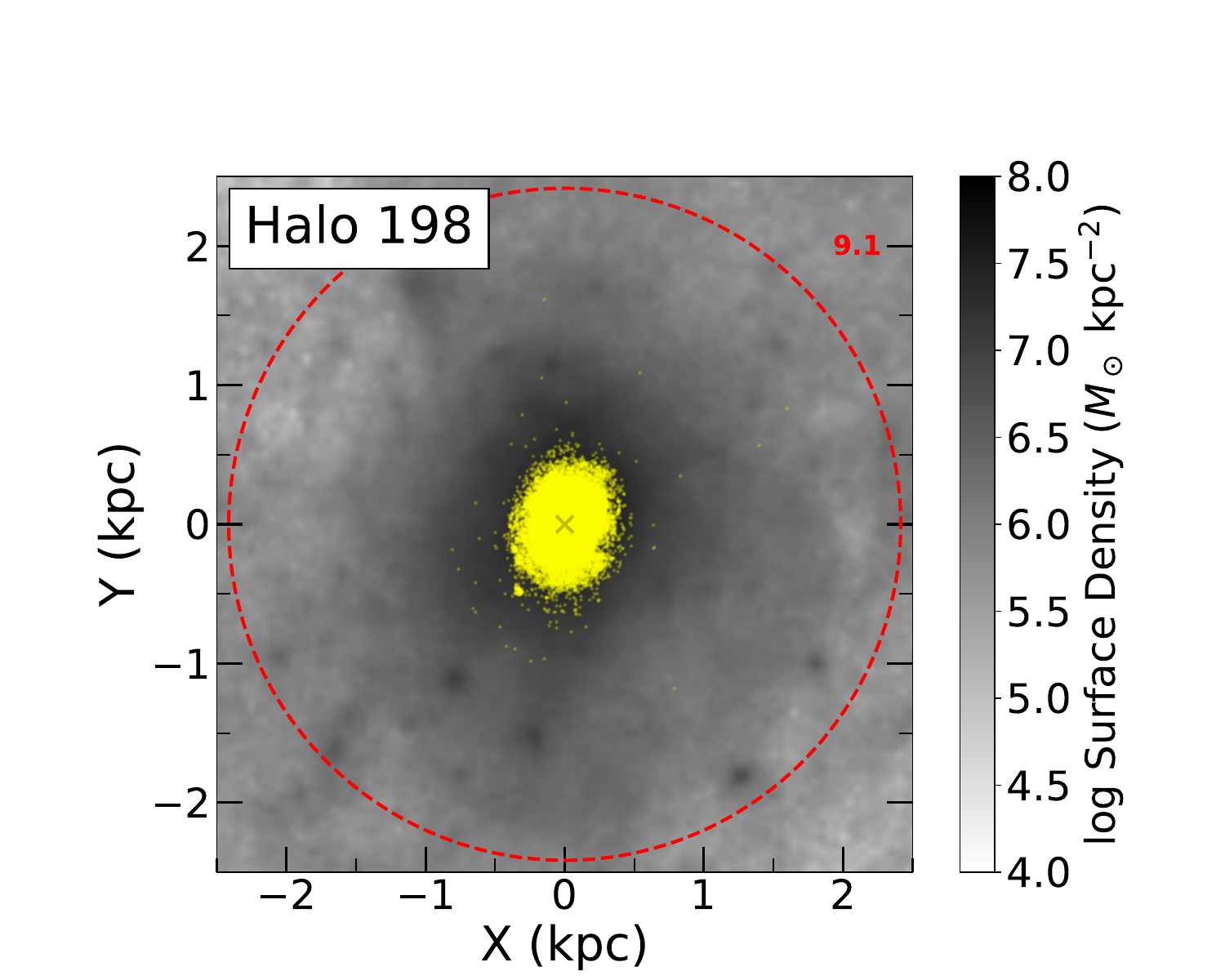}\\
    \includegraphics[width=0.33\textwidth]{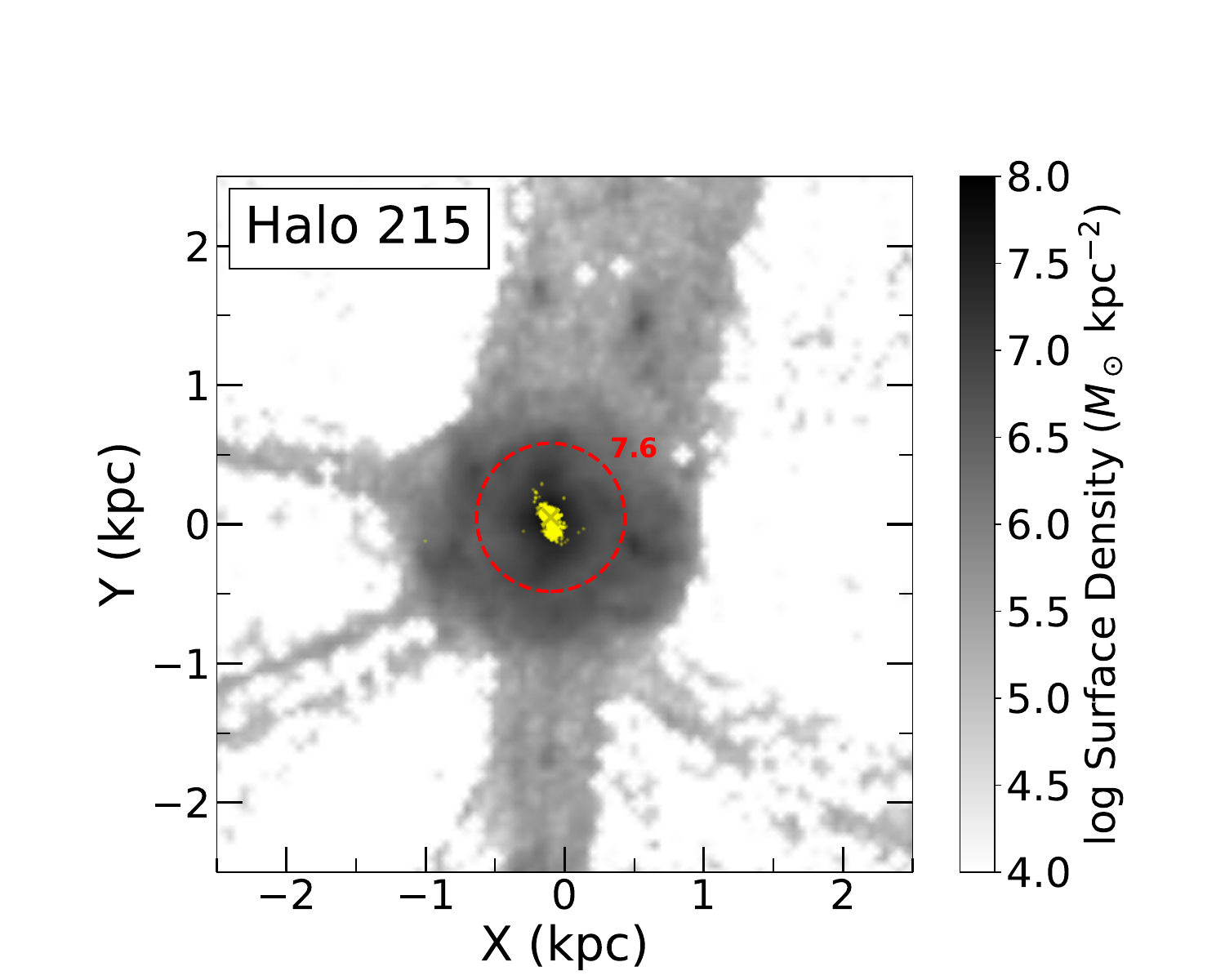}
    \includegraphics[width=0.33\textwidth]{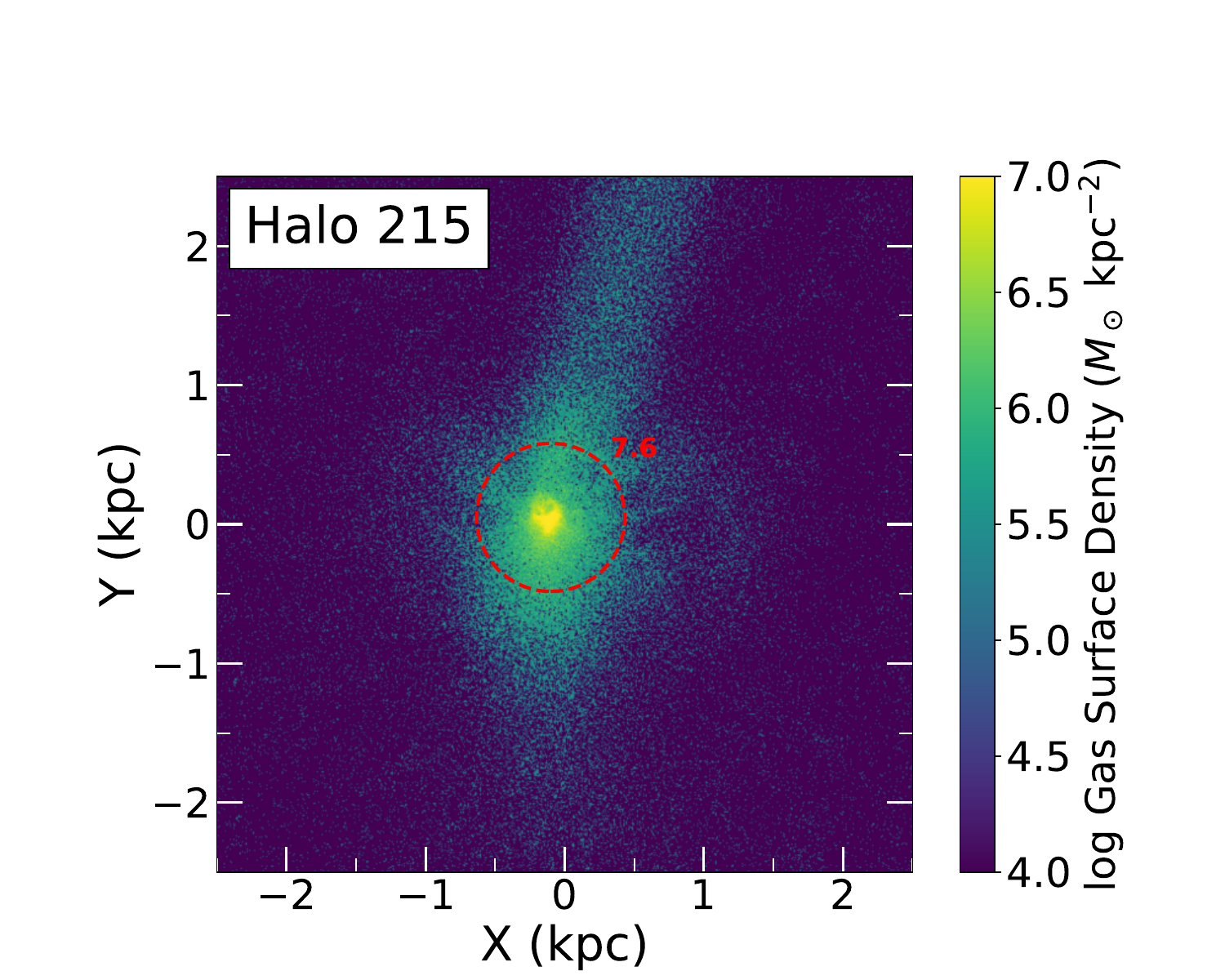}
    \includegraphics[width=0.33\textwidth]{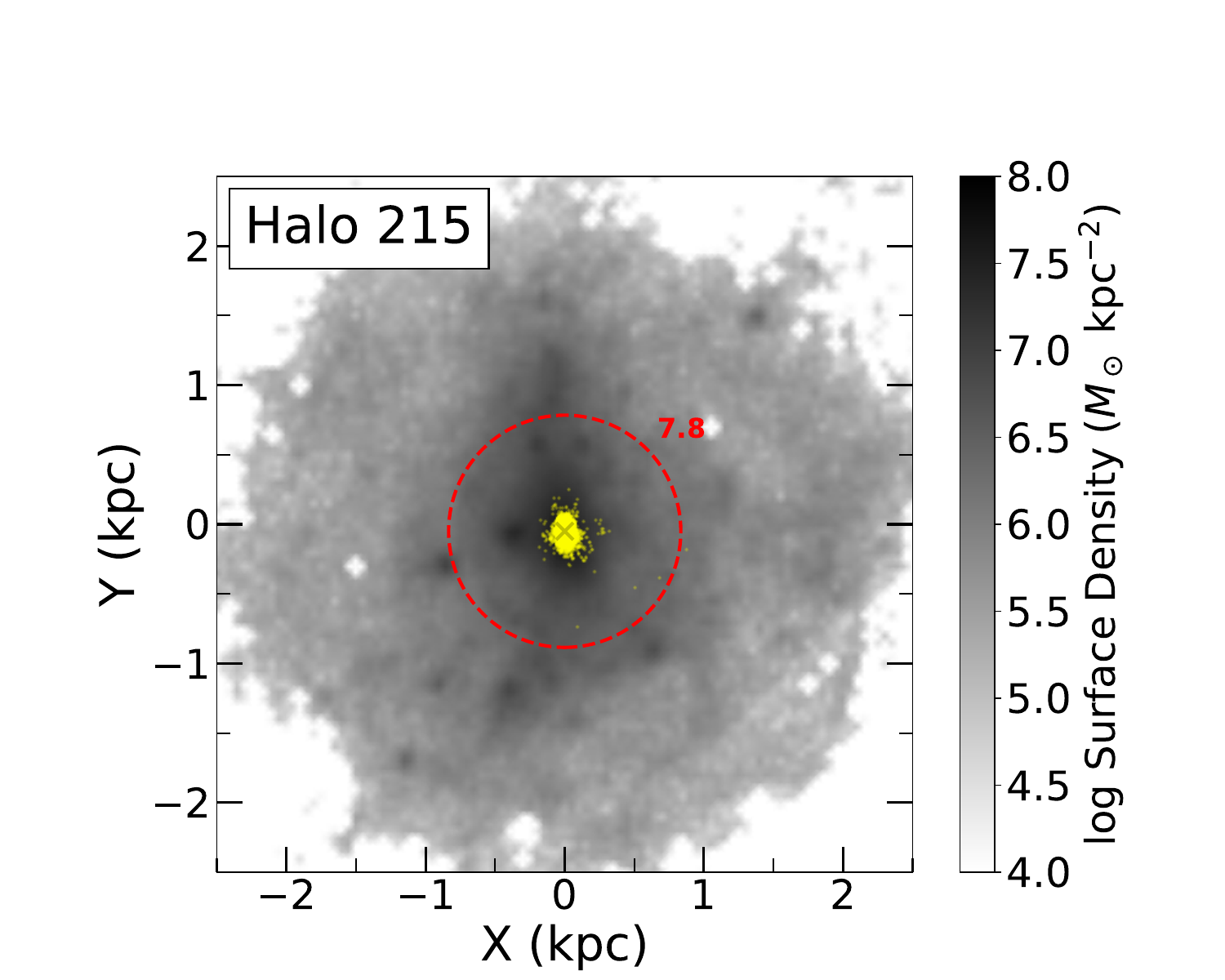}\\
    \includegraphics[width=0.33\textwidth]{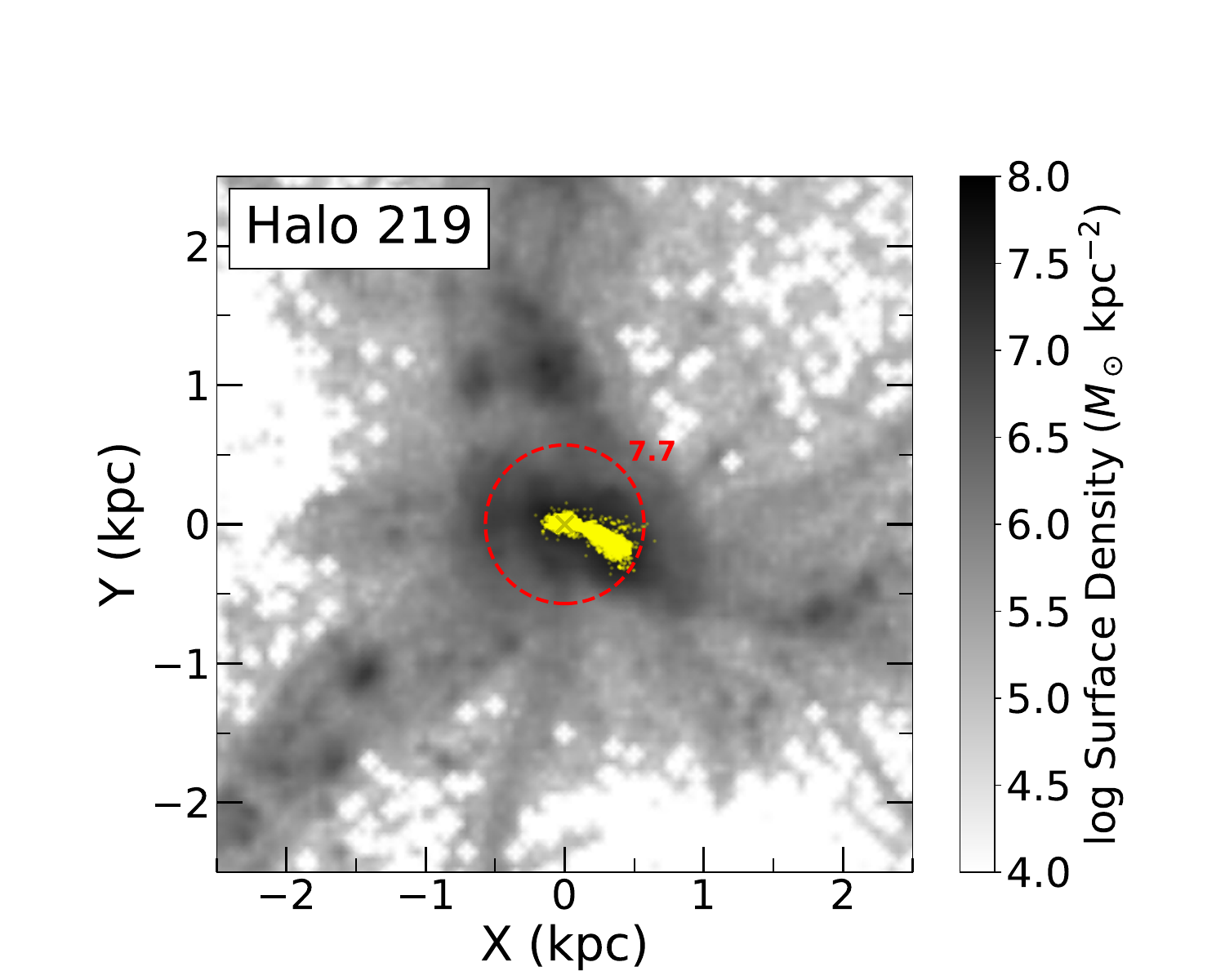}
    \includegraphics[width=0.33\textwidth]{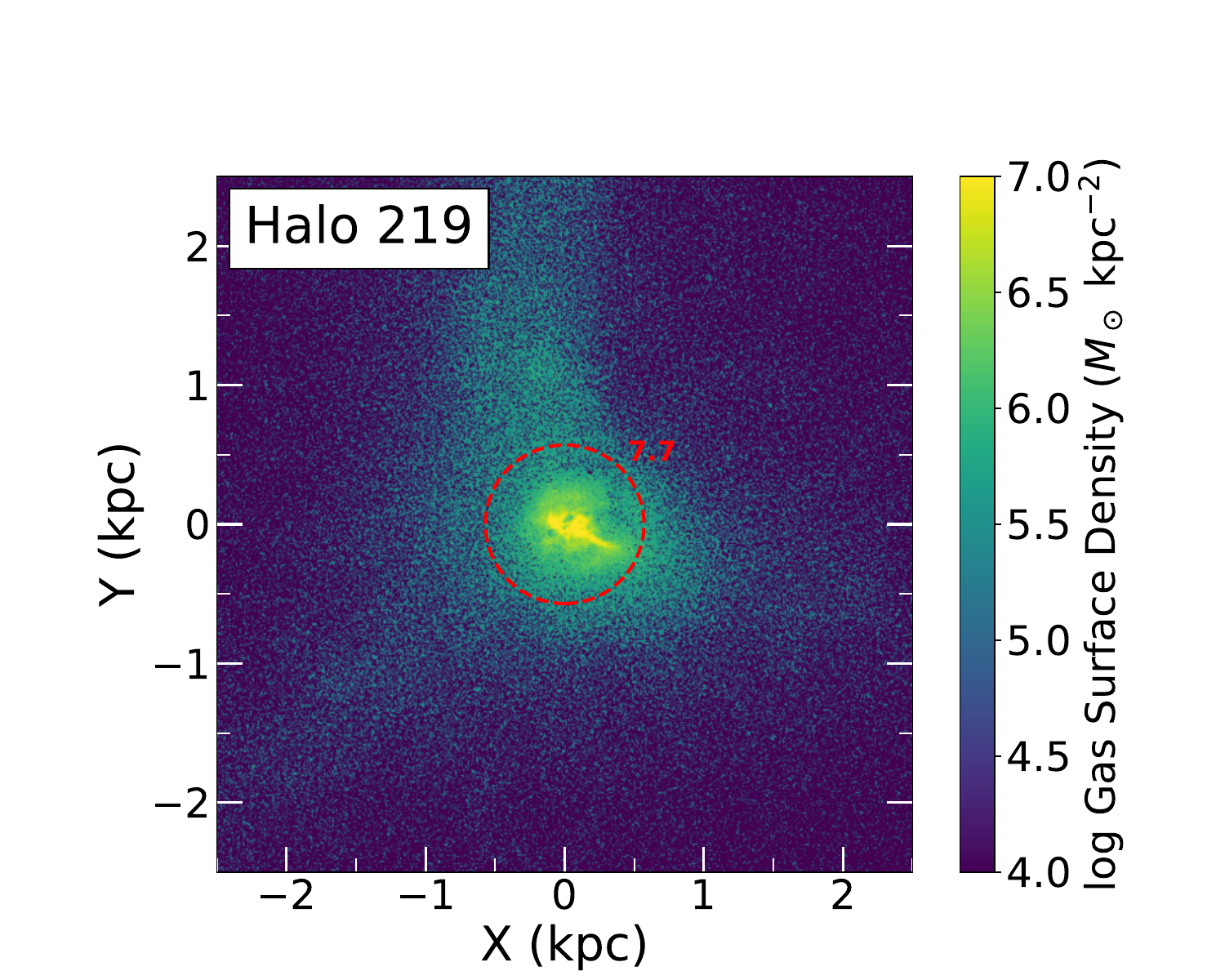}
    \includegraphics[width=0.33\textwidth]{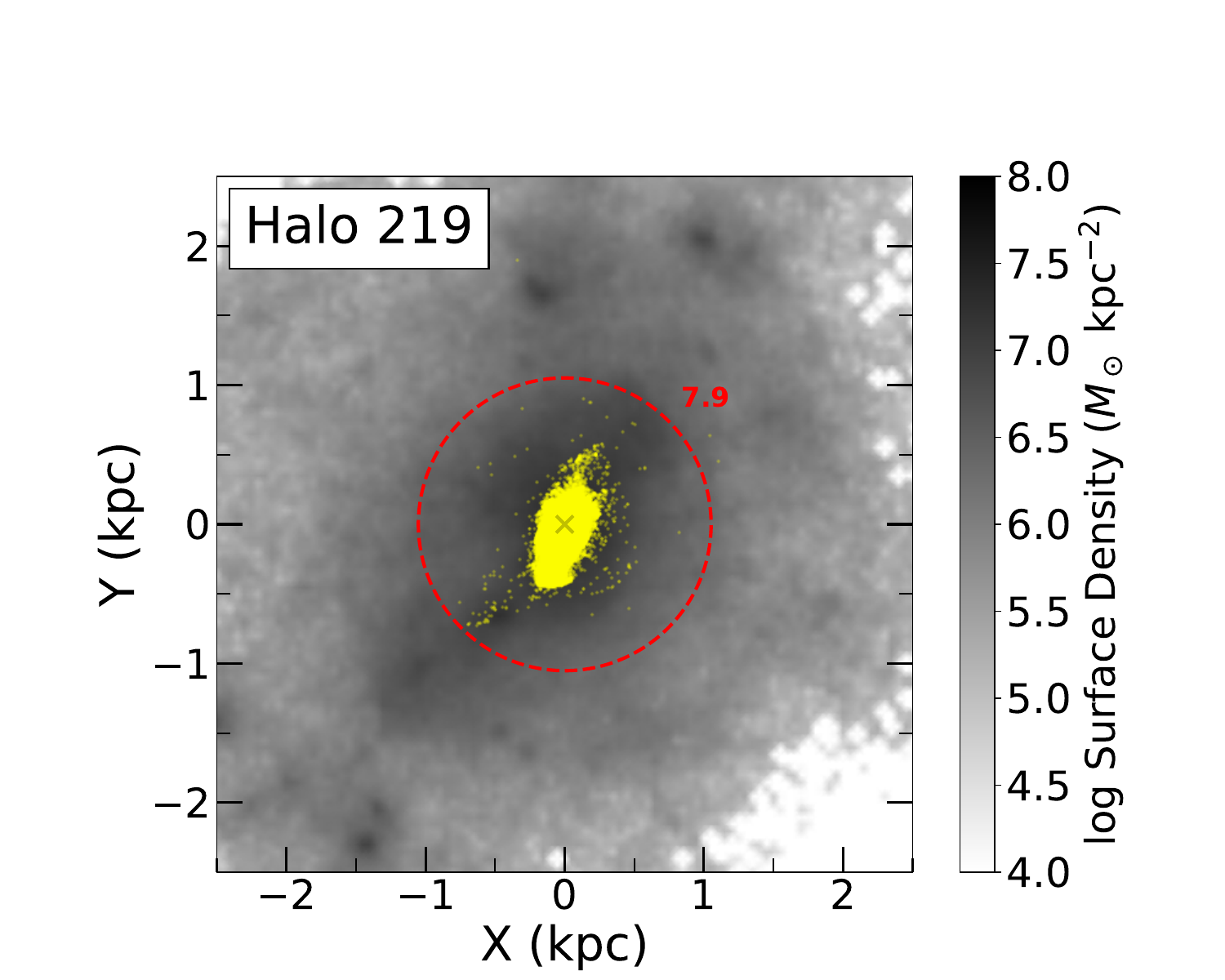}\\
    \includegraphics[width=0.33\textwidth]{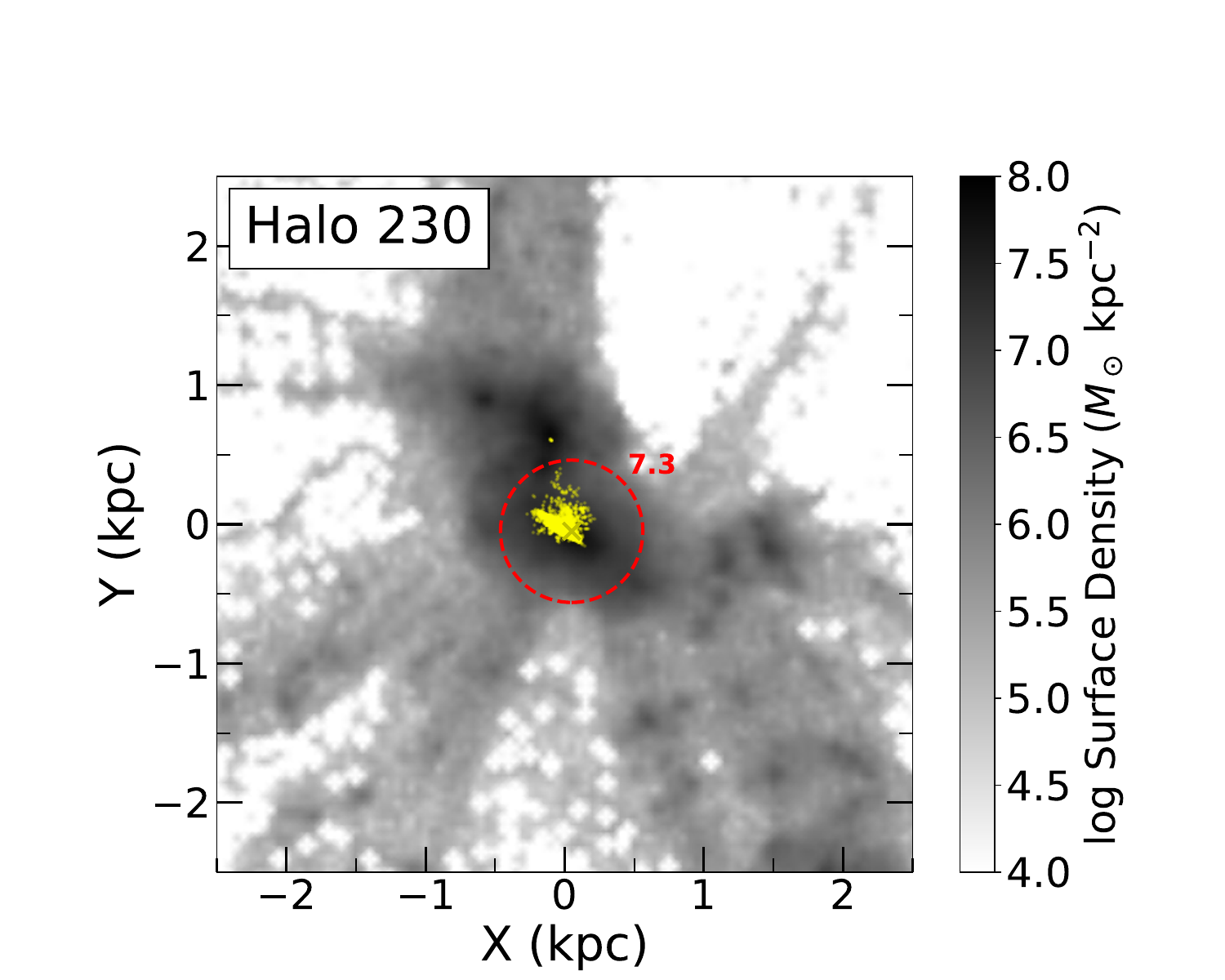}
    \includegraphics[width=0.33\textwidth]{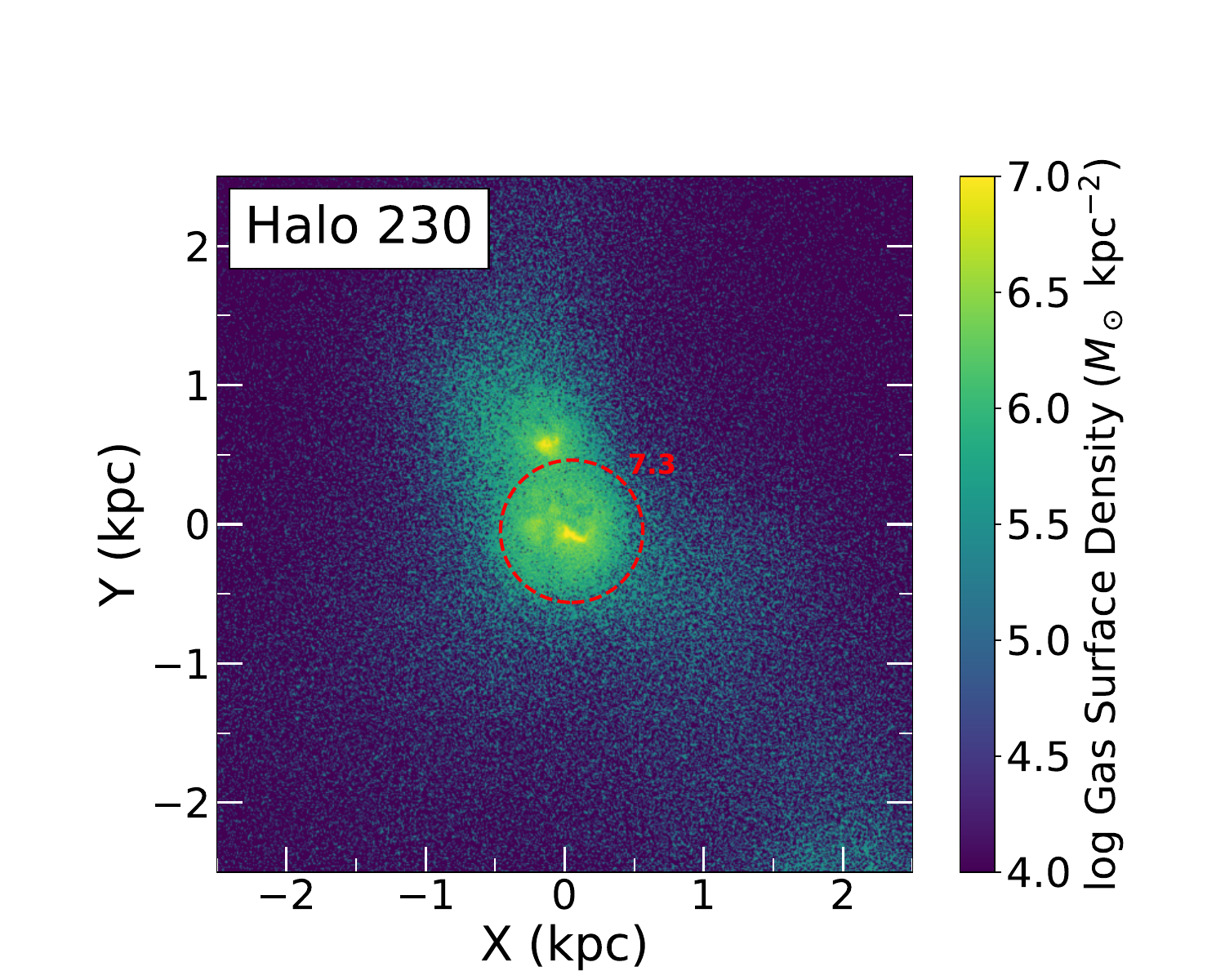}
    \includegraphics[width=0.33\textwidth]{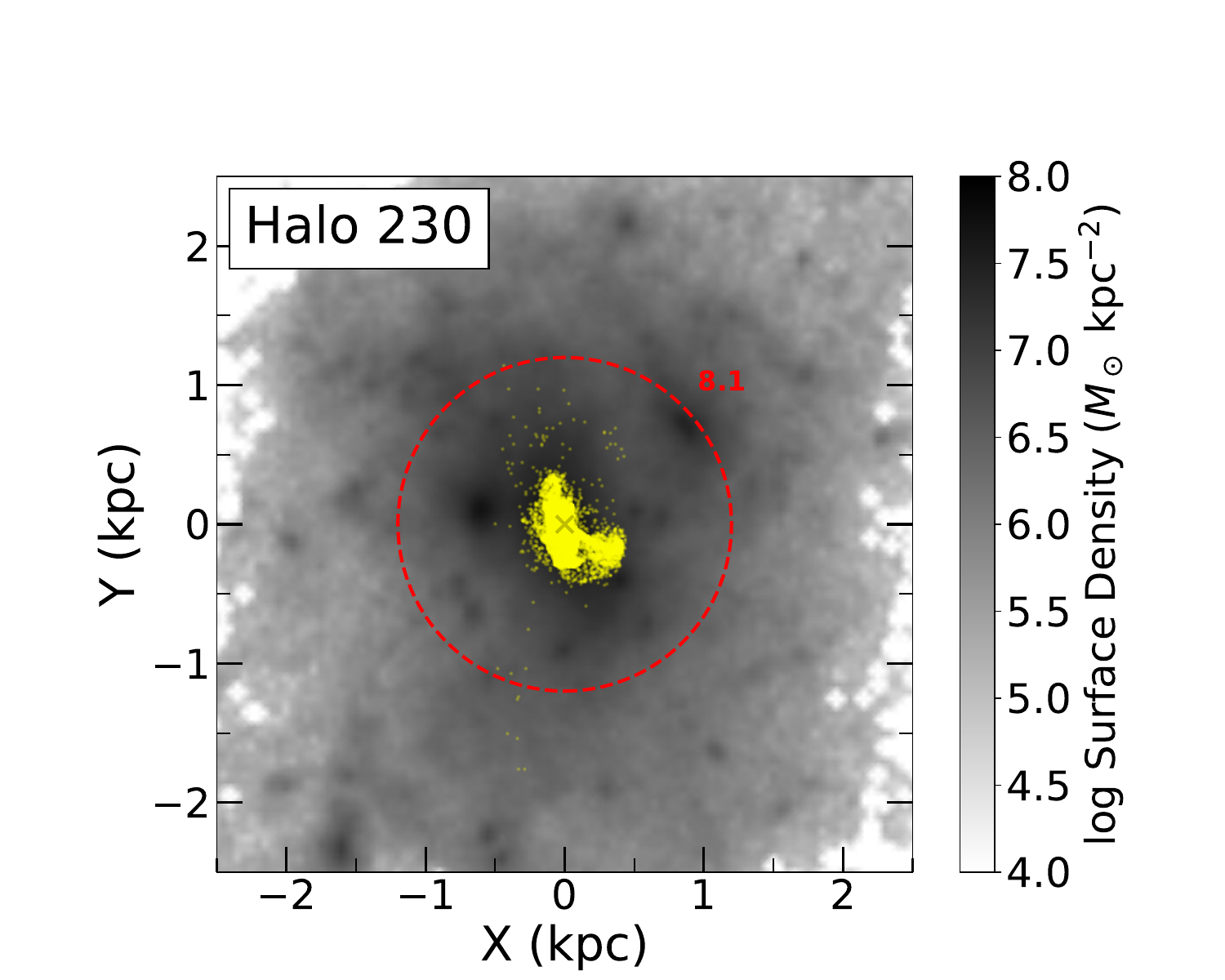}
    \end{center}
  \caption{Snapshots of Halos 198, 215, 219, and 230 (rows from top to bottom). The left and middle panels display the properties at the EoR: DM surface density (grayscale) with stellar distribution (yellow dots), and cold gas surface density ($T<1000$\,K), respectively. The right panels show the DM and stellar distribution at the end of the simulation ($t=1.2$\,Gyr). The red circles and inscribed numbers indicate the virial radius and virial mass, while yellow crosses mark the halo centers.
  {Alt text: Four rows of halo snapshots (Halos 198, 215, 219, 230). Each row shows three panels: left = dark-matter surface density with stellar positions dots; middle = cold gas surface density (T<1000 K) colormap; right = dark-matter and stellar distribution at simulation end (t = 1.2 Gyr). Red dashed circle and number annotate the virial radius and virial mass; yellow cross marks the halo center.}  
  }
    \label{fig:snapshots1}
\end{figure*}

\begin{figure*}
    \begin{center}
    \includegraphics[width=0.33\textwidth]{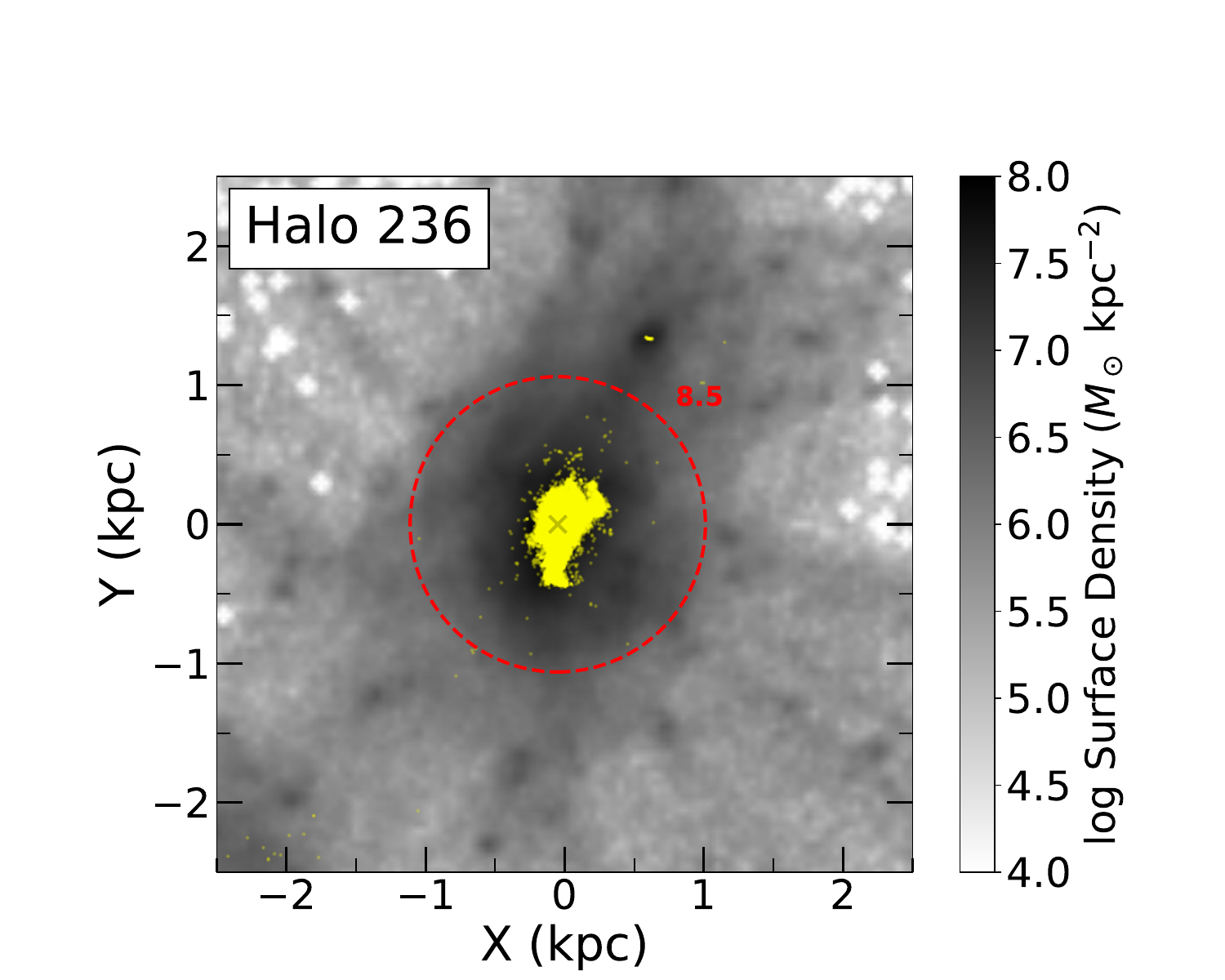}
    \includegraphics[width=0.33\textwidth]{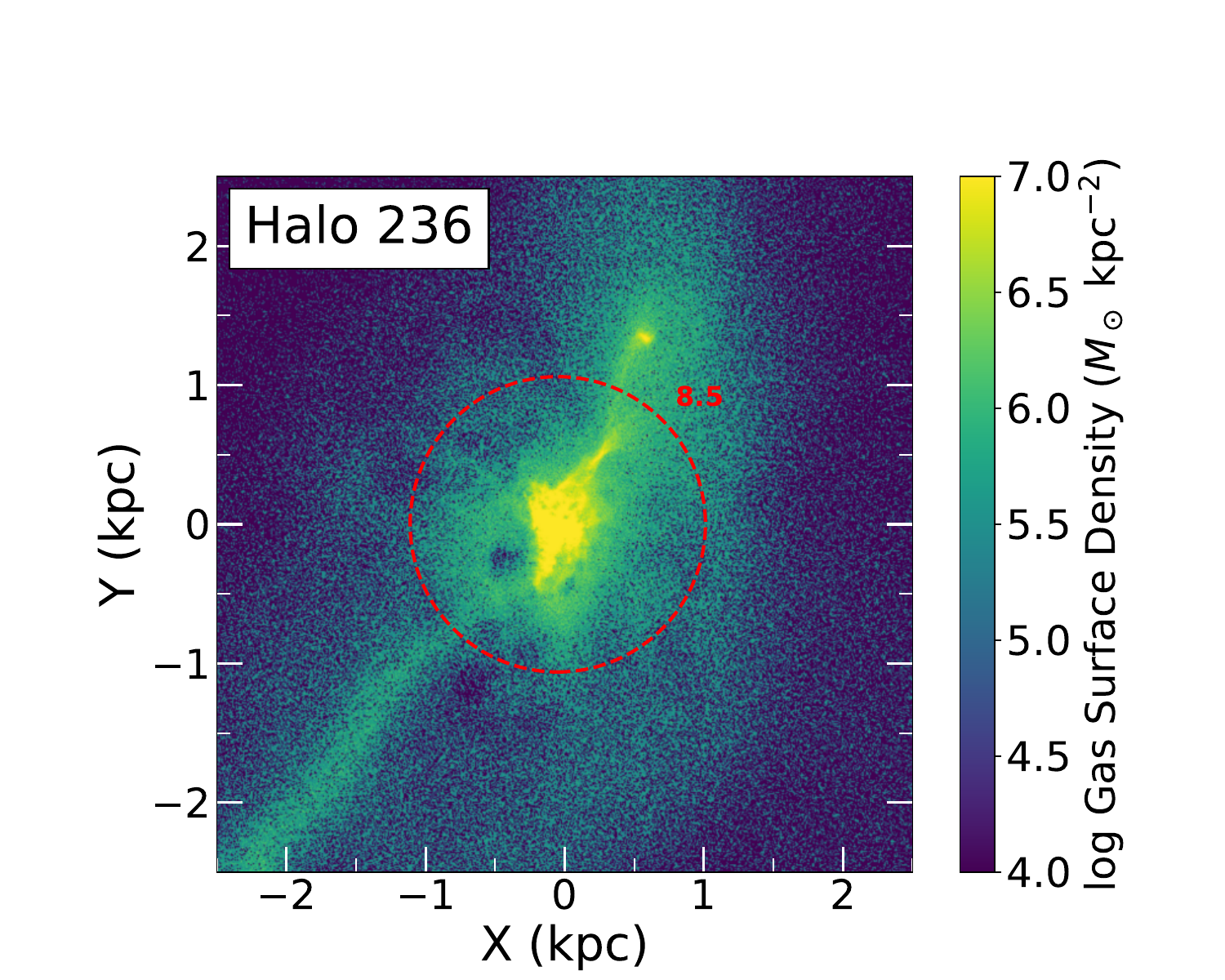}
    \includegraphics[width=0.33\textwidth]{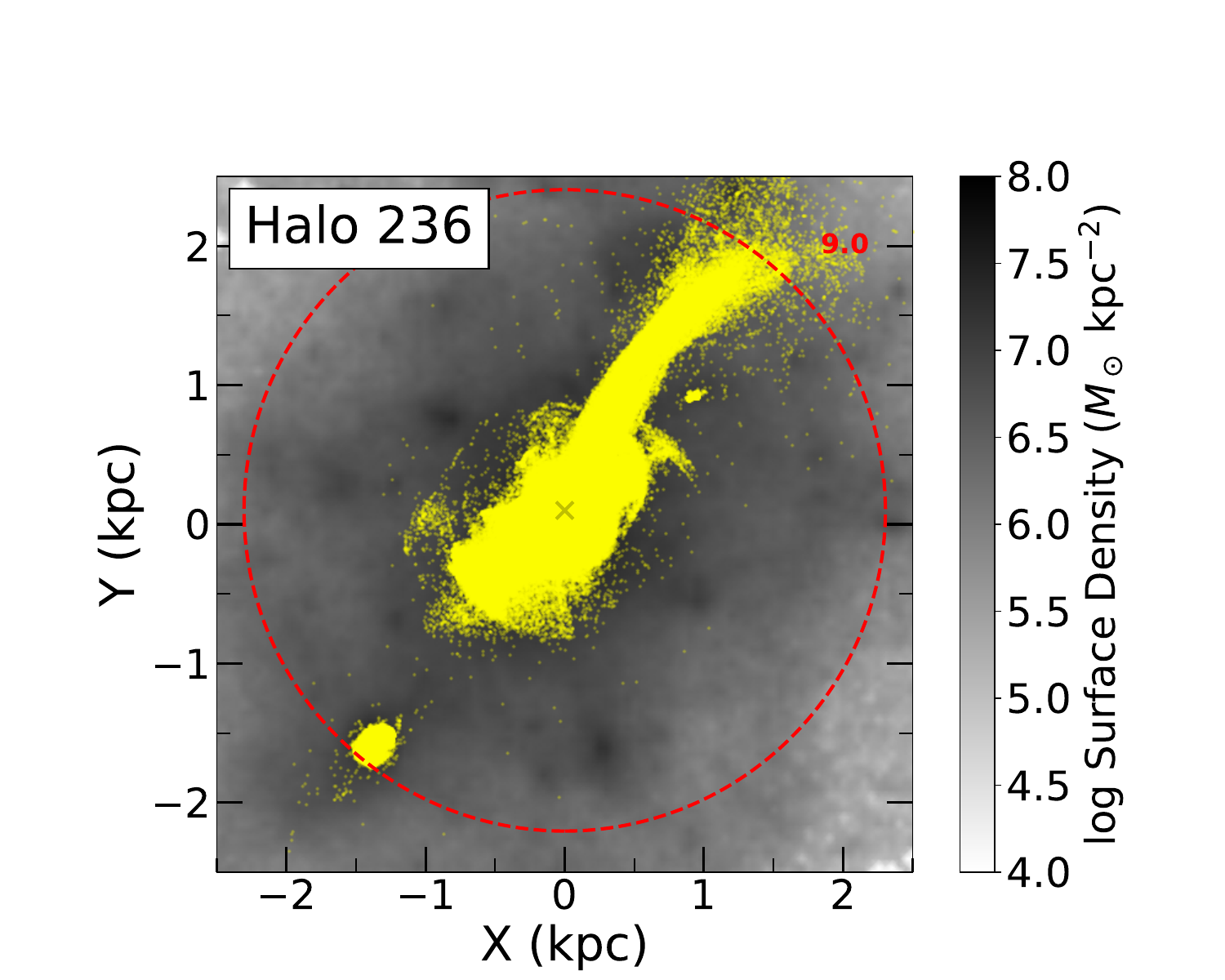}\\
    \includegraphics[width=0.33\textwidth]{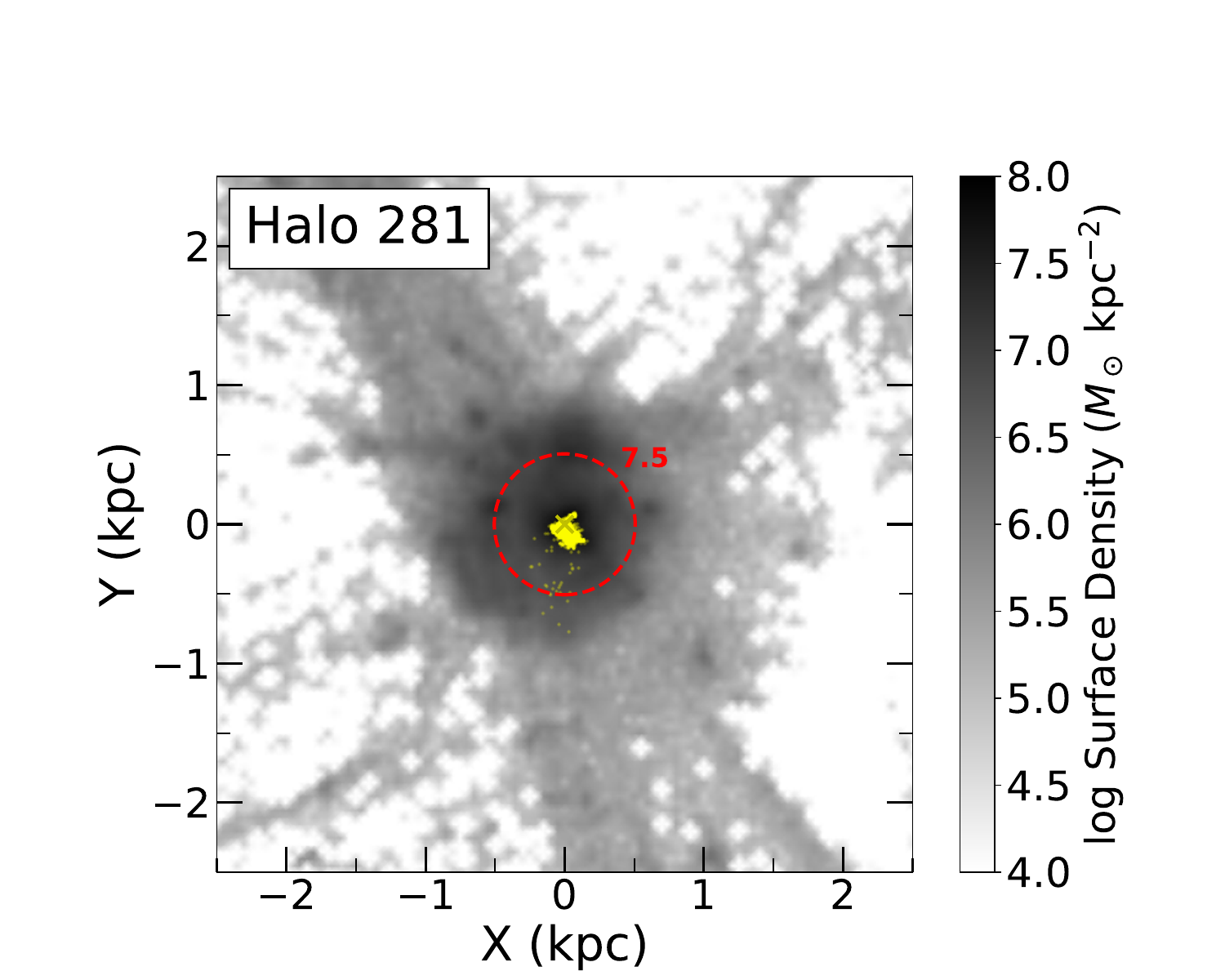}
    \includegraphics[width=0.33\textwidth]{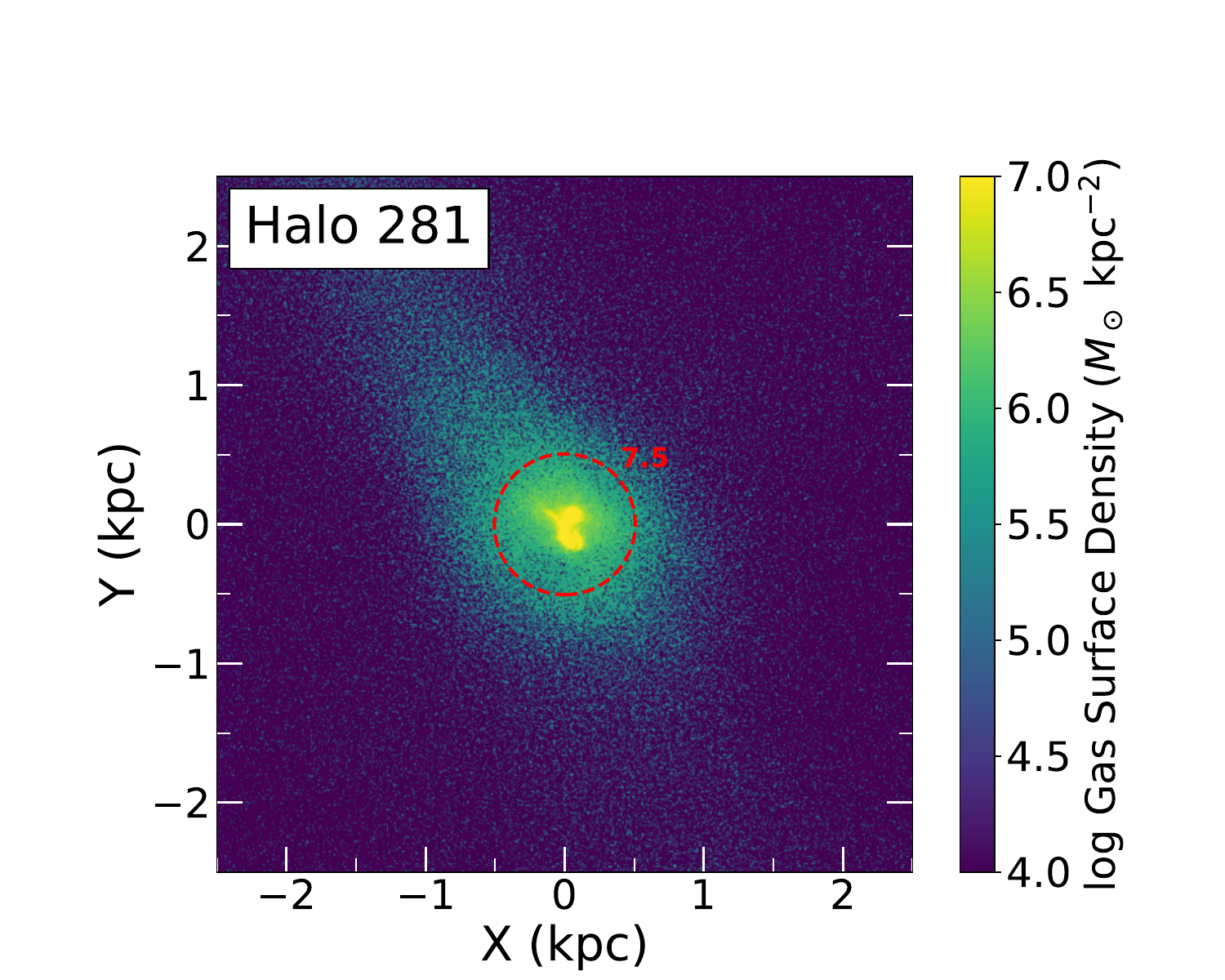}
    \includegraphics[width=0.33\textwidth]{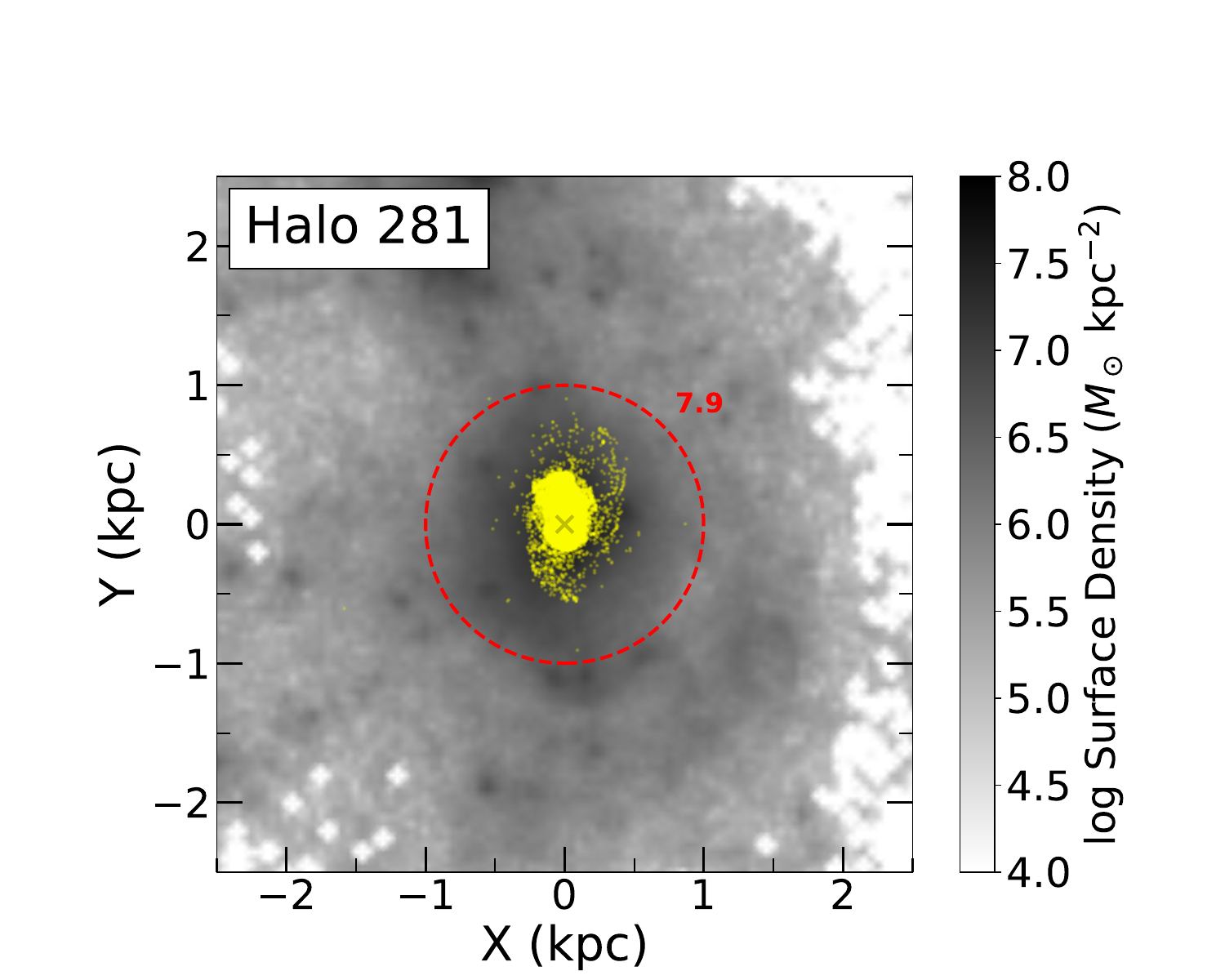}\\
    \includegraphics[width=0.33\textwidth]{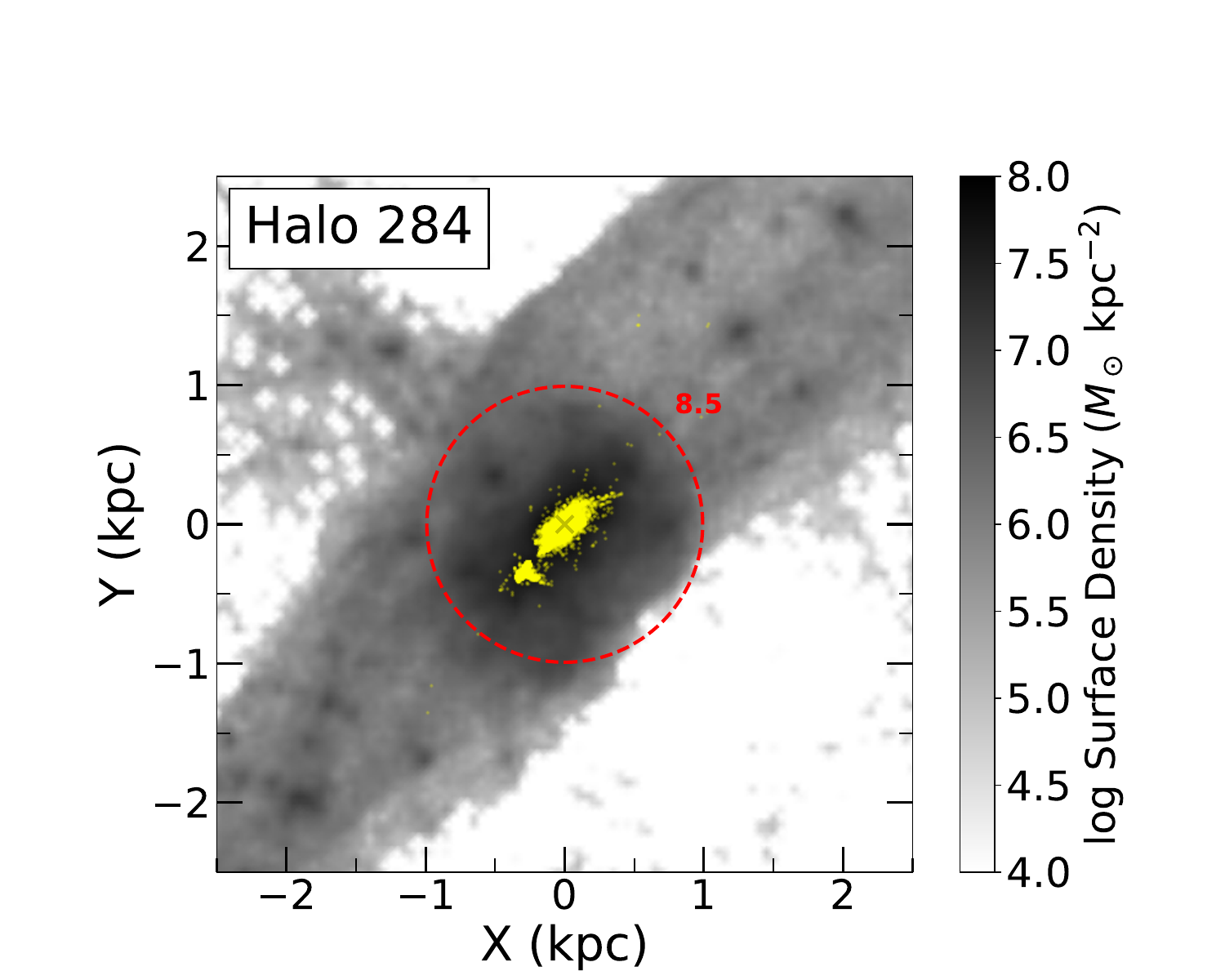}
    \includegraphics[width=0.33\textwidth]{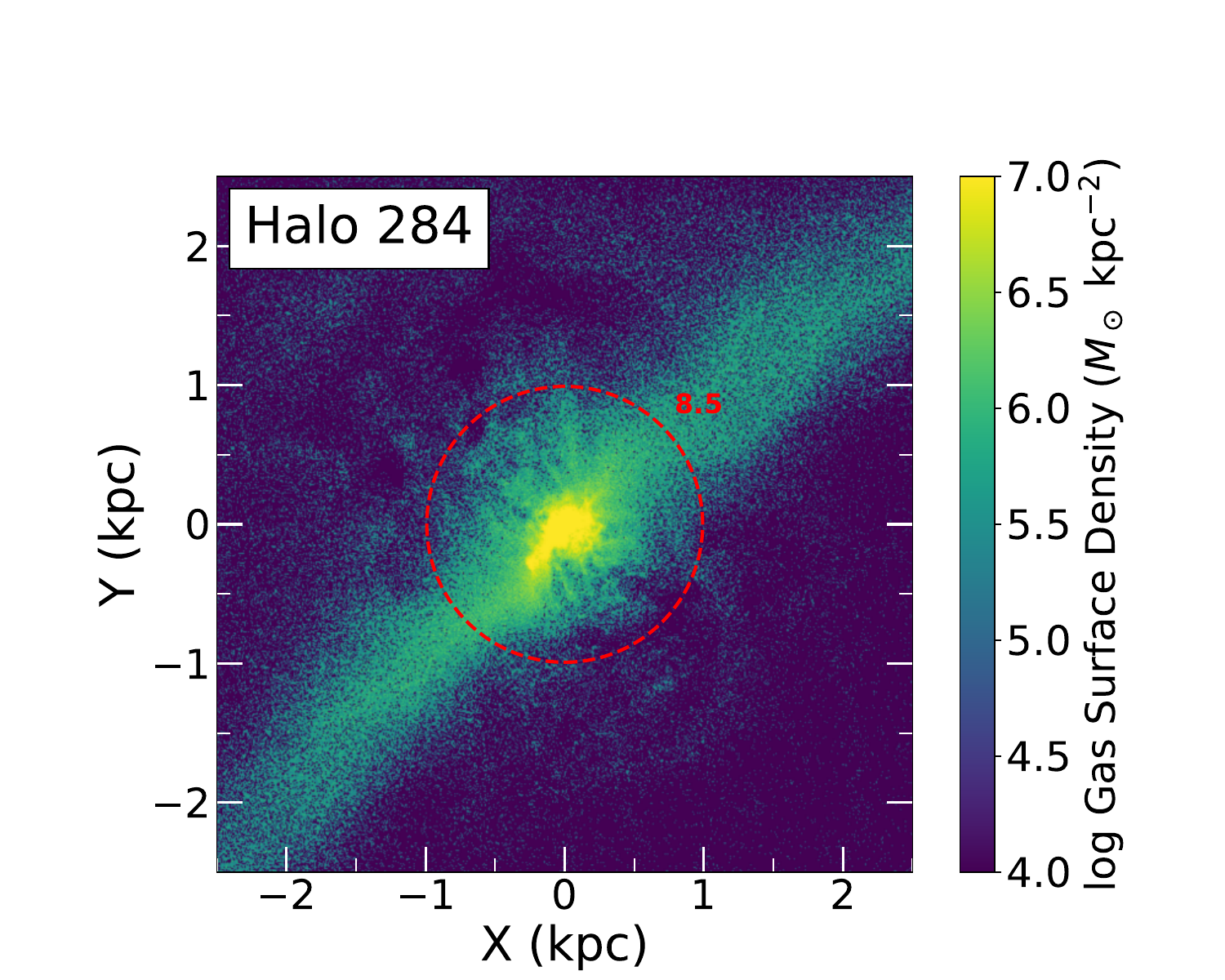}
    \includegraphics[width=0.33\textwidth]{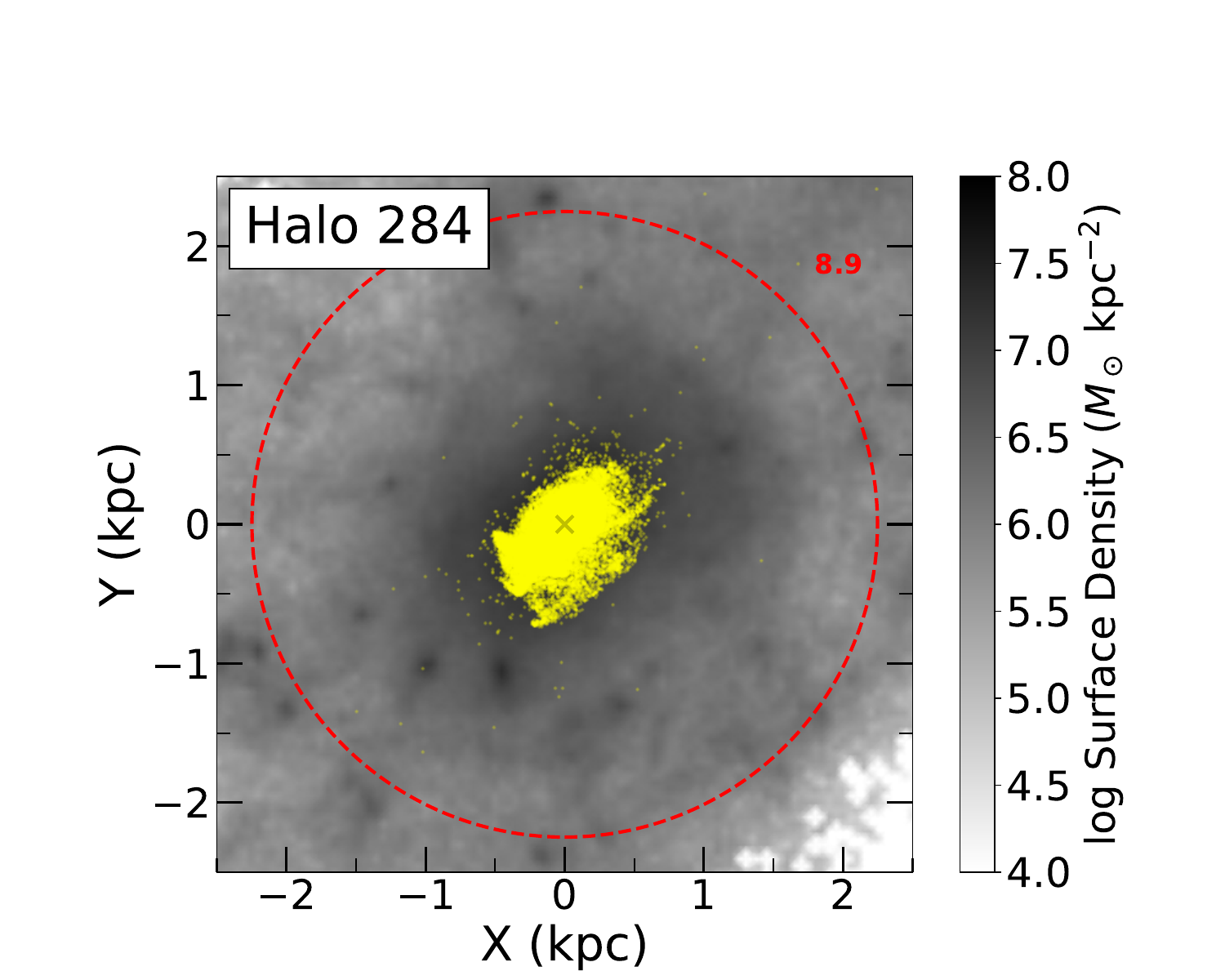}\\
    \includegraphics[width=0.33\textwidth]{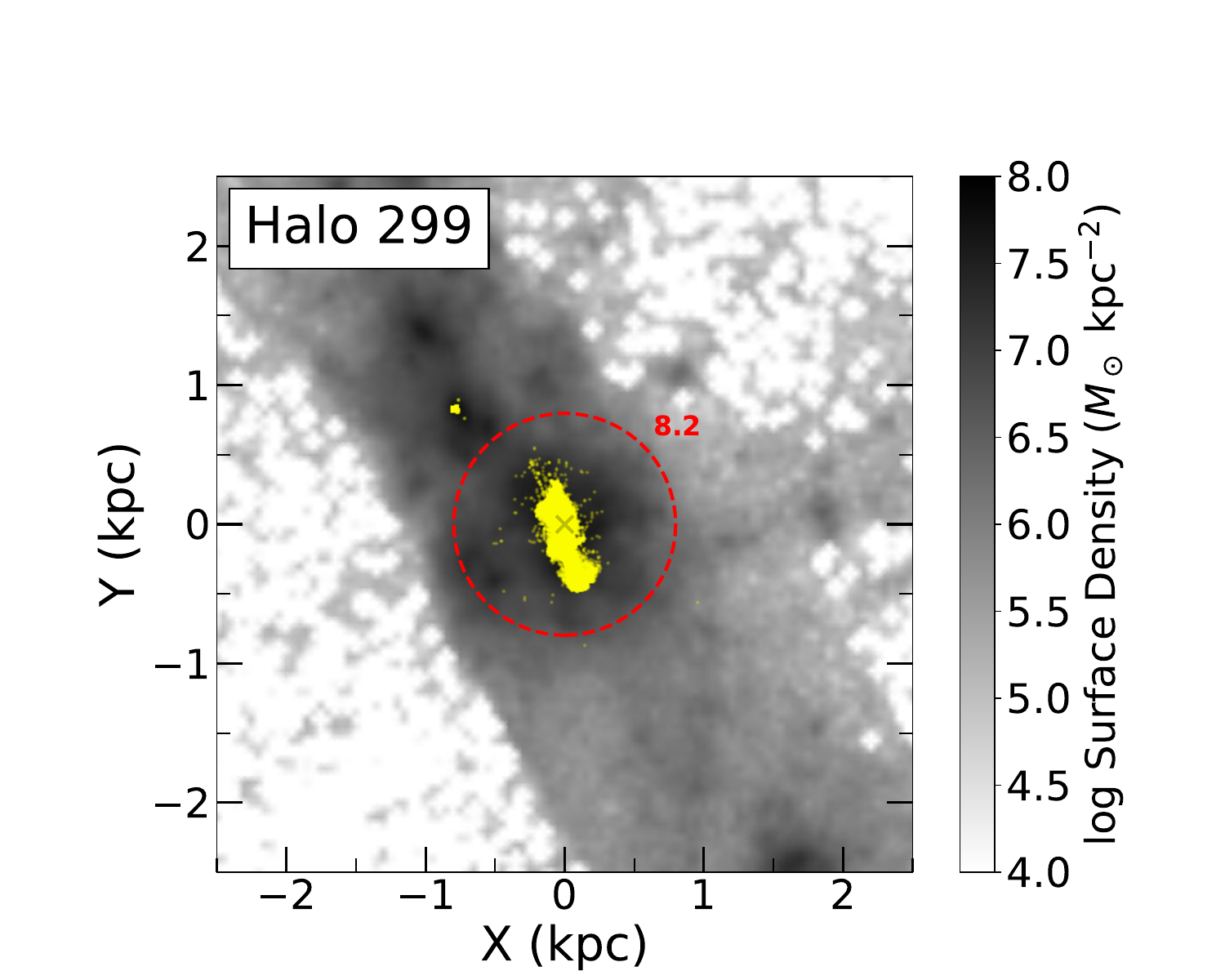}
    \includegraphics[width=0.33\textwidth]{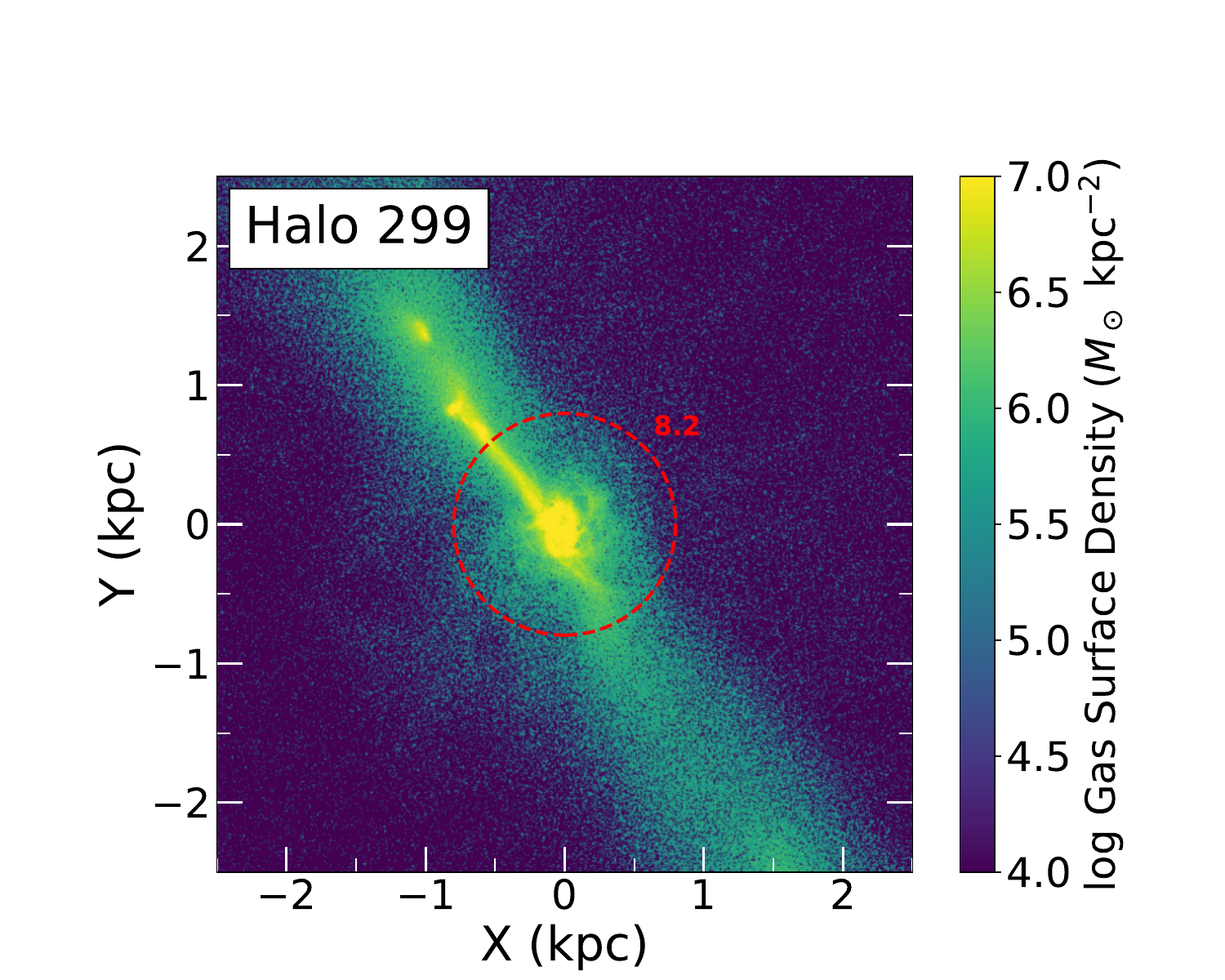}
    \includegraphics[width=0.33\textwidth]{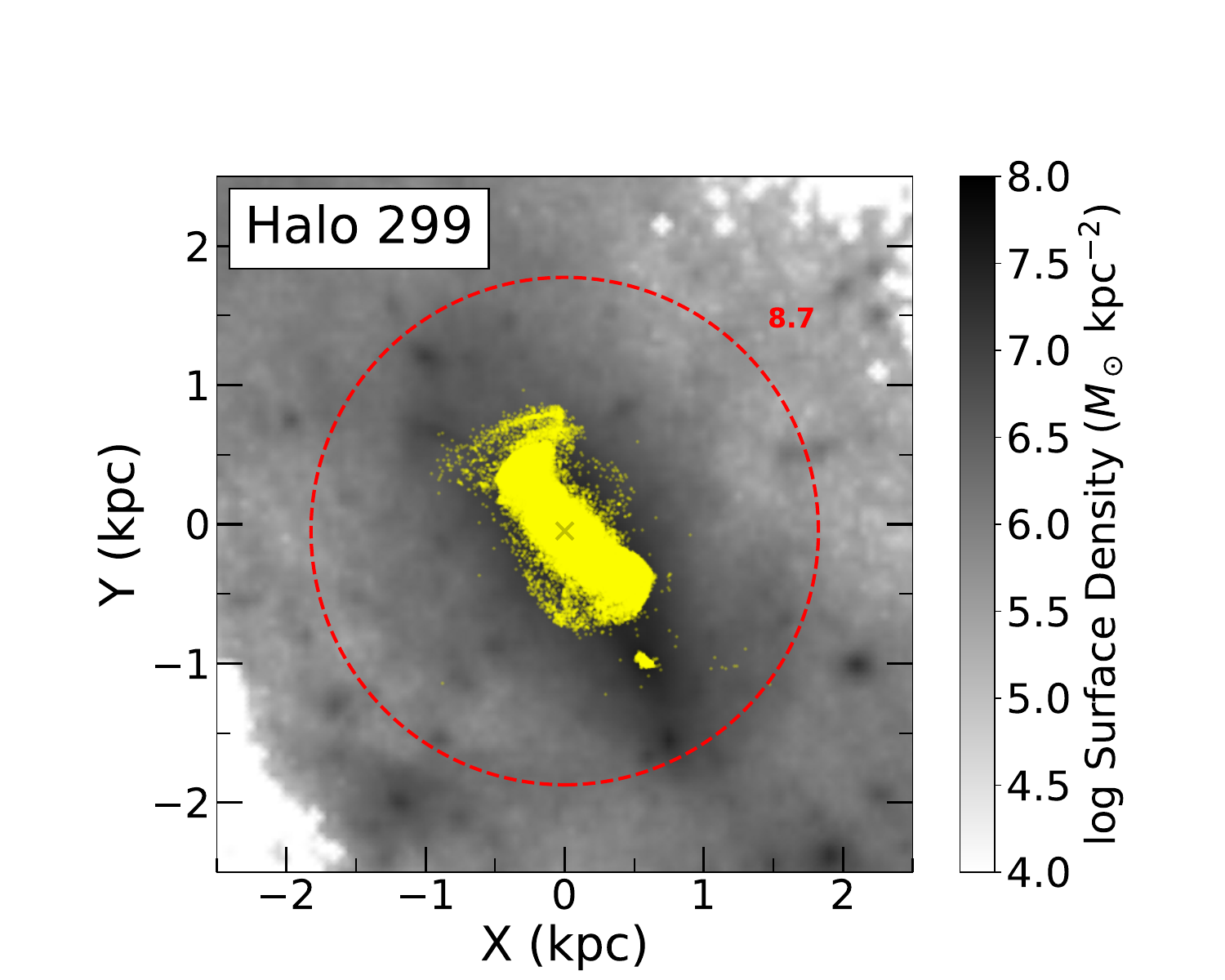}
    \end{center}
    \caption{Same as Fig.~\ref{fig:snapshots1}, but for Halos 236, 281, 284, and 299.
     {Alt text: Four rows of halo snapshots (Halos 236, 281, 284, 299). Each row shows three panels: left = dark-matter surface density with stellar positions dots; middle = cold gas surface density (T<1000 K) colormap; right = dark-matter and stellar distribution at simulation end (t = 1.2 Gyr). Red dashed circle and number annotate the virial radius and virial mass; yellow cross marks the halo center.}}
    \label{fig:snapshots2}
\end{figure*}

Figs.~\ref{fig:snapshots1} and \ref{fig:snapshots2} present the snapshots of the eight halos at the epoch of reionization (EoR) and at the end of the simulations ($t=1.2$\,Gyr). 
Each halo is visualized using two plots: the projected 2D distribution of dark matter (DM) and the 2D distribution of gas. 
Generally, the halos formed from small structures and gradually assembled into larger features such as voids, filaments, and clumps. The gas distribution followed the DM distribution, particularly along the DM filaments, where gas flows toward the central regions of the halo.

\begin{figure}
\begin{center}
\includegraphics[width=0.48\textwidth]{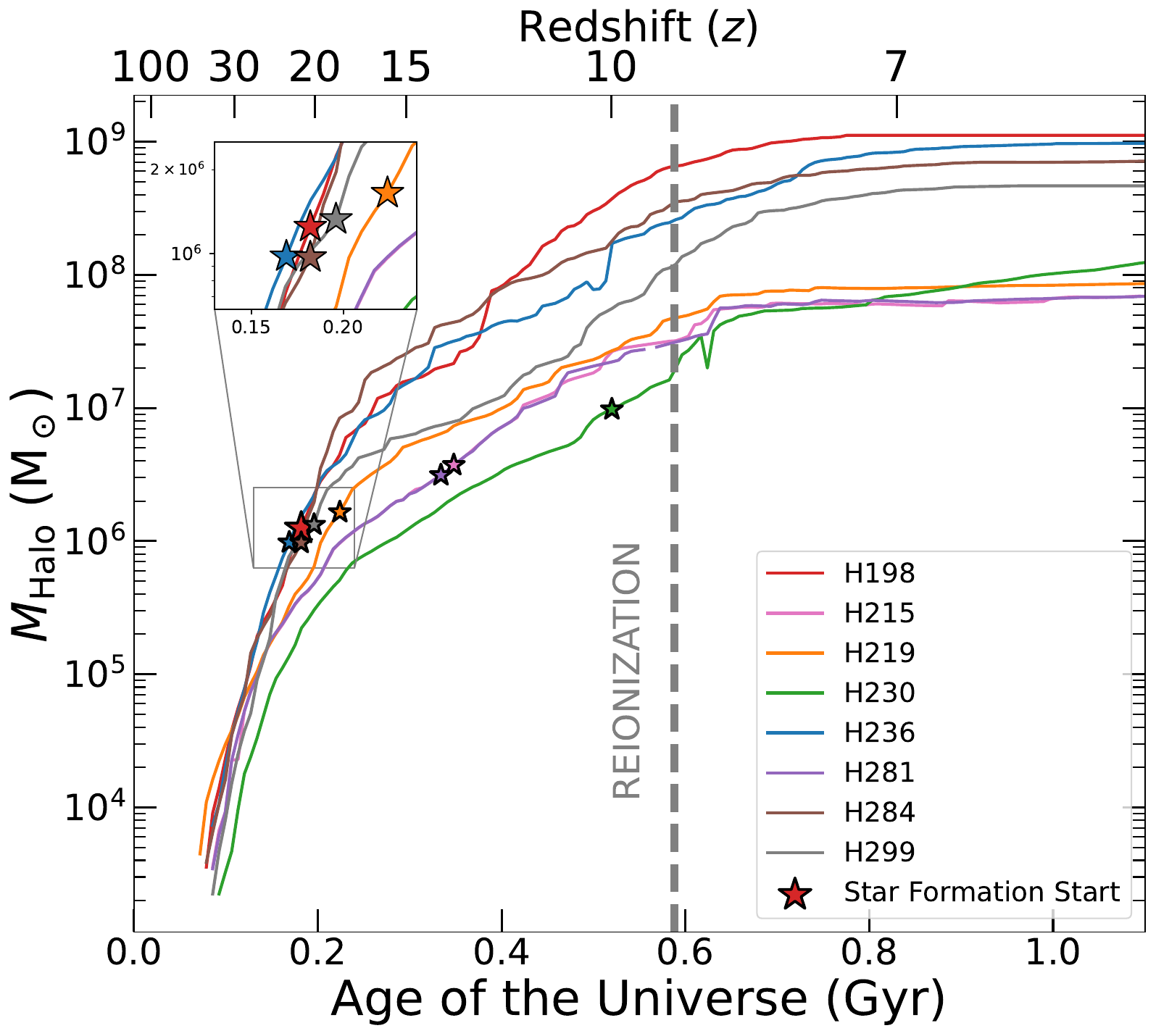}
\end{center}
\caption{Dark matter halo mass evolution. The lower and upper horizontal axes indicate cosmic age in Gyr and redshift ($z$), respectively, while the vertical axis displays halo mass in solar masses ($\rm\ M_{\odot}$). Each colored curve represents the mass assembly history of an individual halo; the color coding is consistent across all figures to facilitate comparison in this paper. Star markers denote the onset of star formation in each halo. The vertical dashed line marks the reference epoch of reionization ($z=8.5$) {Alt text: Line plot of dark-matter halo mass evolution for several halos from cosmic age 0 to about 1.2 Gyr. Each colored curve is one halo’s mass assembly history in $\rm log_10$ solar masses (vertical axis $10^4 \rm\ to\ 10^9$). Star symbols mark the onset of star formation in each halo; a vertical dashed line labeled REIONIZATION indicates the reference epoch ($z = 8.5$). An inset zoom at early times highlights rapid early growth differences. The legend identifies halo IDs (e.g., H198, H215, H219, H230).}}
\label{fig:Halomass}
\end{figure}

\rfig{fig:Halomass} presents the mass assembly histories of the eight halos. The halo mass is defined as the virial mass computed using the \texttt{AHF}. 
The halo mass growth histories exhibit substantial diversity. 
The star formation starts at around the time when the halo mass exceeds $10^6\rm\ M_{\odot}$. In slow assembly halos such as Halos 215, 230, and 281, the mass for the first star formation tends to be delayed. 

Halos 198, 236, 284, and 299 began forming stars relatively early. There are four massive halos at the EoR, and their mass at the EoR was \(\gtrsim10^{8}\ \msun\). The others' halo mass remain \(\lesssim10^{7}\!-\!10^{8}\ \msun\) at the EoR. The halo mass at the end of the simulation looks bimodal, but the halo mass at $z=0$ is expected to be $\sim 10^9\rm\ M_{\odot}$ (see Table ~\ref{tab:HaloPropertyGadget}). Hereafter, we divide the eight halos into two groups: High-Mass Halos (Halos 198, 236, 284, and 299) with $M_{\rm Halo}>10^8\ \msun$ at the reionization and Low-Mass Halos (Halos 215, 219, 230, and 281) with $M_{\rm Halo}<10^8\ \msun$.

\begin{figure*}
\begin{center}
    \includegraphics[width=0.49\textwidth]{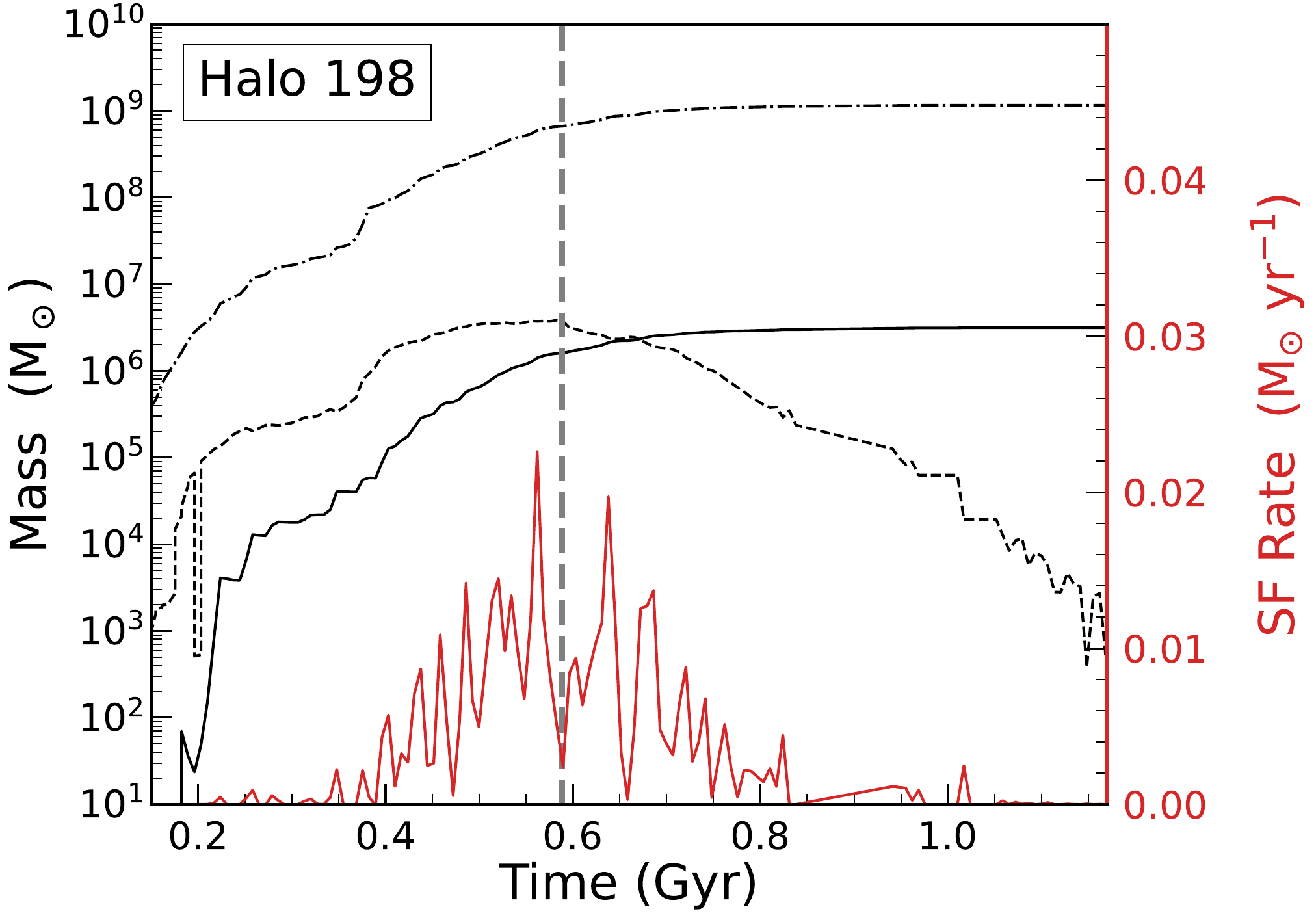}
    \includegraphics[width=0.49\textwidth]{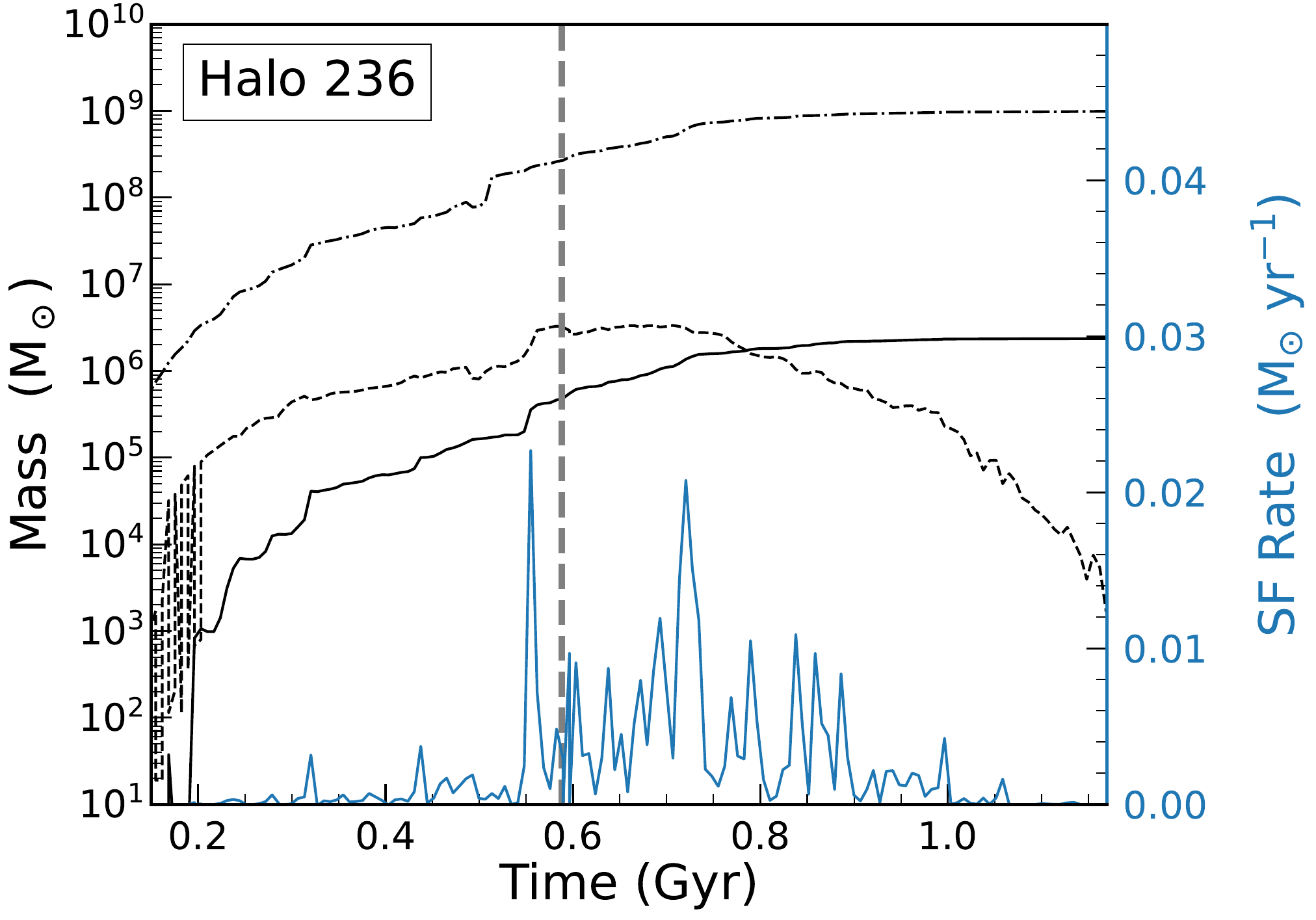}\\
    \includegraphics[width=0.49\textwidth]{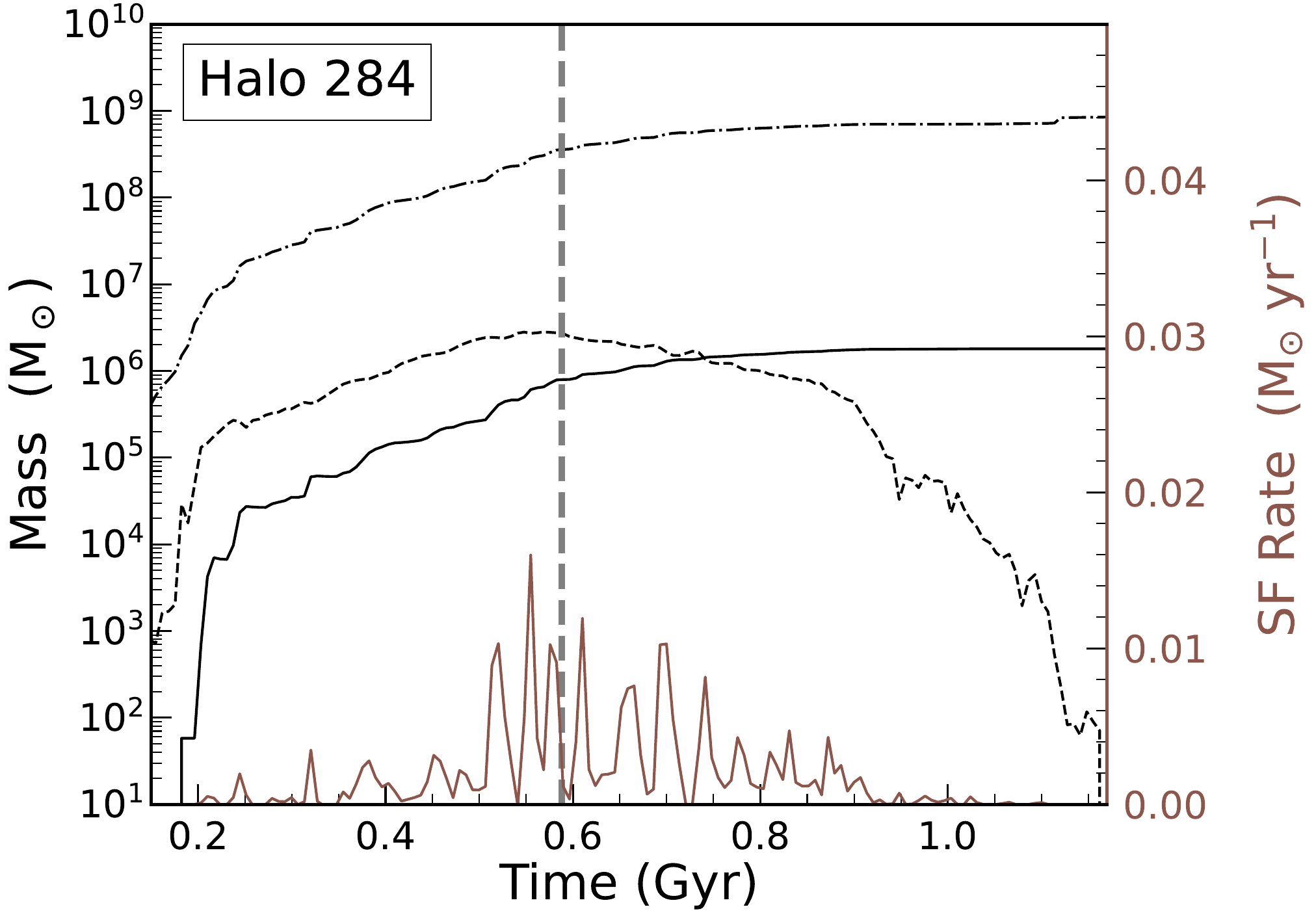}
    \includegraphics[width=0.49\textwidth]{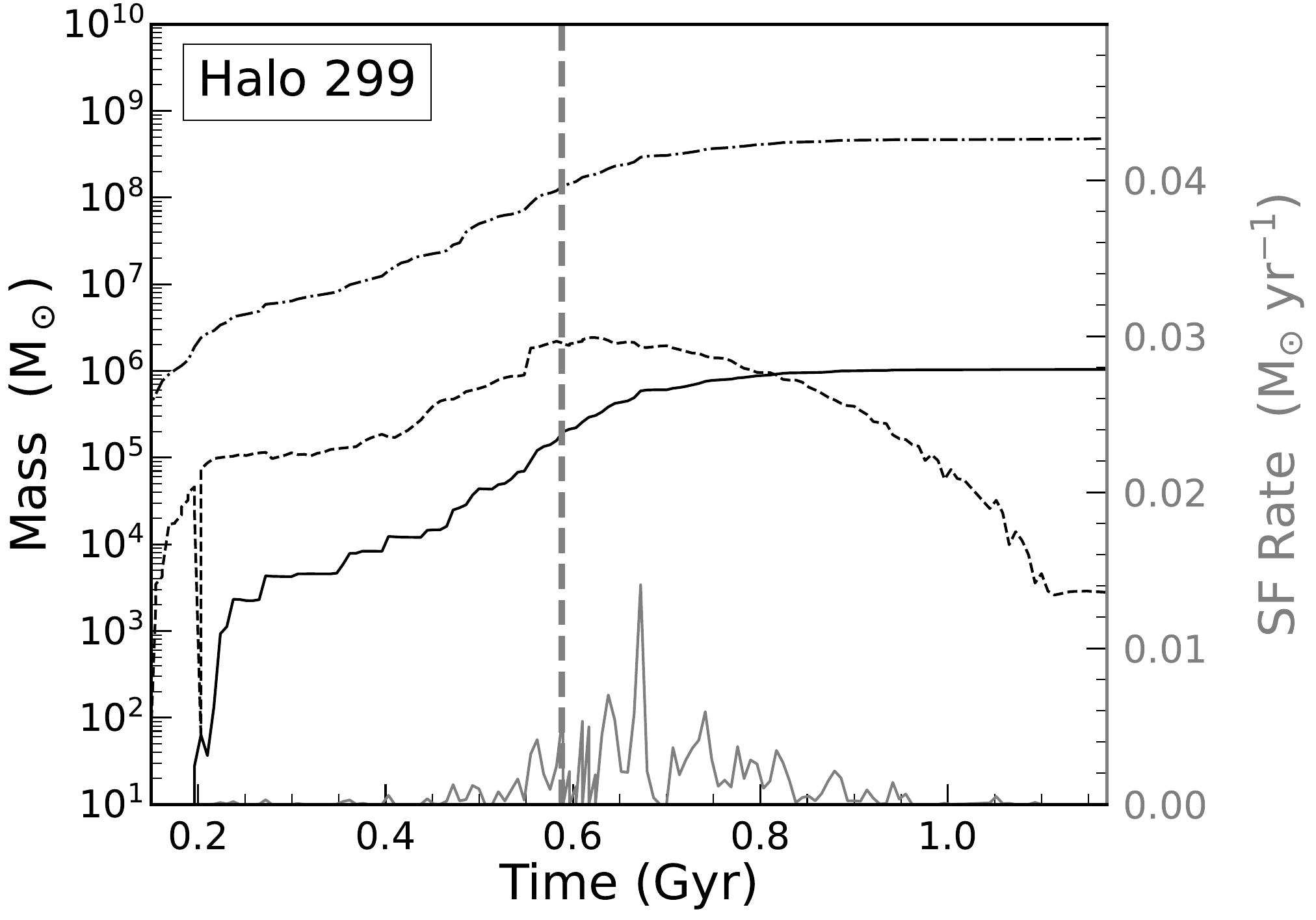}\\
    \end{center}
    \caption{Star formation histories and mass evolution for four High-Mass Halos (Halos 198, 236, 284, and 299, arranged from top-left to bottom-right). 
    The left vertical axis displays cumulative mass ($\rm M_\odot$): black dotted lines represent dark matter, dashed lines indicate cold dense gas ($T<1000$\,K, $n>100\,\mathrm{cm}^{-3}$), and solid lines show stellar mass within the virial radius. 
    The right vertical axis corresponds to the star formation rate (SFR) averaged over 6.9\,Myr, shown by the colored curves. The vertical gray dashed line marks the epoch of reionization ($z=8.5$). {Alt text: Star-formation histories and cumulative mass evolution for four high-mass halos (H198, H236, H284, H299): the left vertical axis shows cumulative mass (dark matter, cold dense gas, and stellar mass within the virial radius, plotted as dotted/dashed/solid lines), the right vertical axis shows the time-varying star-formation rate (colored trace) with bursts as spikes, and a vertical dashed line marks the epoch of reionization ($z = 8.5$).}}
    \label{fig:H_SFH_cont}
\end{figure*}

\begin{figure*}
\begin{center}
  \includegraphics[width=0.49\textwidth]{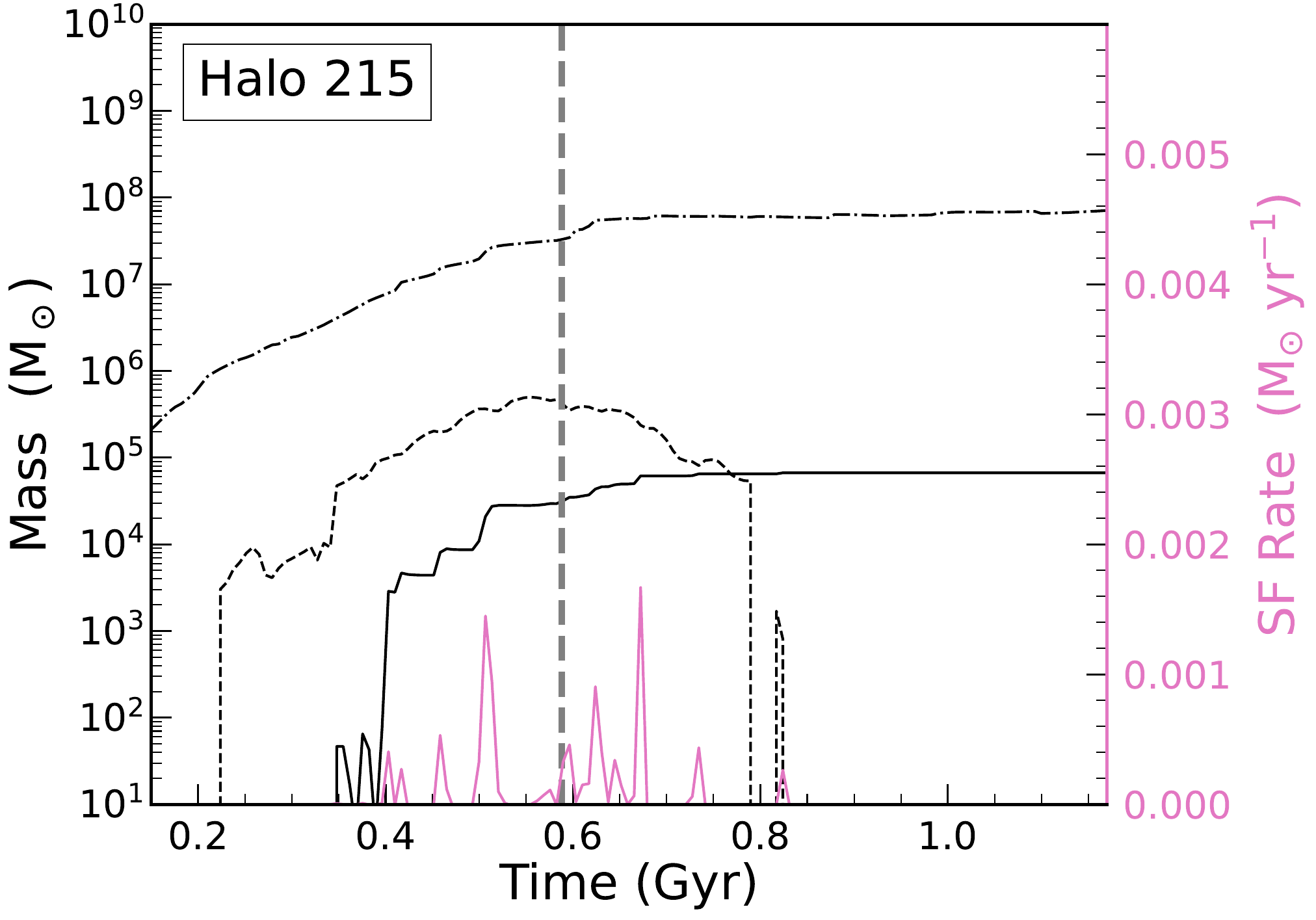}
  \includegraphics[width=0.49\textwidth]{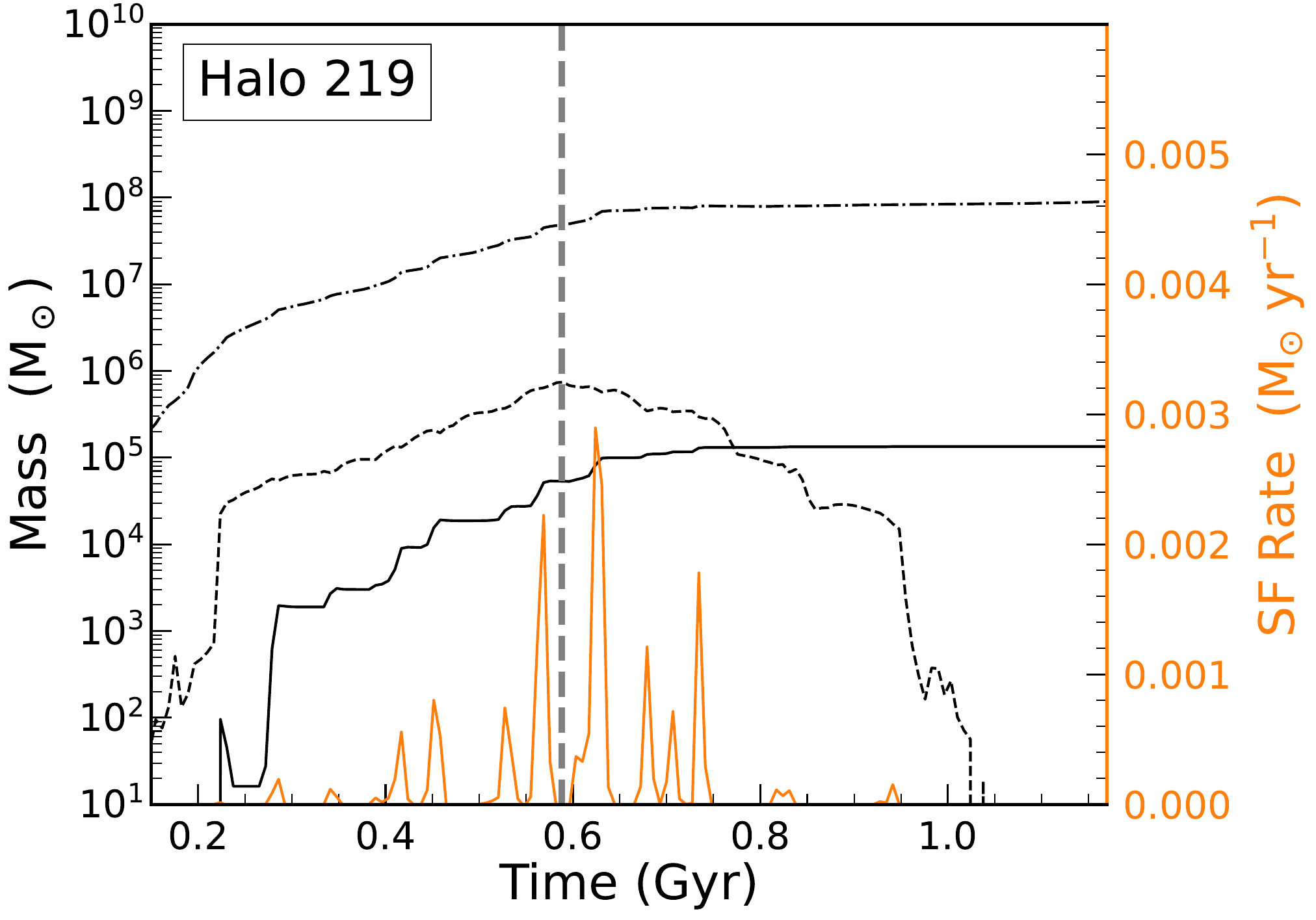}\\
  \includegraphics[width=0.49\textwidth]{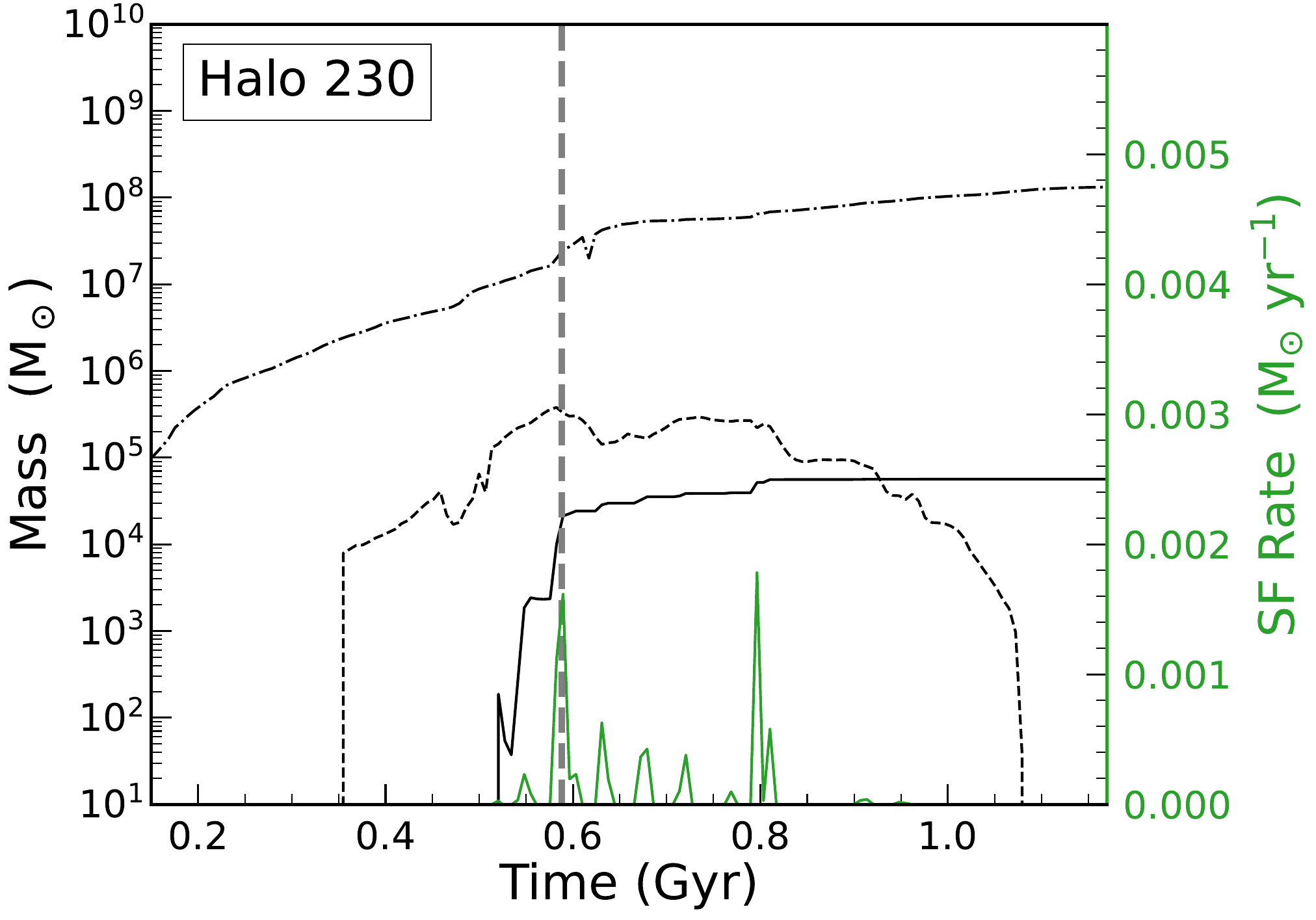}
  \includegraphics[width=0.49\textwidth]{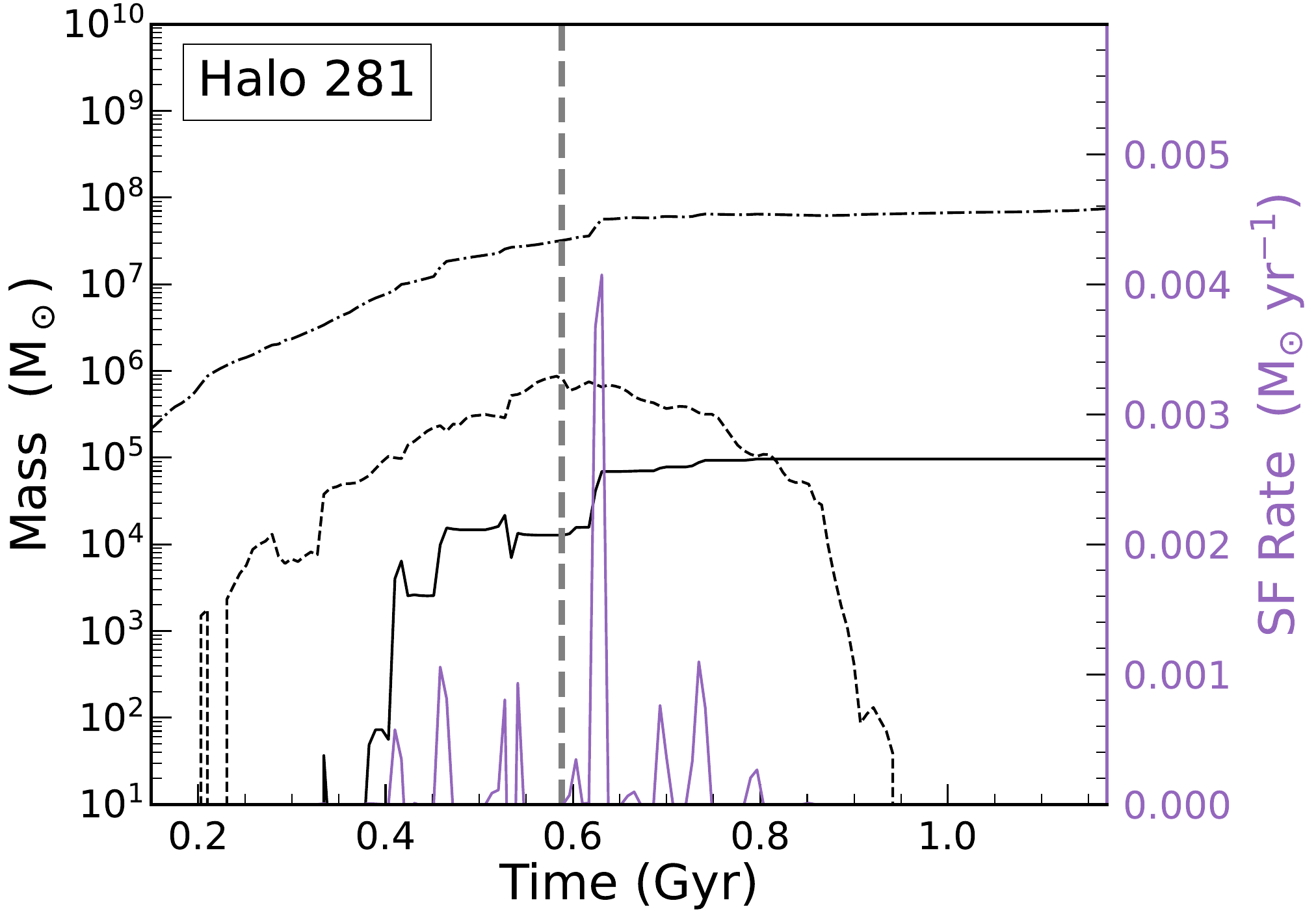}\\
  \end{center}
  \caption{Same as Fig.~\ref{fig:H_SFH_cont} but for the four Low‑Mass Halos, Halos 215, 219, 230, and 281, from top-left to bottom-right. {Alt text: Star-formation histories and cumulative mass evolution for four high-mass halos (H215, H219, H230, H281): the left vertical axis shows cumulative mass (dark matter, cold dense gas, and stellar mass within the virial radius, plotted as dotted/dashed/solid lines), the right vertical axis shows the time-varying star-formation rate (colored trace) with bursts as spikes, and a vertical dashed line marks the epoch of reionization ($z = 8.5$).}}
  \label{fig:L_SFH_cont}
\end{figure*}

In Figs.~\ref{fig:H_SFH_cont} and \ref{fig:L_SFH_cont}, we present the star-formation histories and mass evolution for the High-Mass Halos (Halos 198, 236, 284, and 299) and Low-Mass Halos (Halos 215, 219, 230, and 281). All halos accumulate cold, dense gas with a nearly constant ratio of $M_{\rm gas}/M_{\rm Halo} \sim 0.01$. The maximum mass of cold, dense gas occurs during the reionization epoch; afterward, this gas is heated, leading to a decline in the cold, dense gas mass over time. The cold, dense gas is depleted by $t = 1.2\,\mathrm{Gyr}$, at which point the simulation is stopped.
The High-Mass Halos show an order of magnitude higher star formation rates compared to the Low-Mass Halos. 
The High-Mass Halos rapidly assemble dark matter and quickly replenish central cold, dense gas in their halos. These halos sustained the reservoir gas for longer than Low-Mass Halos after reionization.

The star formation rates (SFR) of the Low-Mass Halos are an order-of-magnitude lower than those of the High-Mass Halos. 
Compared to the High-Mass Halos, the star formation histories of the Low-Mass Halos show brief, sharp SFR spikes, producing stepwise increases in cumulative stellar mass.
The interval between the peaks of SFR is longer than that of the High-Mass Halos.
After the reionization, the cold gas is exhausted earlier than the High-Mass Halos, typically within $\sim 200$\,Myr.
The failure of Low-Mass Halos to retain or re-accrete cold gas after the reionization underscores the role of their shallow gravitational potentials.

\subsection{Diversity in the stellar-to-halo-mass ratio}

\begin{figure*}
\begin{center}
    \includegraphics[width=0.49\textwidth]{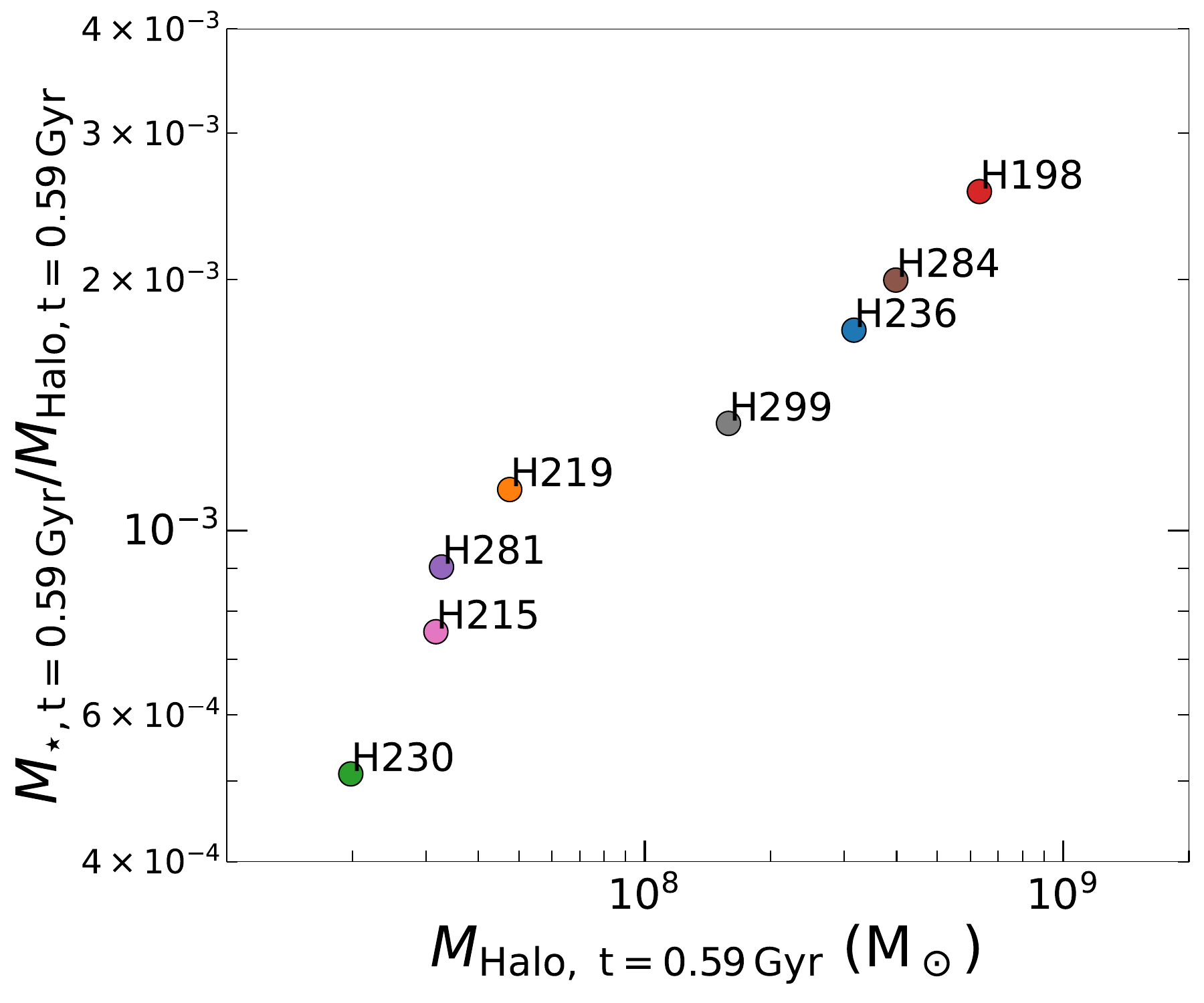}
    \includegraphics[width=0.49\textwidth]{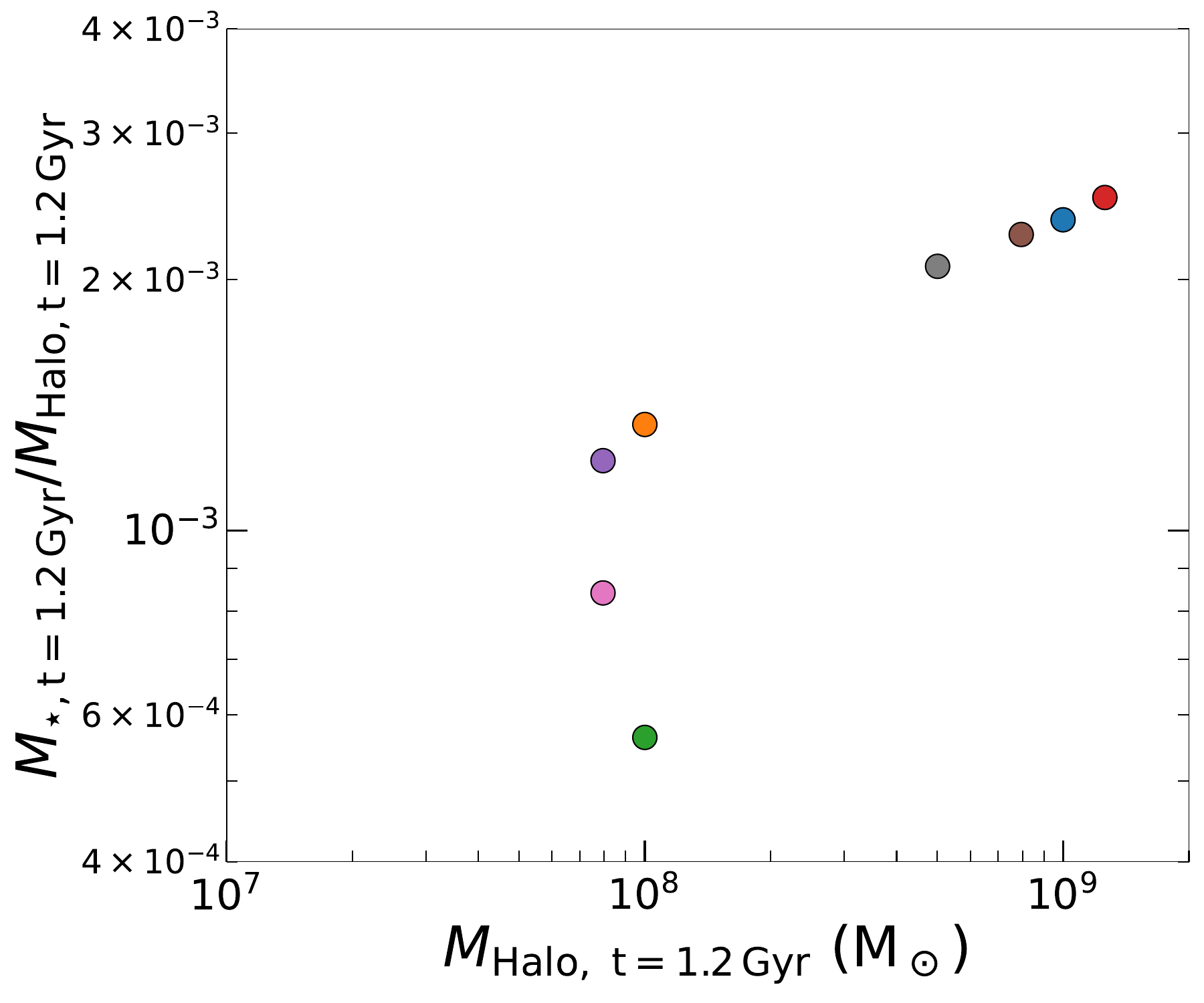}
    \end{center}
    \caption{Stellar-to-halo mass ratio ($M_\star/M_{\rm Halo}$) as a function of halo mass ($M_{\rm Halo}$) at the epoch of reionization ($t=0.59$\,Gyr; left panel) and at the end of the simulation ($t=1.2$\,Gyr; right panel). Each data point represents an individual halo, with color coding consistent with Figures~\ref{fig:H_SFH_cont} and \ref{fig:L_SFH_cont}. {Alt text: Stellar-to-halo mass ratio versus halo mass for individual halos at t=0.59 Gyr (left) and t=1.2 Gyr (right); colored points show each halo (IDs labeled), illustrating larger stellar fractions in higher-mass halos by the simulation end.}}
\label{fig:Correlations}
\end{figure*}

We investigate the stellar-to-halo mass ratios of our simulated galaxies. In previous simulations, they varied widely. 
Fig.~\ref{fig:Correlations} shows the stellar-to-halo mass ratio ($M_\star/M_{\rm Halo}$) versus halo mass at the EoR ($t=0.59$ Gyr) and the end of our simulation ($t=1.2$ Gyr).
At the EoR, the stellar-to-halo mass ratios ($M_\star/M_{\rm Halo}$) clearly correlate with the halo mass ($M_{\rm Halo}$). 
At $t=1.2$\,Gyr, the High-Mass Halos still show the correlation between the stellar-to-halo mass ratio and halo mass. However, the order of the stellar-to-halo mass ratio slightly changes; the stellar-to-halo mass ratio of Halo 236 exceeds that of Halo 284 at $t=1.2$\,Gyr. As seen in Fig.~\ref{fig:H_SFH_cont}, Halo 236 shows star formation more active after the EoR. In contrast, the halo mass of the Low-Mass Halos at $t=1.2$\,Gyr are all $\sim 10^8 \rm\ M_{\odot}$.
Their stellar-to-halo mass ratios at the EoR are almost maintained at $t=1.2$\,Gyr.

We expect no subsequent gas supply and star formation, because they are in isolated environments. In \citet{2026arXiv260113765K}, we continued simulations of Halo 230 and 284 to $z=0.5$ ($t\simeq8.5$\,Gyr). Only Halo 284 experienced a second star formation at $t\simeq1.2$\,Gyr, but the forming stellar mass was negligible compared to the total stellar mass. 
We therefore assume that the stellar mass does not change to $z=0$. We do not have the exact halo mass at $z=0$, but we assume that it does not differ much from that in the dark-matter only simulation that we performed for the target halo selection. 

\begin{figure}[htbp] 
    \begin{center}       
    \includegraphics[width=0.48\textwidth]{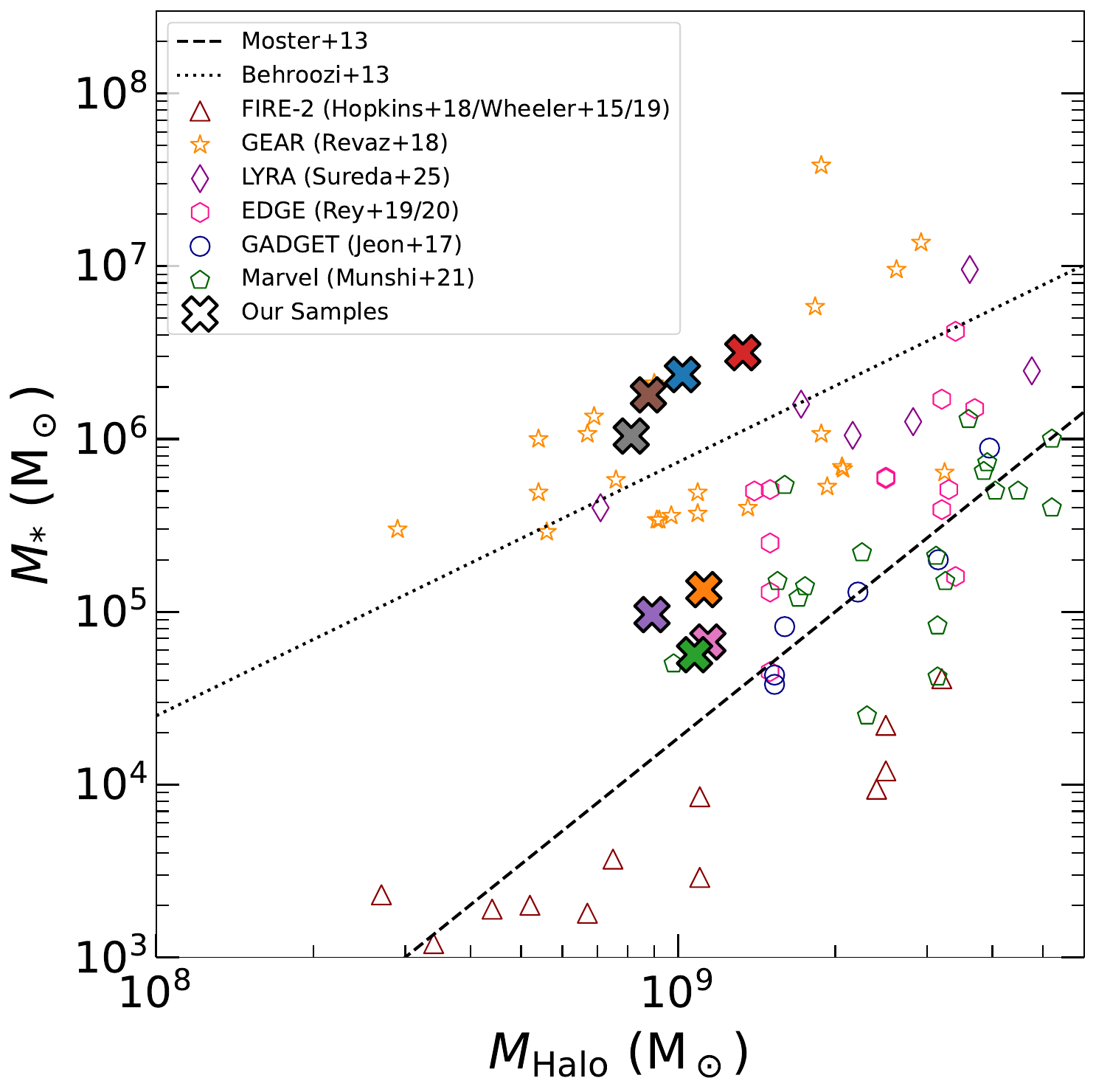}
    \end{center}
   \caption{The stellar mass-halo mass (SMHM) relation of our simulations and previous studies. Our eight simulated halos are marked with crosses and colored individually to match previous figures. Here, we adopt the stellar mass at $t=1.2$\,Gyr as the stellar mass and the halo mass at $z=0$ in the dark-matter only simulation (see Table~\ref{tab:HaloPropertyGadget}) as the halo mass. 
    The dotted and dashed lines represent the abundance-matching relations obtained by \citet{behroozi2013average} and \citet{2013MNRAS.428.3121M}, respectively; note that these lines are extrapolated into the low-mass regime shown here. Comparison samples from previous simulations are shown as follows: brown triangles (FIRE-2; \citealt{wheeler2015sweating,hopkins2018fire,Wheeler_2019}), yellow stars (GEAR; \citealt{2018A&A...616A..96R}), purple diamonds (LYRA; \citealt{sureda2025co}), pink hexagons (EDGE; \citealt{rey2019edge,2020MNRAS.497.1508R}), blue circles (GADGET; \citealt{jeon2017connecting}), and green pentagons (Marvel; \citealt{munshi2021quantifying}). {Alt text: Stellar mass versus halo mass comparing eight simulated halos (large colored crosses) to literature samples and abundance-matching relations: colored markers represent comparison datasets, dotted and dashed lines show Behroozi et al. 2013, and Moster et al. 2013 relations for context, and the figure demonstrates how the simulated halos sit relative to observed and modeled SMHM trends at t = 1.2 Gyr.}}
    \label{fig:SMHM}  
\end{figure}

In Fig.~\ref{fig:SMHM}, we present the stellar mass-halo mass (SMHM) relation of our simulated halos. We also plot samples from previous simulations and the relations obtained from abundance matching \citep{2013MNRAS.428.3121M,behroozi2013average}, although these observational relations are extrapolations from those for more massive galaxies. 
Our high-mass halos show stellar masses slightly higher than the relation obtained by \citet{behroozi2013average} and consistent with the GEAR results \citep{2018A&A...616A..96R}. LYRA results \citep{sureda2025co} may also distribute similarly, but the number of their samples with a similar halo mass is only two. Our Low-Mass Halos are located on the relation obtained by \citet{behroozi2013average} and \citet{2013MNRAS.428.3121M}. EDGE \citep{rey2019edge} and Marvel \citep{munshi2021quantifying} results are distributed at a similar location in the plot, but their halo masses are larger than ours. 
Thus, our simulation results show a diversity in the SMHM relation. In the stellar-to-halo mass ratio, our results show a diversity of two orders of magnitude between $5\times 10^{-5}$ to $2\times 10^{-3}$.

\subsection{Mass-radius relation of stellar distribution}

\begin{figure}
\begin{center}
\includegraphics[width=0.48\textwidth]{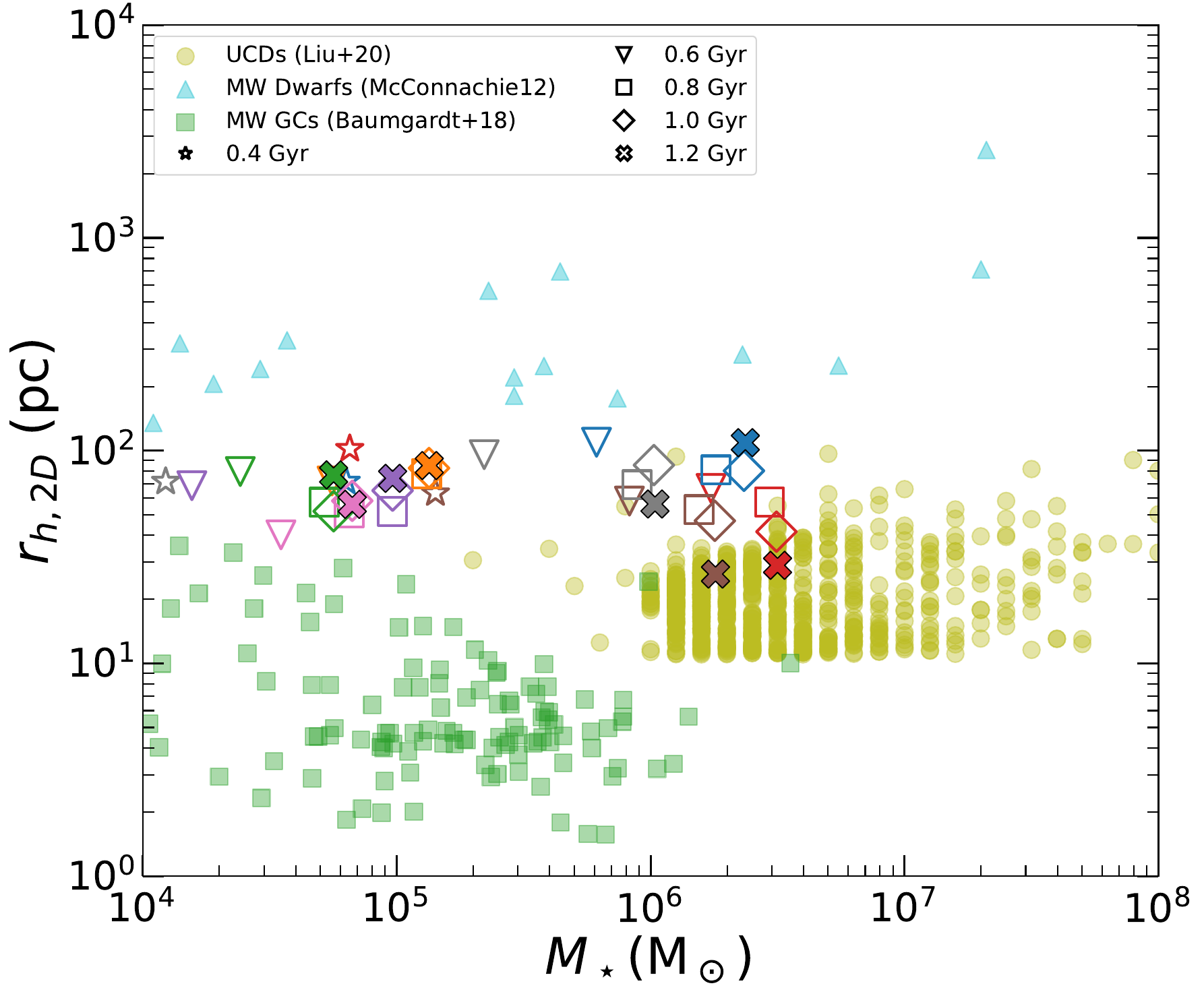}    
\end{center}
\caption{Projected half-mass (or half-light) radius ($r_{h,2D}$) versus stellar mass ($M_{\star}$) for observed compact and diffuse stellar systems compared to our simulated halos. 
Observational data include Milky Way globular clusters (green squares;
\citealt{baumgardt2018catalogue}), Milky Way dwarf galaxies (blue triangles;
\citealt{mcconnachie2012observed}), and ultra-compact dwarfs (yellow circles;
\citealt{2020ApJS..250...17L}). For the simulations, colors correspond to individual halos as in \rfig{fig:Halomass}. We trace the size evolution using different symbols for different epochs (see legend): open symbols represent earlier times, while the solid crosses mark the final state at the end of the simulation ($t=1.2$\,Gyr). {Alt text: Projected half-mass (half-light) radius versus stellar mass comparing observed compact and diffuse stellar systems (globular clusters, dwarf galaxies, ultra-compact dwarfs) with the simulated halos across epochs: marker shapes and colors indicate observational samples and simulation epochs (open symbols for earlier times, solid crosses for final states), showing that simulated halo sizes occupy the intermediate region between compact clusters and more diffuse dwarfs.
}}
\label{fig:Obs}
\end{figure}

Another form of diversity is seen in the compactness. 
We measure the two-dimensional half-mass radius of stellar distribution (\(r_{\rm h,2D}\)) and stellar mass within the virial radius (\(M_\star\)) every 0.2\,Gyr from 0.4 to 1.2\,Gyr, and the results are shown in \rfig{fig:Obs}. 
For comparison, we also show the observational data of MW globular clusters \citep{baumgardt2018catalogue}, MW dwarf galaxies \citep{mcconnachie2012observed}, and UCDs \citep{2020ApJS..250...17L}. 
In this mass-radius plane, Halos 198 and 284 occupy the UCD locus, and Halo 299 is also close to them. These systems exhibit highly concentrated stellar distributions formed in their deep potentials and efficient central gas cooling and star formation. From their evolutionary tracks in this plane, these compact systems contract (decrease their half-mass radii) as they accumulate mass through star formation. This is because the star formation proceeded in the innermost region of the galaxy. The other halos, which are similar to MW dwarf galaxies, evolve without such contraction; their stellar mass increases, while their radii remain.

One may think that the half-mass radii are a few factors smaller than those of the observed MW dwarfs. 
We stopped our simulations and measured the radii at \(t=1.2\) Gyr, which is far earlier than observed systems (\(t\sim13.8\) Gyr). In addition, our samples are all isolated halos. 
In contrast, Milky Way (MW) dwarf galaxies are typically much older and are gravitationally bound to their host halo. Several mechanisms can increase the radius of the stellar distribution of these dwarfs. When dwarf galaxies are accreted onto their host halos, tidal stripping by the host can remove outer stars and dark matter, weakening the gravitational potential and consequently increasing the radius of the dwarf galaxy \citep{sales2010effect}. Tidal shocks or tidal heating can further inject energy into the inner regions of the dwarf, leading to an expansion of the effective radius \citep{gnedin1999tidal,fielder2024all}. Additionally, merger events can cause significant expansion (or "puffing up") of the stellar system, resulting in a larger half-mass radius \citep{revaz2023compactness}.

Intriguingly, Halos 284 and 299 are not the highest-SFR systems. The peak SFR of Halo 236 is higher (see Fig.~\ref{fig:H_SFH_cont}). This result suggests that compact morphology does not depend solely on the peak star-formation rate. We further investigate the special properties of these UCD-like galaxies in the subsequent sections.

\subsection{Dynamical time of gas}
\begin{figure*}
    \begin{center}
    \includegraphics[width=0.49\textwidth]{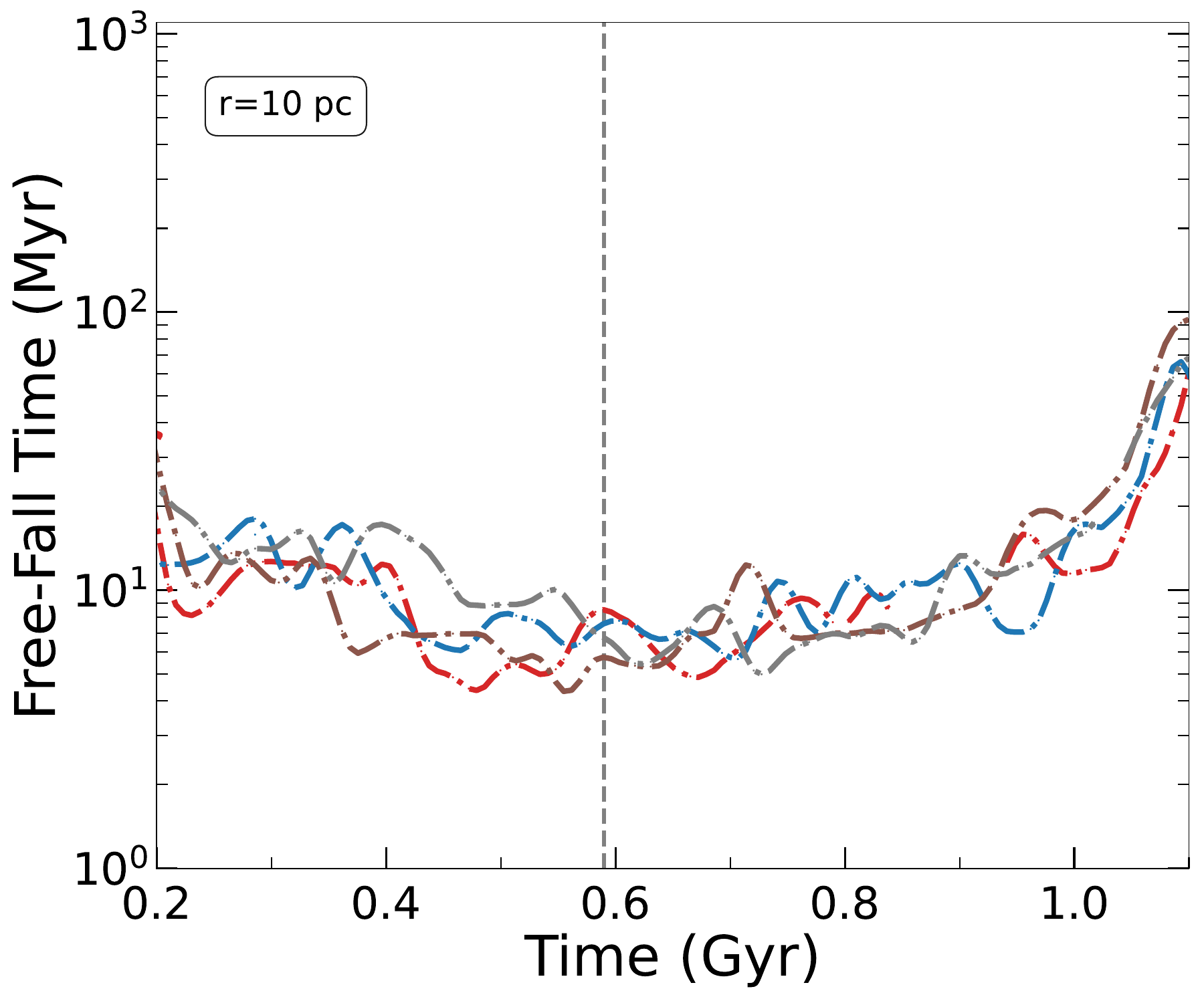}
    \includegraphics[width=0.49\textwidth]{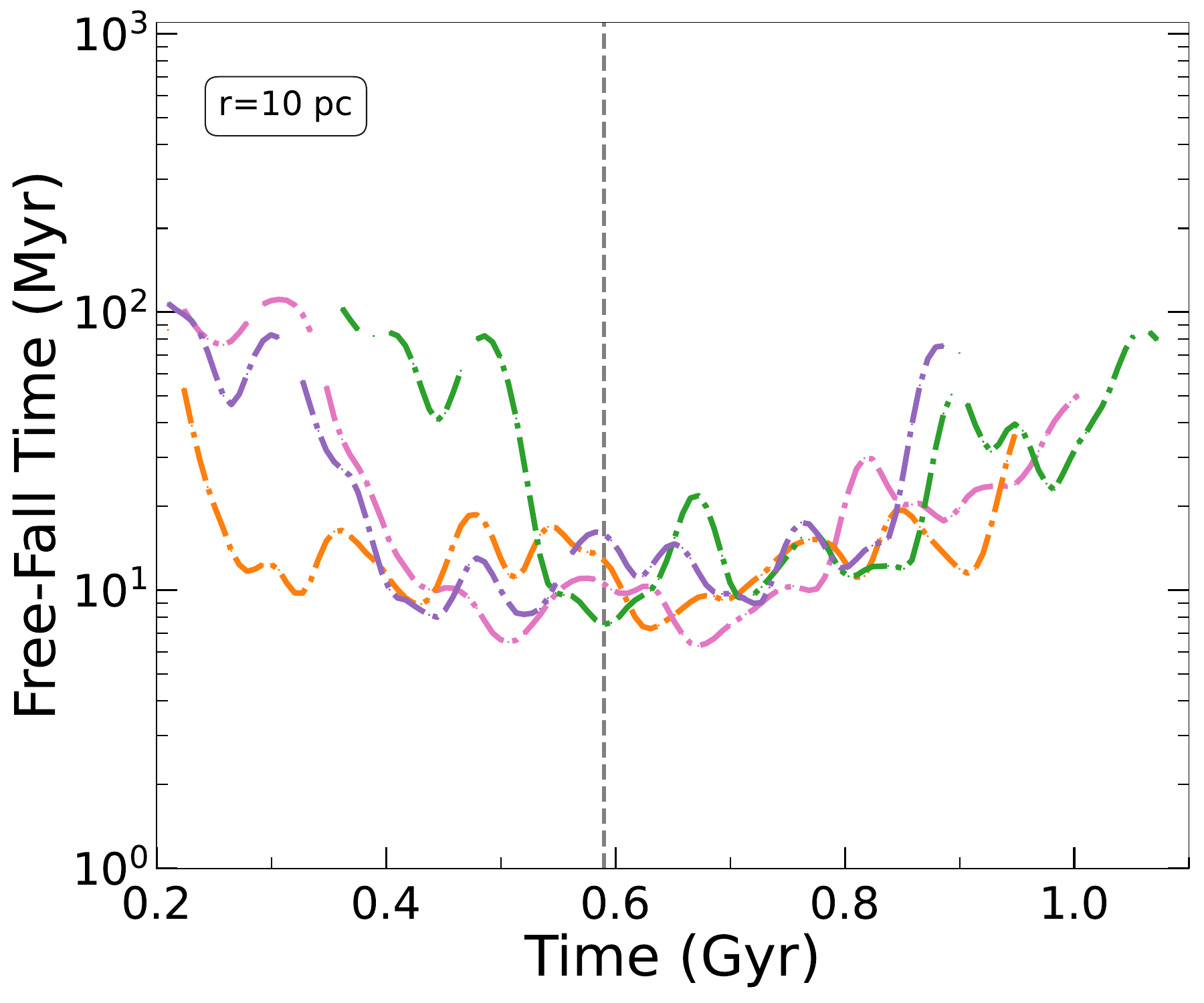}
    \end{center}
    \caption{Evolution of the gas free-fall time for the simulated dwarf galaxies. The left panel shows the results for the High-Mass Halos, while the right panel displays the Low-Mass Halos. The free-fall time is calculated based on the gas mass enclosed within the central $<10$\,pc region. The vertical dashed line indicates the epoch of reionization at $t \sim 0.59$\,\ Gyr. {Alt text: Gas free-fall time at r = 10 pc for simulated dwarf galaxies (left: high-mass, right: low-mass) versus cosmic time (0.2–1.1 Gyr); vertical dashed line marks reionization ($t \approx$ 0.59 Gyr).}}
    \label{fig:freefall_time}
\end{figure*}

We begin by analyzing the gas free-fall time ($\tau_{\rm ff}$) as an indicator of the typical star formation timescale in our simulated halos. The free-fall time provides a direct estimate of how fast gas can collapse to form stars in the absence of feedback. Fig.~\ref{fig:freefall_time} shows the time evolution of the gas free-fall time within the central $r=10$\,pc region for all our samples.
In both the High-Mass and Low-Mass Halos, the free-fall time initially decreases, reaches a minimum around the epoch of reionization (EoR), and then increases at later times. The minimum free-fall time is approximately $\sim 5$ Myr for the High-Mass Halos and $\sim 8$\ Myr for the Low-Mass Halos. Thus, we did not find any large gap in the free-fall time between the High-Mass and Low-Mass Halos. 

\citet{dekel2023efficient} proposed the feedback-free starburst (FFB) scenario proposed by \citet{dekel2023efficient}, in which rapid collapse allows massive star formation before feedback can stop the process. Their criterion for the FFB scenario is $\tau_{\rm ff} < 1$\,Myr.
However, our simulated halos do not reach such short free-fall times even at the innermost regions ($<10$\,pc). Moreover, the star formation rates (SFRs) in our simulations show multiple peaks rather than a rapid FFB. 
Thus, the FFB scenario cannot account for the formation criteria of our UCD-like galaxies.
We note that the halo masses considered in \citet{dekel2023efficient}, $M_{\rm Halo} \sim 10^{11}\,\rm M_\odot$, are much higher than those in our sample, $M_{\rm Halo} \sim 10^{9}\,\rm M_\odot$. The difference in the halo mass might be the reason why a short free-fall time of $<1$\,Myr did not occur in our simulations.

\subsection{Compact stellar system formation as an analog of star cluster formation}

Next, we consider an analog of star cluster formation. Star clusters have masses similar to our galaxies, although they are dark-matter free. In
\cite{fukushima2021radiation} and \cite{fukushima2022far}, they performed a series of star cluster formation simulations with initial molecular cloud masses of $10^5$-- $10^6 \rm\ M_{\odot}$. Since their simulation includes radiation feedback from massive stars, the gas clouds are finally heated up and evaporated. They investigated the star formation efficiency (the ratio of the formed stellar mass to the initial gas mass), bound stellar mass, and radius.  
They found that the gas surface density of the molecular cloud is critical for forming dense star clusters similar to globular clusters. They also constructed an one-dimensional semi-analytic model to understand their simulation results. 
In their model, they considered the balance between the gravitational force toward the molecular cloud center ($F_{\rm grav}$) and the thermal pressure from massive stars outward ($F_{\rm th}$). If the gravitational force is stronger than the thermal pressure, the cluster can continue efficient star formation, overcoming the feedback from massive stars.  

We expand this model by adding the gravitational force from dark matter and the pressure from supernova explosions.
In \citet{fukushima2021radiation}, the gravitational force (acceleration) is estimated as
\begin{equation}
\label{eq:FGrav_simplified_fixed}
F'_{\mathrm{grav}} \;\simeq\; -\frac{G\,M_{\rm gas}}{2\,R_{\rm gas}^2},
\end{equation}
where $M_{\rm gas}$ and $R_{\rm gas}$ are the mass and radius of the gas cloud.
Adding a factor for the relative contribution of dark matter to the enclosed mass ($f_{\rm DM}$), the gravitational force can be written as
\begin{equation}
\label{eq:FGrav_simplified_mod}
F_{\mathrm{grav}} \;\simeq\; -\,f_{\rm DM}\,\frac{G\,M_{\rm gas}}{2\,R_{\rm gas}^2}.
\end{equation}
The value of $f_{\rm DM}$ can differ in each halo, and the radius of the halo that we consider. 
Throughout this section, we adopt the force-per-unit-mass (acceleration) form following the description in \citet{fukushima2022far}; thus, $F$ denotes an effective acceleration.

For the thermal pressure, we use the definition of \citet{fukushima2022far}:
\begin{equation}
\label{eq:Fthermal_fixed}
F_{\mathrm{th}} \;=\; k_{\rm B} T \left(\frac{12\pi f_{\rm ion}\,\epsilon_\ast\,s_{\rm EUV}\,R_{\rm gas}}{\alpha_B\,M_{\rm gas}}\right)^{1/2},
\end{equation}
where $\alpha_B(T)=2.6\times10^{-13}(T/10^4\ \mathrm{K})^{-4/5}\ \mathrm{cm^3\,s^{-1}}$
\citep{osterbrock2006astrophysics} and we adopt fiducial values $T=10^4\ \mathrm{K}$
and $f_{\rm ion}=0.73$ \citep{mckee1997luminosity,krumholz2009dynamics}.
Here, $\epsilon_\ast$ denotes the star formation efficiency, representing the fraction of the initial gas mass converted into stars. In this work, we adopt the following value for $\epsilon_\ast$ derived in \citet{fukushima2021radiation}.
They assumed that the star formation duration is equal to the timescale for the molecular cloud to be fully ionized by the radiation from formed stars and derived the star formation efficiency when this condition is satisfied. For the ionization timescale, they adopted the propagation time of the ionizing front in an expanding H\textsc{ii} region. 
The resulting efficiency, $\epsilon_\ast$, is given as a function of the cloud gas mass and surface density as follows:
\begin{equation}
\label{eq:SFE_multiline_fixed}
\begin{split}
\epsilon_{\ast} &= 0.09 
\left(\frac{\epsilon_{\rm ff}}{0.1}\right)^{4/5}
\left(\frac{\Sigma_{\rm gas}}{80\ \mathrm{M_\odot\ pc^{-2}}}\right)^{1/2} 
\left(\frac{M_{\rm gas}}{10^{5}\ \mathrm{M_\odot}}\right)^{1/10}\\
&\quad\times
\left(\frac{T}{8000\ \mathrm{K}}\right)^{-14/25}
\left(\frac{s_{\rm EUV}}{7.5\times10^{46}\ \mathrm{s^{-1}\ M_\odot^{-1}}}\right)^{-1/5},
\end{split}
\end{equation}
where $\Sigma_{\rm gas}\equiv M_{\rm gas}/(\pi R_{\rm gas}^2)$, and
$s_{\rm EUV}$ is given by a fitting formula \citep{inoguchi2020factories,fukushima2022far}:
\begin{equation}
\label{eq:EUV_fit_fixed}
\log_{10}\!\left(\frac{s_{\rm EUV}}{\mathrm{s^{-1}\ M_\odot^{-1}}}\right)
= \frac{46.7\,\chi^6}{3.66 + \chi^6}, \qquad
\chi \equiv \log_{10}\!\left(\frac{M_\ast}{\mathrm{M_\odot}}\right),
\end{equation}
where $M_\ast$ is the stellar mass and obtained from $M_\ast\equiv\epsilon_\ast M_{\rm gas}$.

We estimate the outward
acceleration contributed by SNe by modeling CCSNe feedback in the steady momentum-rate approximation as:
\begin{eqnarray}
\label{eq:FSNe_fixed}
F_{\rm SN}=f_{\rm coup}\,\frac{p_{\rm SN}\,N_{\rm SN}}{\tau_{\rm SN}M_{\rm gas}},
\end{eqnarray}
because $M_{\ast}\equiv \epsilon_{\ast}M_{\rm gas}$, 
\begin{eqnarray}
F_{\rm SN}= f_{\rm coup}\,\frac{p_{\rm SN}\,\epsilon_\ast \,N_{\rm SN}}{\tau_{\rm SN}M_{\ast}},
\end{eqnarray}
where $f_{\rm coup}\in[0,1]$ is the fraction of injected SN momentum that couples to the shell (we adopt $f_{\rm coup}=0.1$, \citealt{walch2015energy,2015ApJ...802...99K}),  
$p_{\rm SN}$ is the terminal momentum per SN (we adopt $p_{\rm SN}=2\times10^{5}\ \mathrm{M_\odot\ km\ s^{-1}}$; \citealt{oku2022osaka}), 
$N_{\rm SN}/M_{\ast}$ is the number of SNe per unit stellar mass (for a Salpeter IMF $N_{\rm SN}/M_{\ast}\sim 0.01\ \mathrm{M_\odot^{-1}}$), 
and $\tau_{\rm SN}$ is the effective SN injection timescale (we adopt $\tau_{\rm SN}=30\ \mathrm{Myr}$, consistent with the $\sim30$--$40\ \mathrm{Myr}$ lifetime of an $8\rm\ M_\odot$ progenitor).

\begin{figure}
\begin{center}
\includegraphics[width=0.48\textwidth]{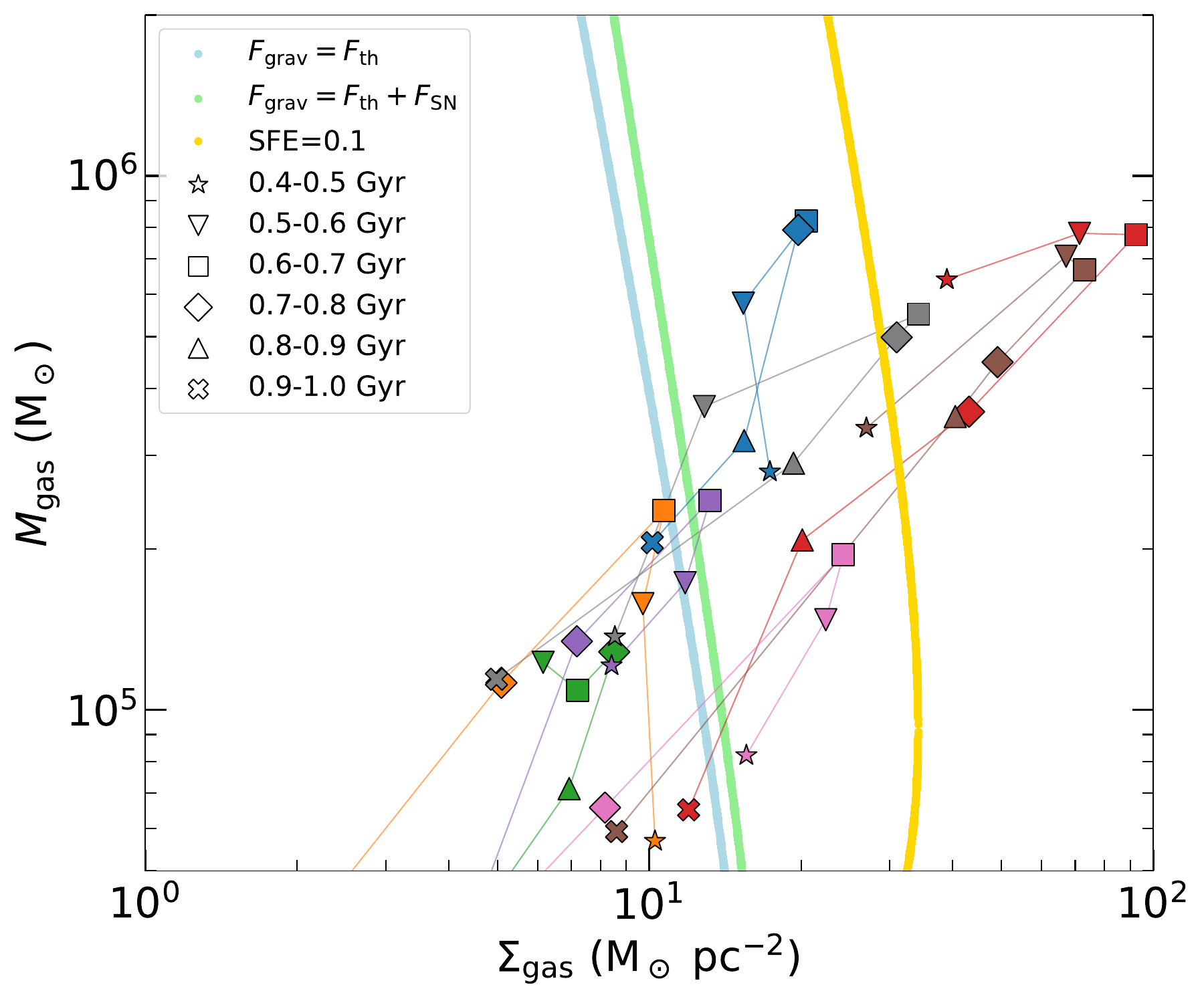}
\end{center}
\caption{Evolution of the simulated galaxies in the $\Sigma_{\text{cl}}$--$M_{\text{cl}}$ plane. The vertical axis shows the gas mass enclosed ($M_{\rm gas}$) within the stellar half-mass radius ($r_{\rm h}$), and the horizontal axis represents the gas surface density defined as $\Sigma_{\rm gas}=M_{\rm gas}/(\pi r_{\rm h}^2)$. 
The blue curve indicates the equilibrium condition $F_{\text{grav}} = F_{\text{th}}$, while the green curve represents $F_{\text{grav}} = F_{\text{th}} + F_{\text{SN}}$. 
The yellow line defines the threshold where the expected star formation efficiency (SFE) reaches 0.1, based on star cluster formation simulations \citep{fukushima2021radiation,fukushima2022far}.
The simulation results are color-coded individually, with symbols indicating different evolutionary stages. 
Halos 198, 284, and 299, evolving into ultra-compact dwarf (UCD)-like structures by the end of the simulation, cross into the high-SFE regime. {Alt text: Gas surface density vs enclosed gas mass tracks for simulated galaxies, showing equilibrium and SN‑supported curves and the SFE=0.1 threshold; three halos (H198, H284, H299) enter the high‑SFE (UCD‑like) regime.}}
\label{fig:equ}
\end{figure}

By numerically solving these equations, we can obtain the relation between gas mass and surface density satisfying $F_{\rm grav}=F_{\rm th}+F_{\rm SN}$. 
In Fig.~\ref{fig:equ}, we present the gas mass and density relations for $F_{\rm grav}=F_{\rm th}$ and $F_{\rm grav}=F_{\rm th}+F_{\rm SN}$. Here, we adopt $f_{\rm DM}=50$, which is the value measured in our simulations.

In \citet{fukushima2022far}, they showed that compact star clusters can be formed when $F_{\rm grav}>F_{\rm th}$ and the critical surface density is $\Sigma_{\rm gas}>200$--$300 \rm\ M_{\odot}$\,pc$^{-2}$ for $M_{\rm gas}=10^5$--$10^6 \rm\ M_{\odot}$. Compared to their results, the critical density derived from $F_{\rm grav}>F_{\rm th}$ or $F_{\rm grav}>F_{\rm th}+F_{\rm SN}$ condition is $\Sigma_{\rm gas}\gtrsim 10 \rm\ M_{\odot}$\,pc$^{-2}$, which is an order of magnitude lower than their critical surface density, because of the gravity from the dark matter component in our cases. 
We also plot the relation between gas mass and the surface density taken from our simulated galaxies in Fig.~\ref{fig:equ}. All of the galaxies exceeded this criterion at least once during the simulation. 

Using eq.~(\ref{eq:SFE_multiline_fixed}), we can also draw a curve for a given SFE. In \citet{fukushima2021radiation}, they also discussed the critical SFE to form dense star clusters. In their simulations, the SFE also depends on the gas surface density, and the final bound mass of the cluster suddenly jumps up when SFE ($\epsilon_\ast$) exceeds 0.1. We also plot the relation satisfying $\epsilon_\ast=0.1$ in Fig.~\ref{fig:equ}. As shown in the figure, all the galaxies that formed UCD-like compact stellar distribution (Halos 198, 284, and 299) satisfied the criterion of $\epsilon_\ast>0.1$. If we convert this criterion to the critical surface density, it is $\Sigma\gtrsim 30\,\rm\ M_\odot\,{\rm pc}^{-2}$.

\subsection{Density profiles}

\begin{figure*}

  \begin{center}
    \begin{overpic}[width=0.4\textwidth]{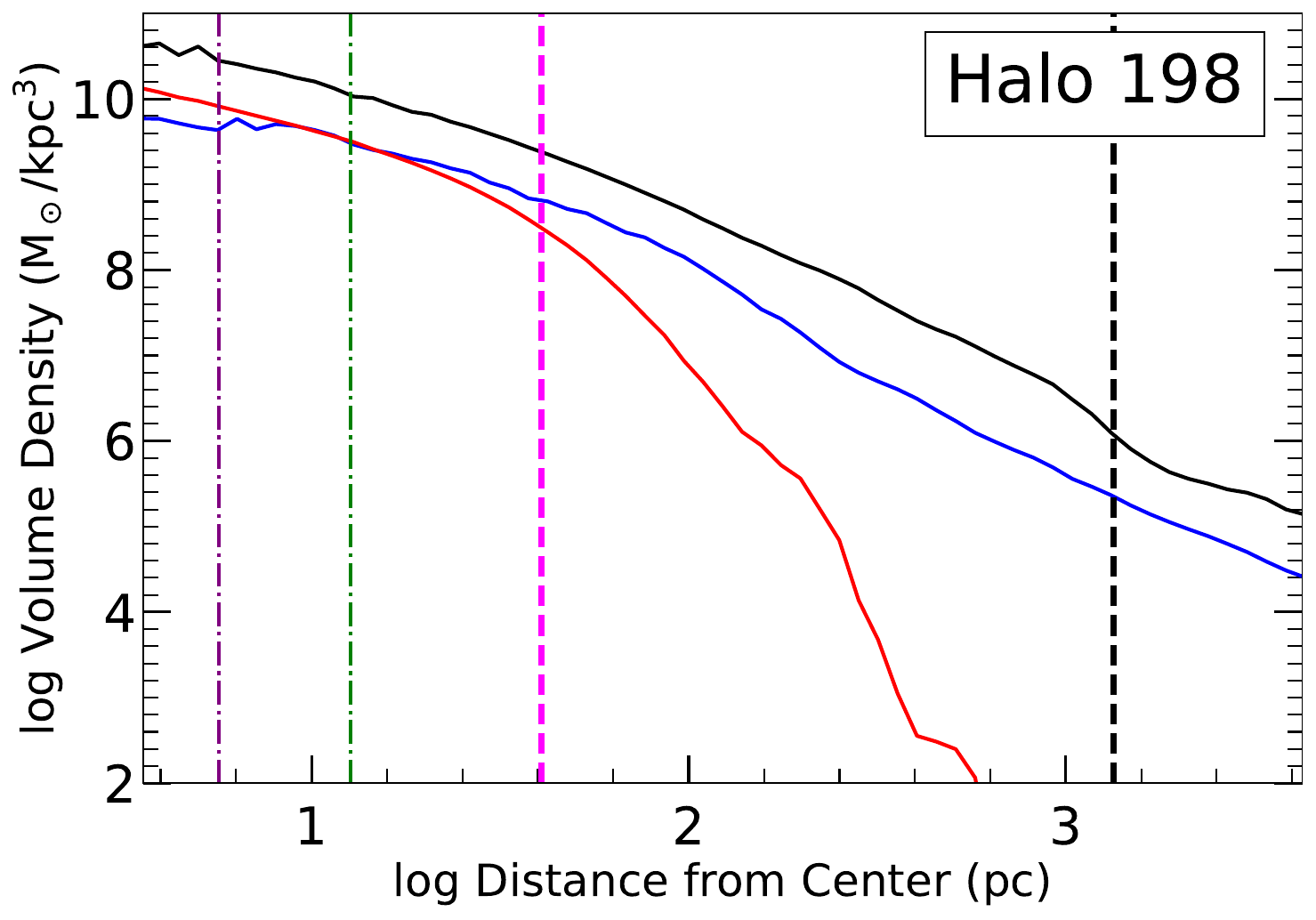}
      \put(65,52){\normalsize  \textbf{EoR}}
    \end{overpic}
    \begin{overpic}[width=0.4\textwidth]{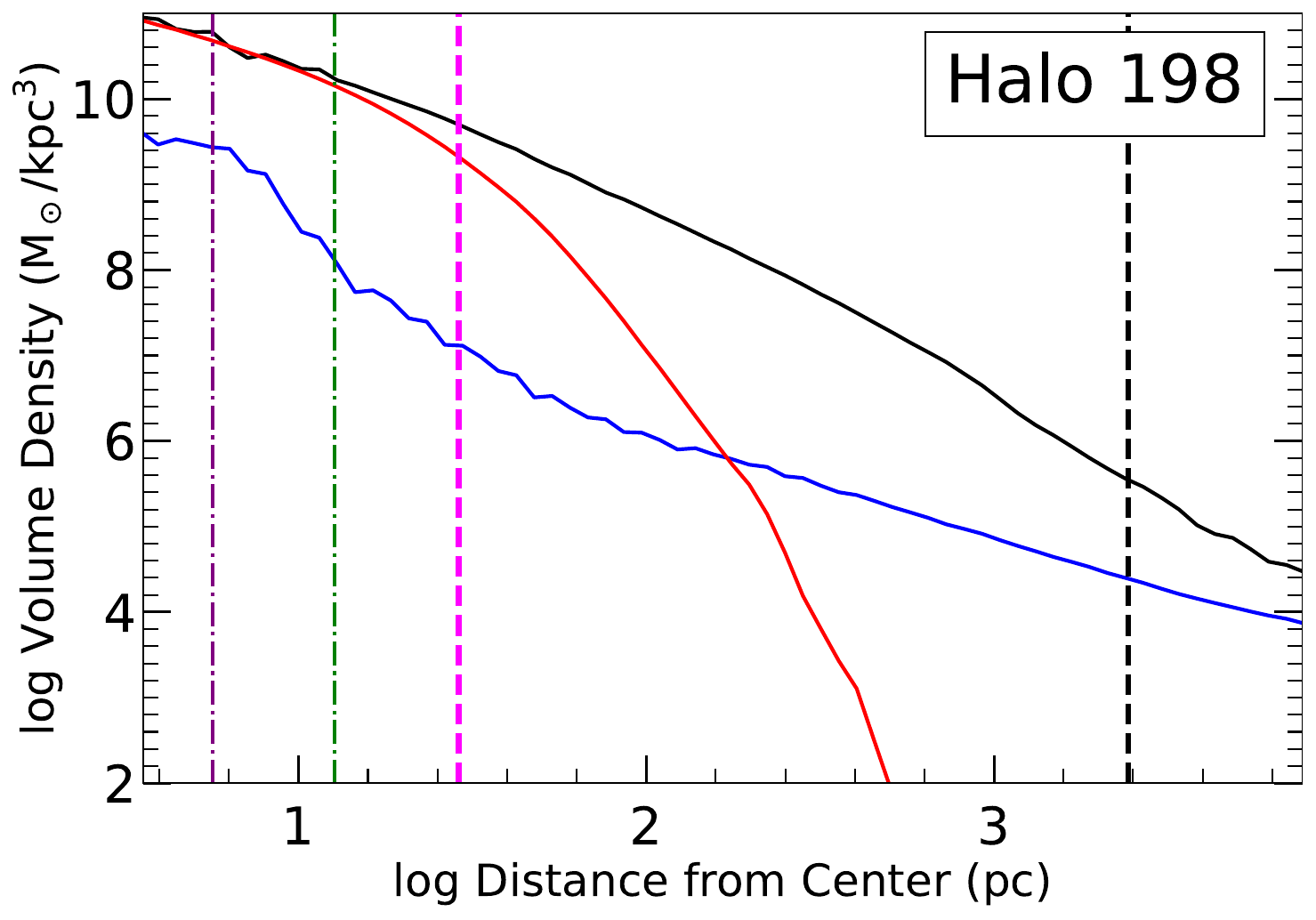}
      \put(60,52){\normalsize  \textbf{$t=1.2$ Gyr}}
    \end{overpic}\\

    \begin{overpic}[width=0.4\textwidth]{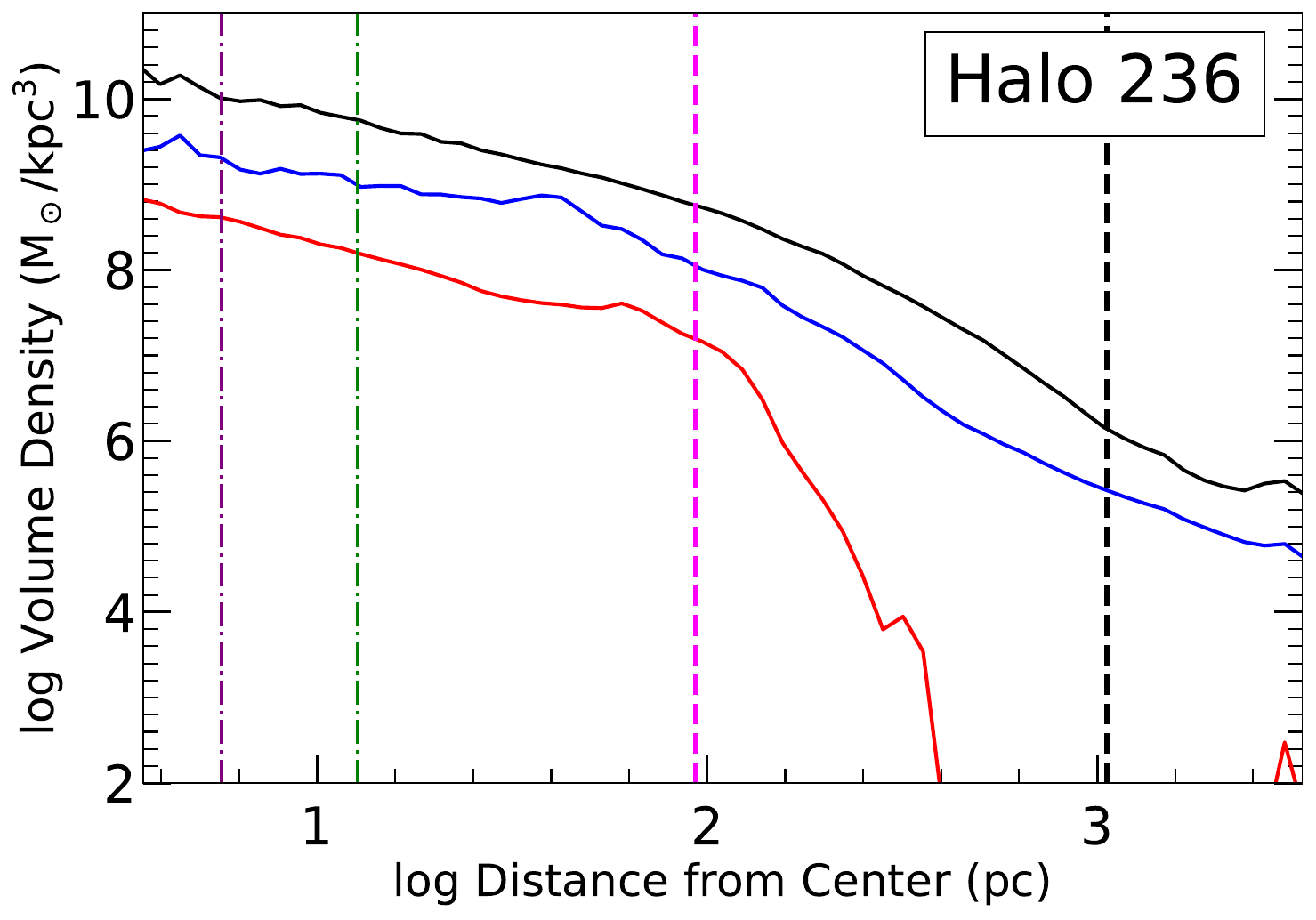}
      \put(65,52){\normalsize  \textbf{EoR}}
    \end{overpic}
    \begin{overpic}[width=0.4\textwidth]{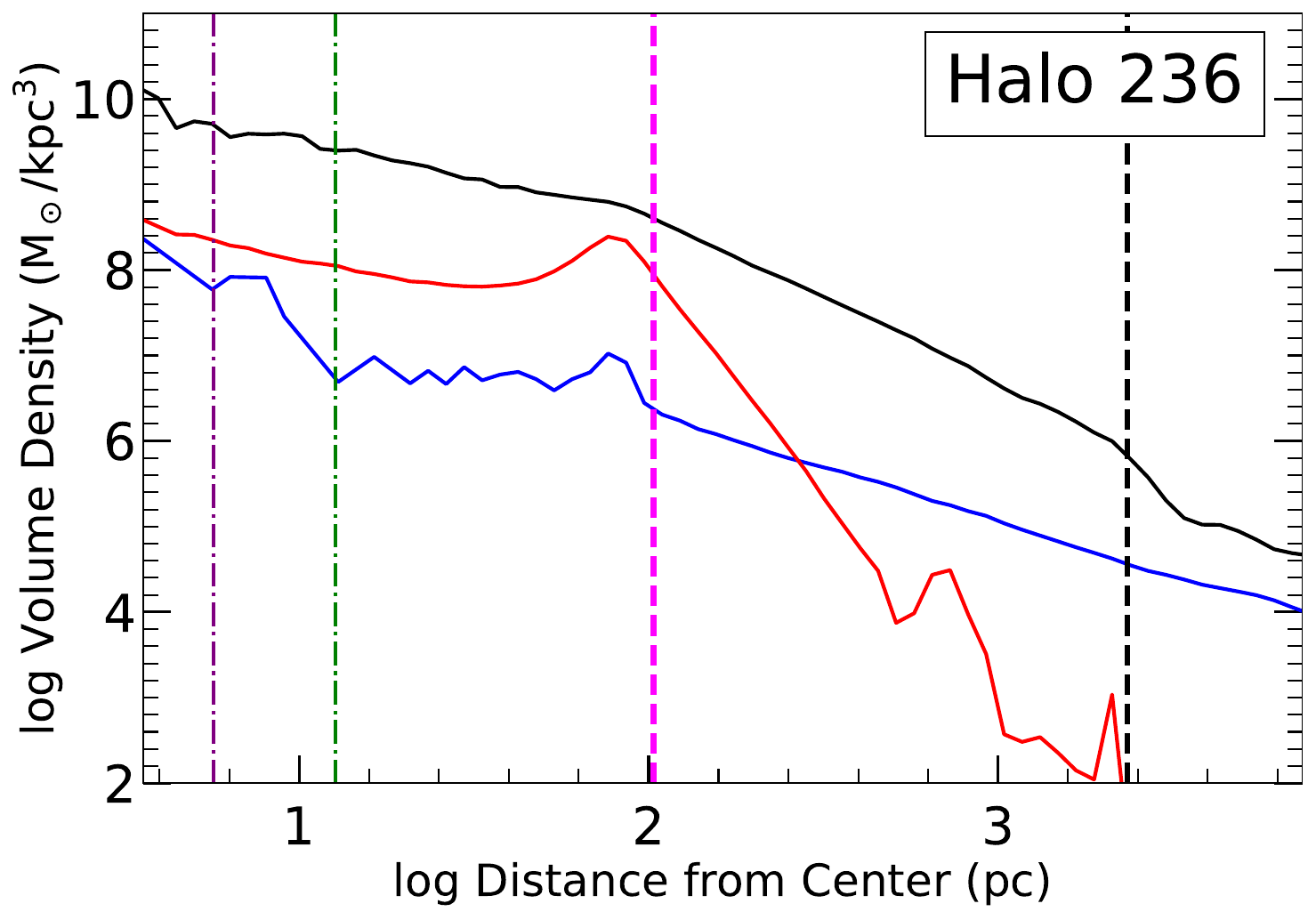}
      \put(60,52){\normalsize  \textbf{$t=1.2$ Gyr}}
    \end{overpic}\\

    \begin{overpic}[width=0.4\textwidth]{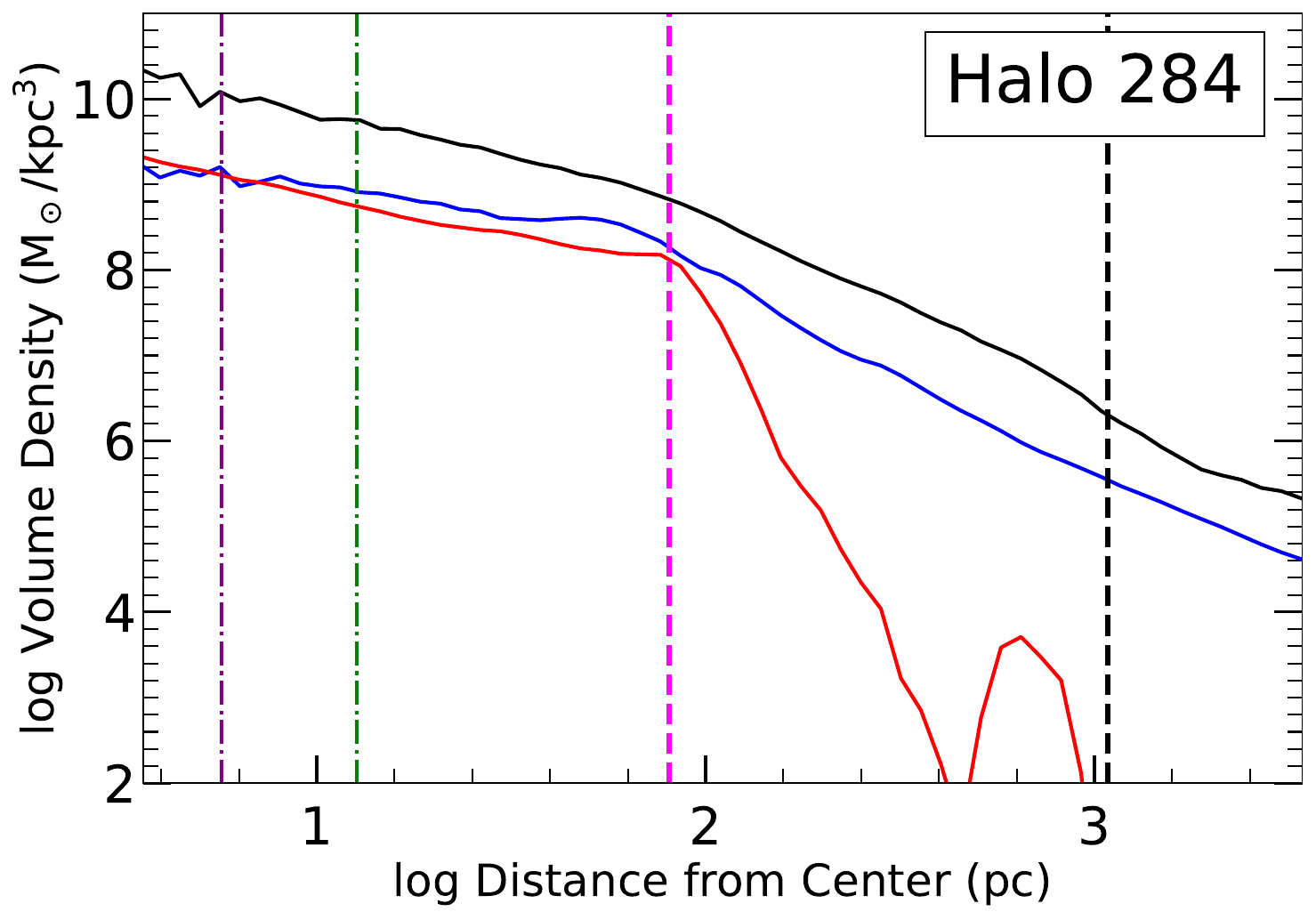}
      \put(65,52){\normalsize  \textbf{EoR}}
    \end{overpic}
    \begin{overpic}[width=0.4\textwidth]{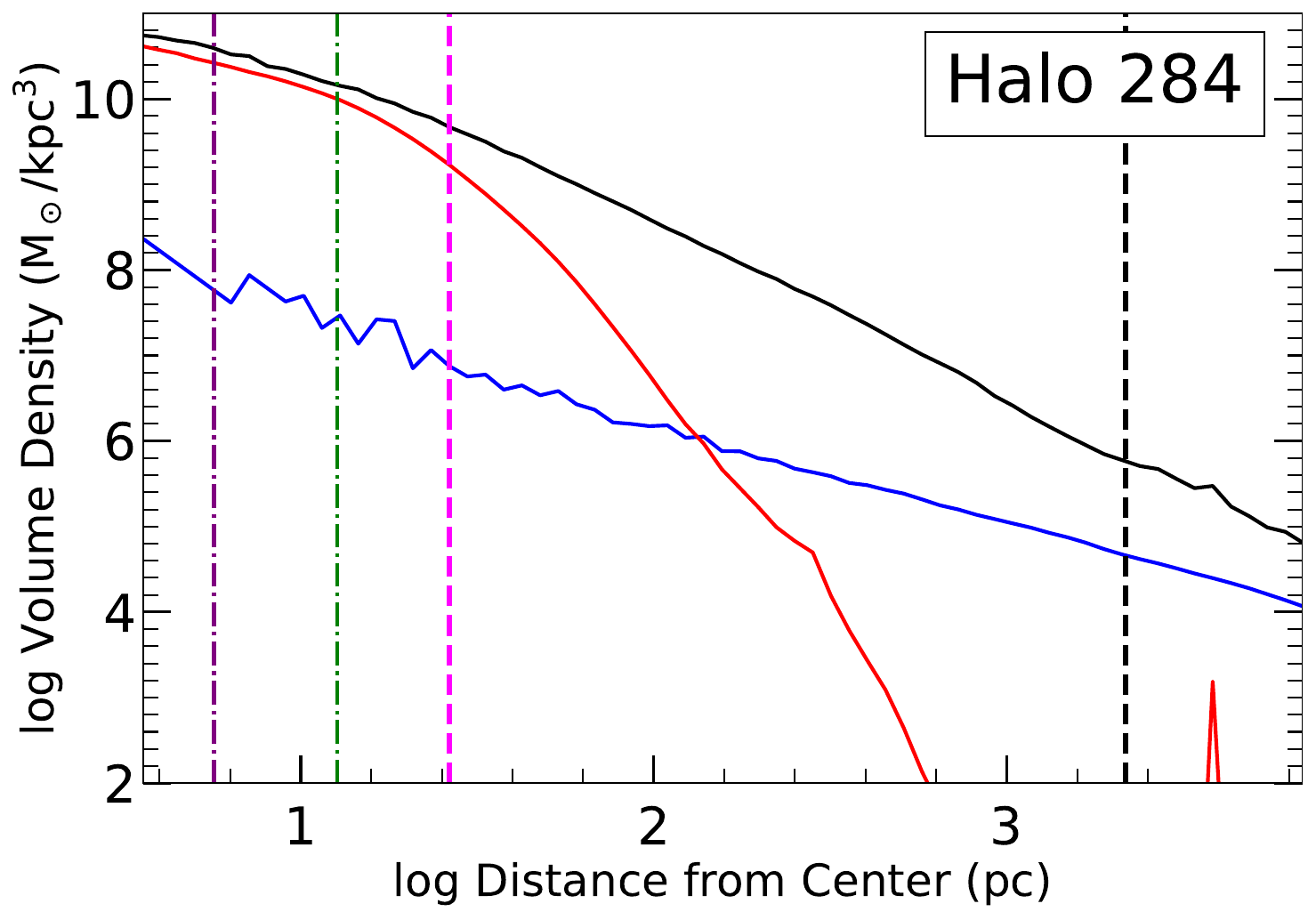}
      \put(60,52){\normalsize  \textbf{$t=1.2$ Gyr}}
    \end{overpic}\\

    \begin{overpic}[width=0.4\textwidth]{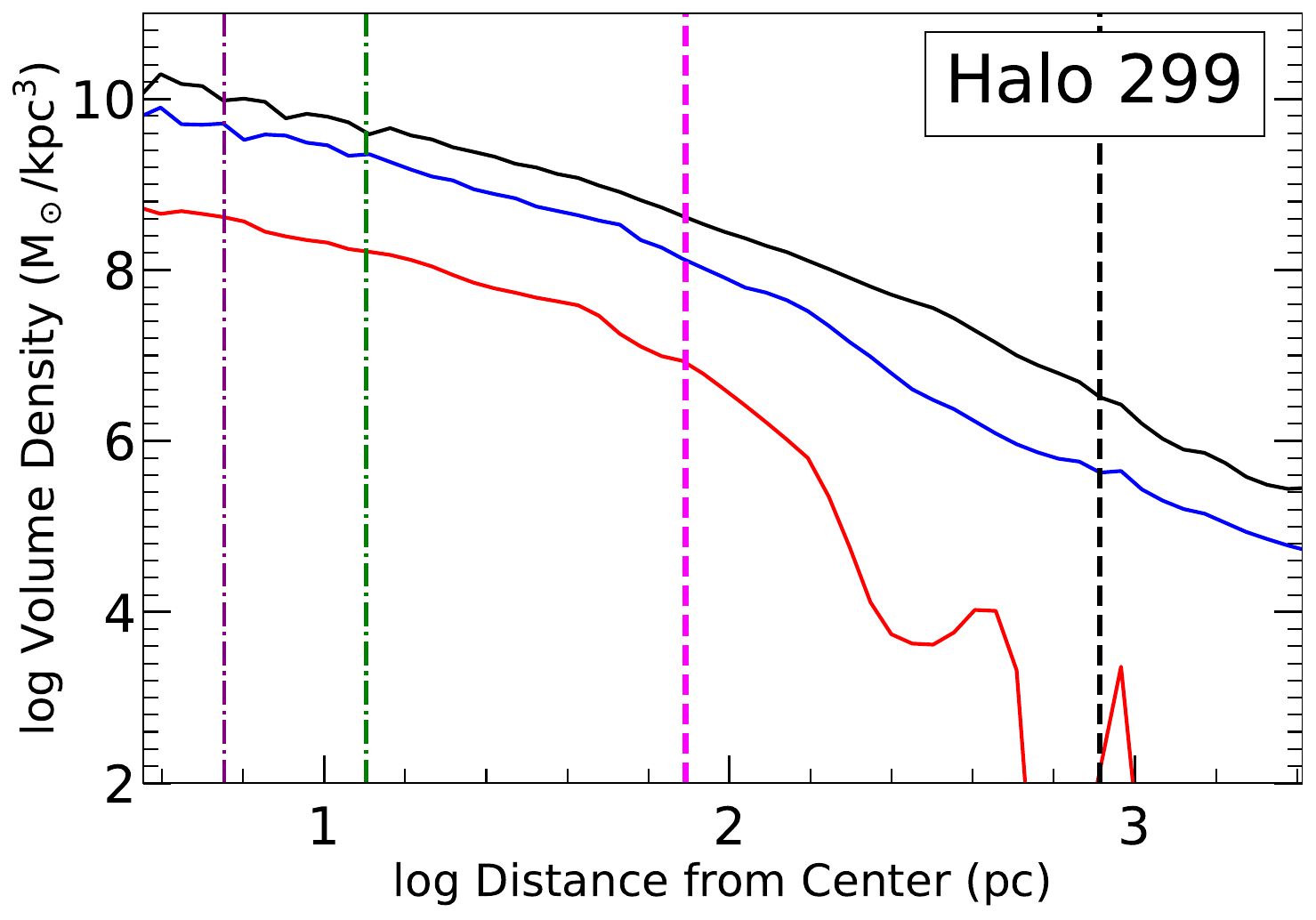}
      \put(65,52){\normalsize  \textbf{EoR}}
    \end{overpic}
    \begin{overpic}[width=0.4\textwidth]{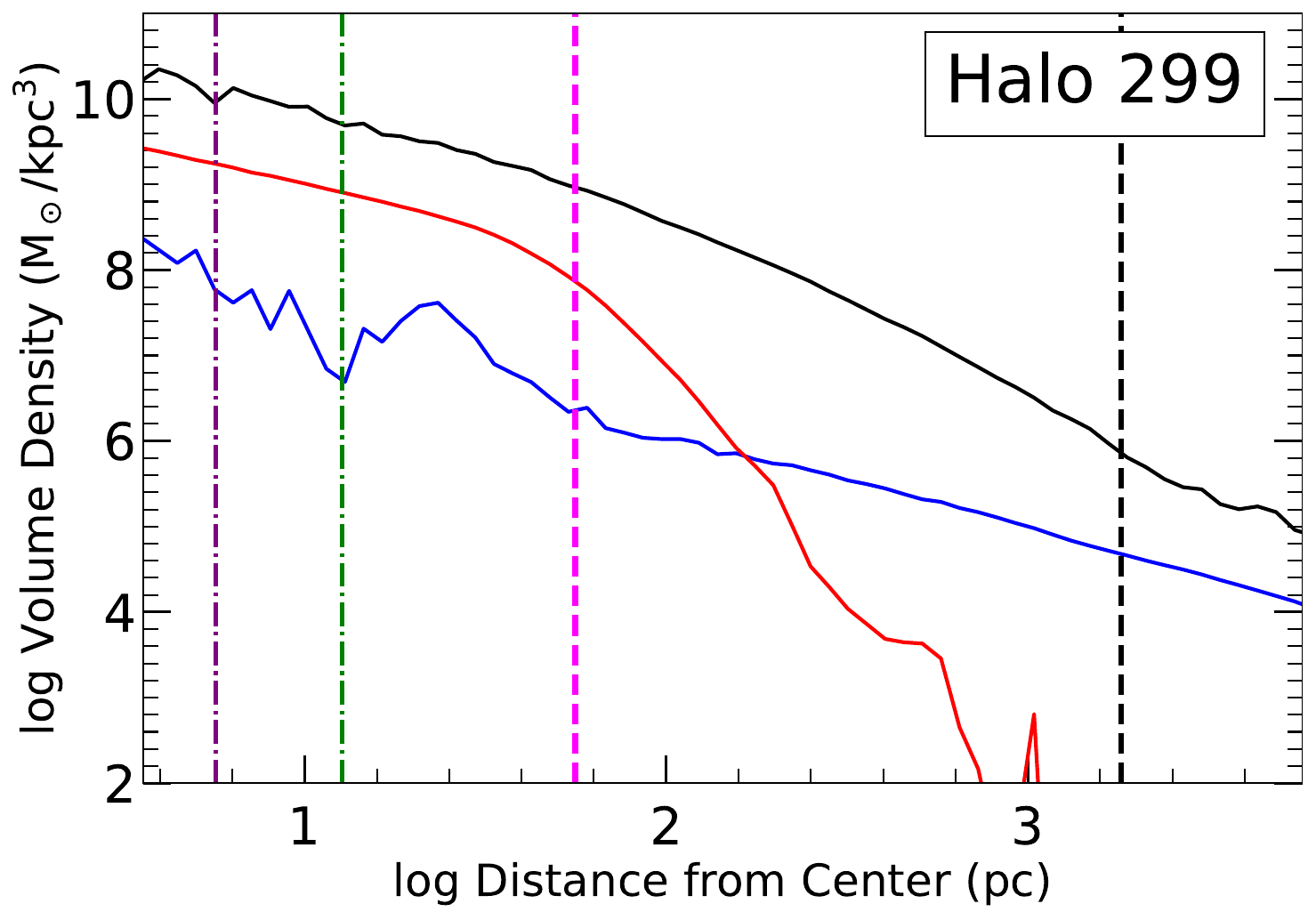}
      \put(60,52){\normalsize  \textbf{$t=1.2$ Gyr}}
    \end{overpic}\\
  \end{center}
  \caption{Radial density profiles of the high-mass halos (Halos 198, 236, 284, and 299 from top to bottom) at the epoch of reionization (left) and the end of the simulation $t=1.2$ Gyr (right). Black, red, and blue curves indicate dark matter, stars, and gas, respectively. The vertical dashed magenta, black, green, and purple lines indicate the half-mass radius of the stellar distribution, the virial radius of the dark matter halo, dark matter softening length (12.7 pc), and the stellar softening length (5.68 pc). {Alt text: Radial density profiles (DM in black, stars in red, gas in blue) for halos H198, H236, H284, H299 at EoR (left) and t=1.2 Gyr (right); vertical lines mark half‑mass, virial, and softening radii.}}
  \label{fig:ReRadial_high}
\end{figure*}

\begin{figure*}
  \begin{center}
    \begin{overpic}[width=0.4\textwidth]{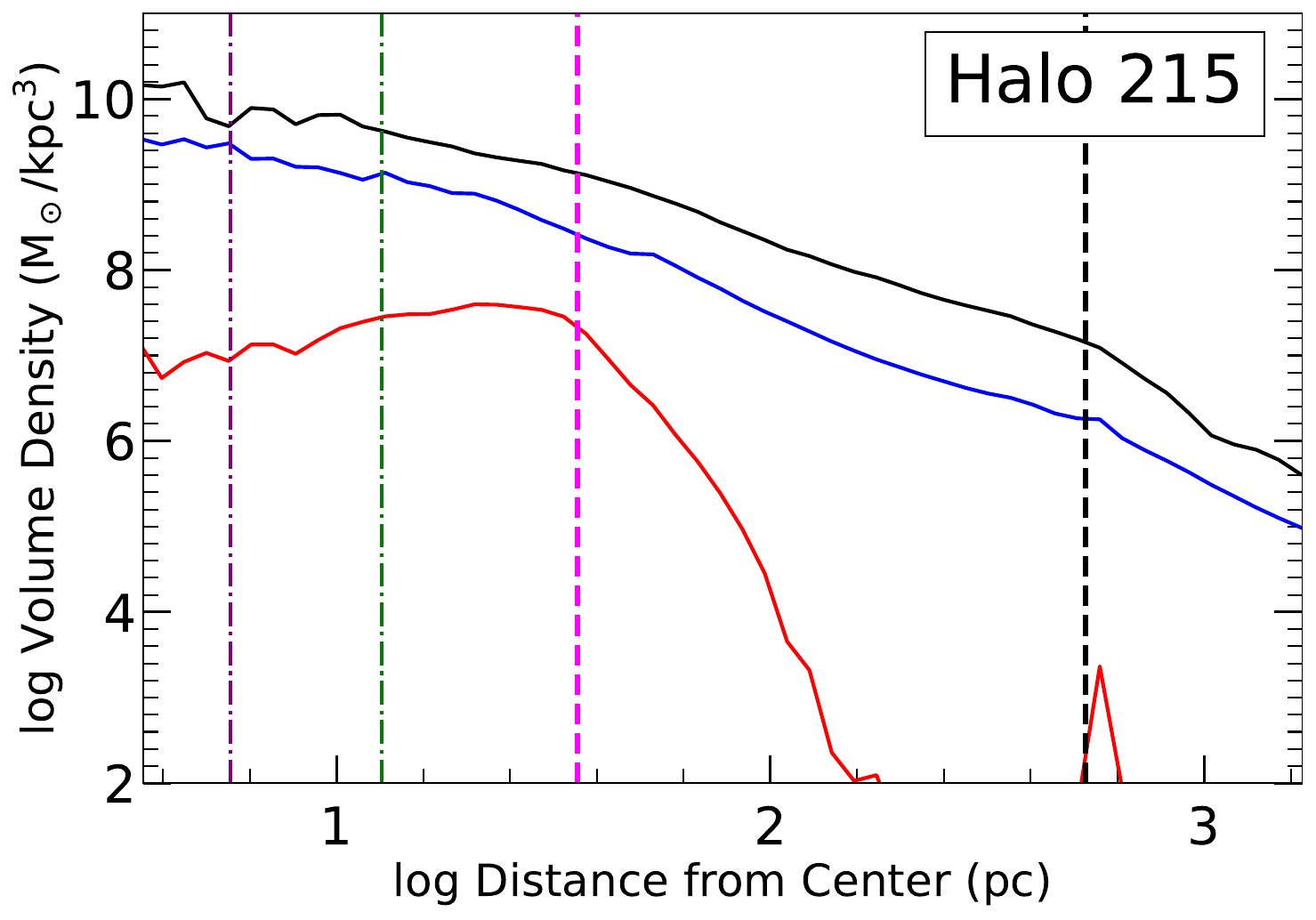}
      \put(65,52){\normalsize  \textbf{EoR}}
    \end{overpic}
    \begin{overpic}[width=0.4\textwidth]{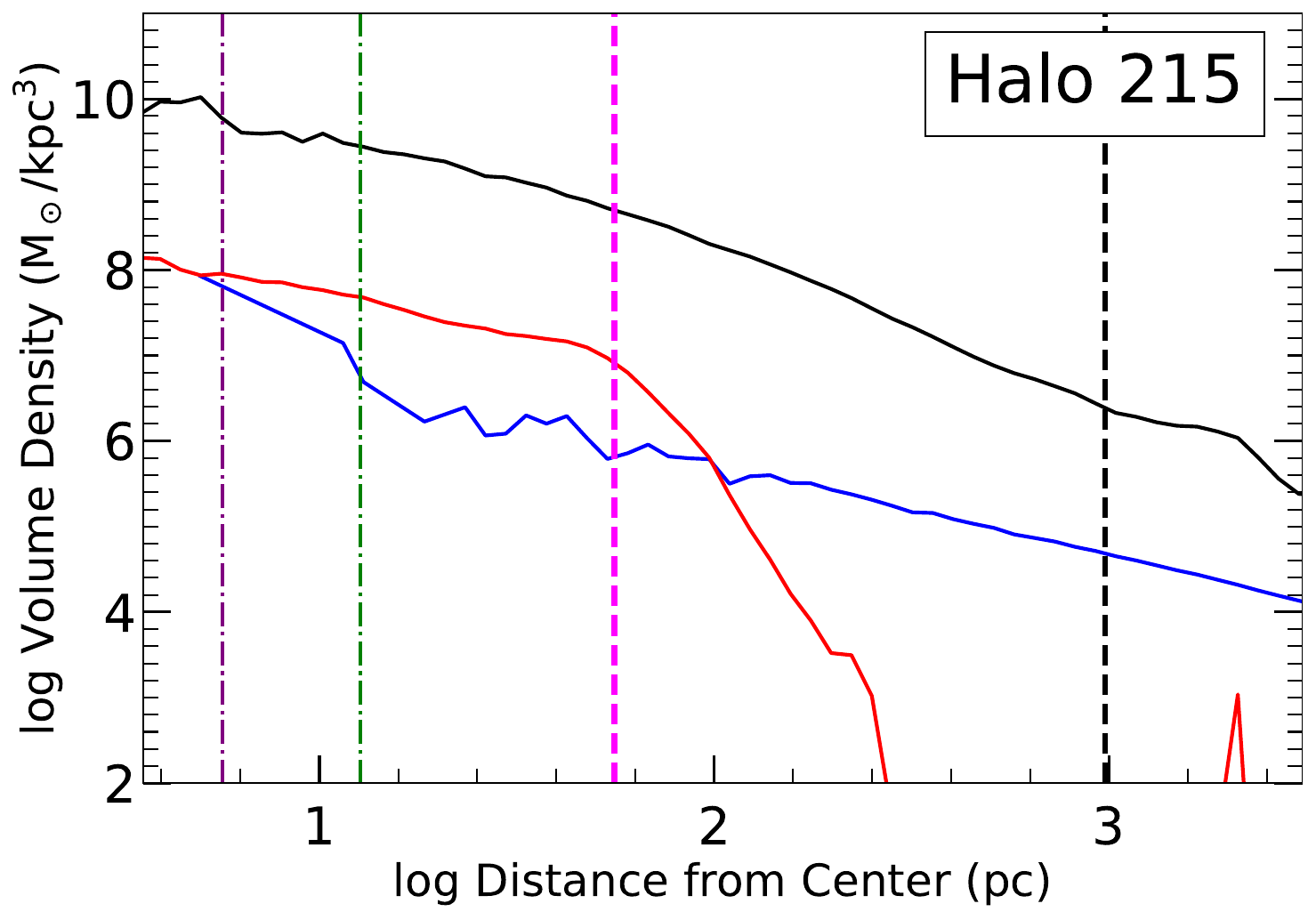}
      \put(60,52){\normalsize  \textbf{$t=1.2$ Gyr}}
    \end{overpic}\\
      \begin{overpic}[width=0.4\textwidth]{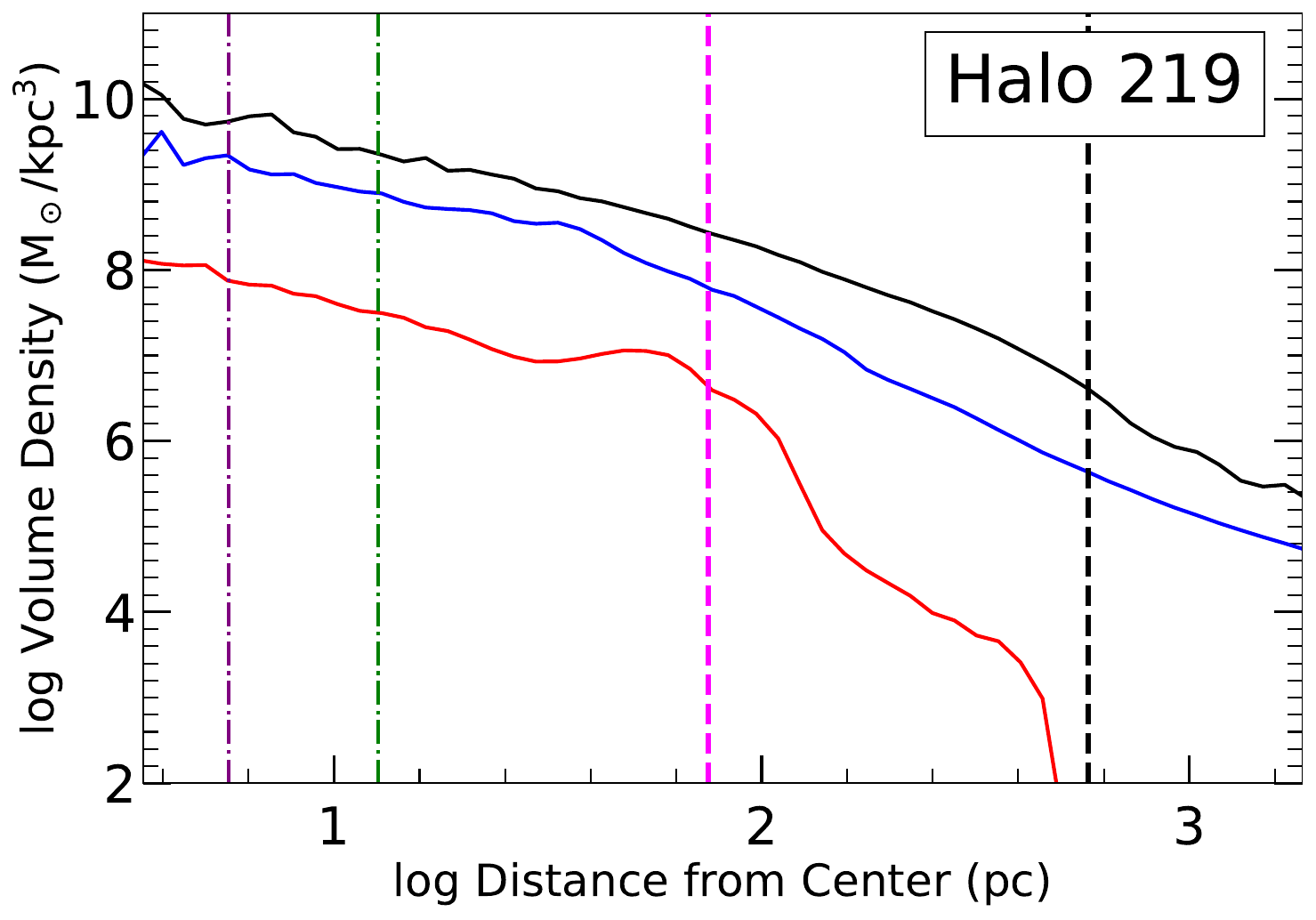}
      \put(65,52){\normalsize  \textbf{EoR}}
    \end{overpic}
    \begin{overpic}[width=0.4\textwidth]{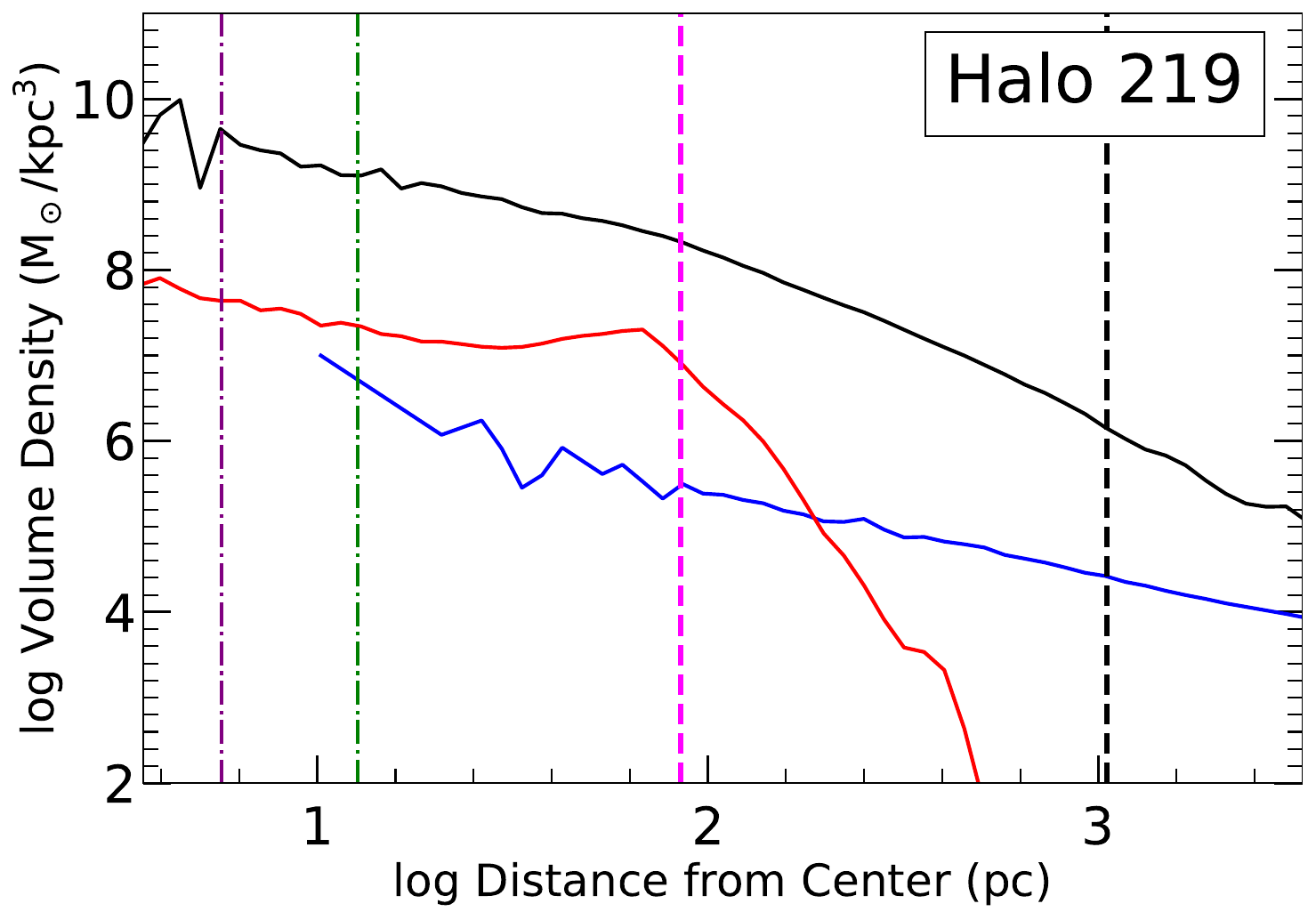}
      \put(60,52){\normalsize  \textbf{$t=1.2$ Gyr}}
    \end{overpic}\\
    \begin{overpic}[width=0.4\textwidth]{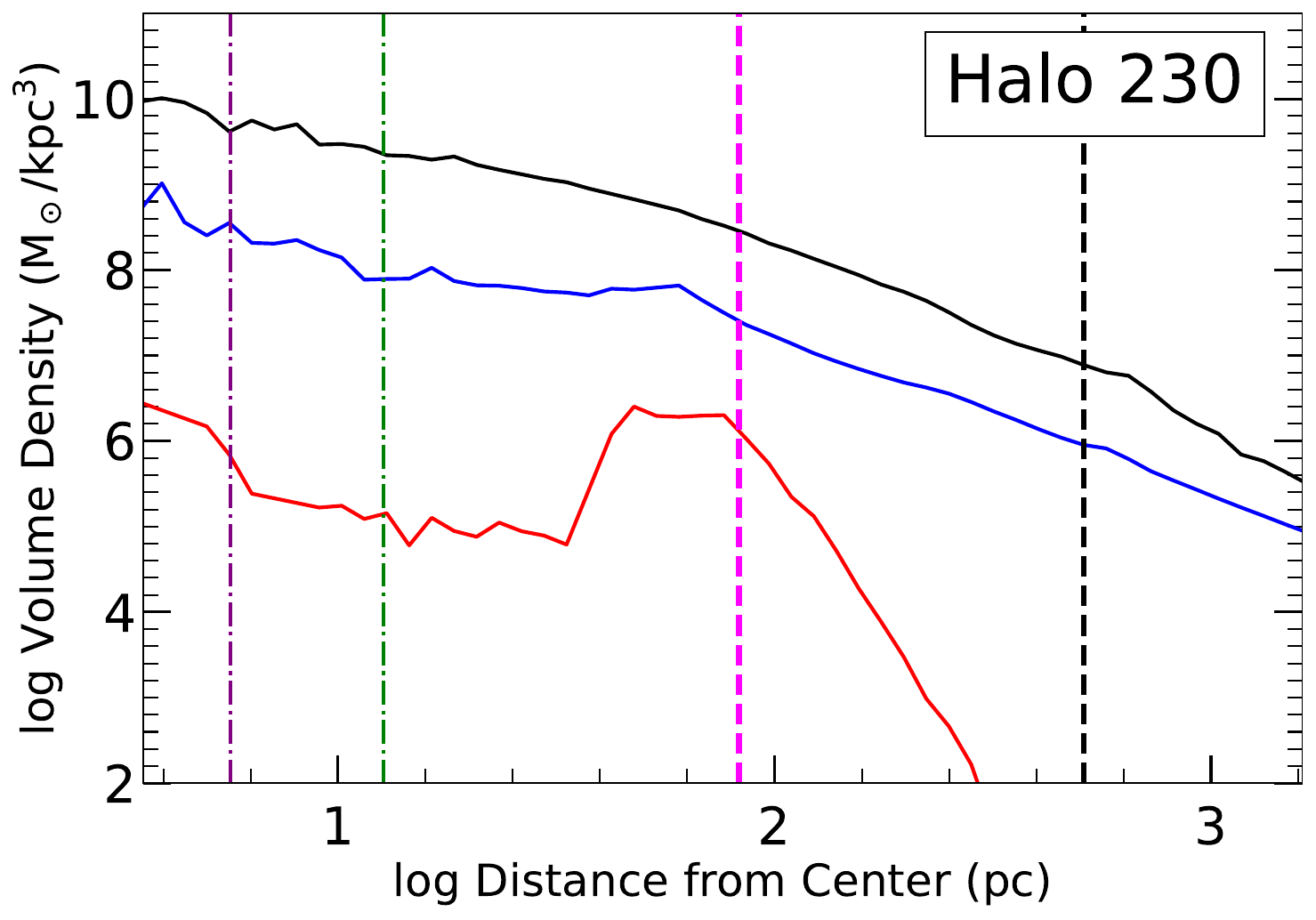}
      \put(65,52){\normalsize  \textbf{EoR}}
    \end{overpic}
    \begin{overpic}[width=0.4\textwidth]{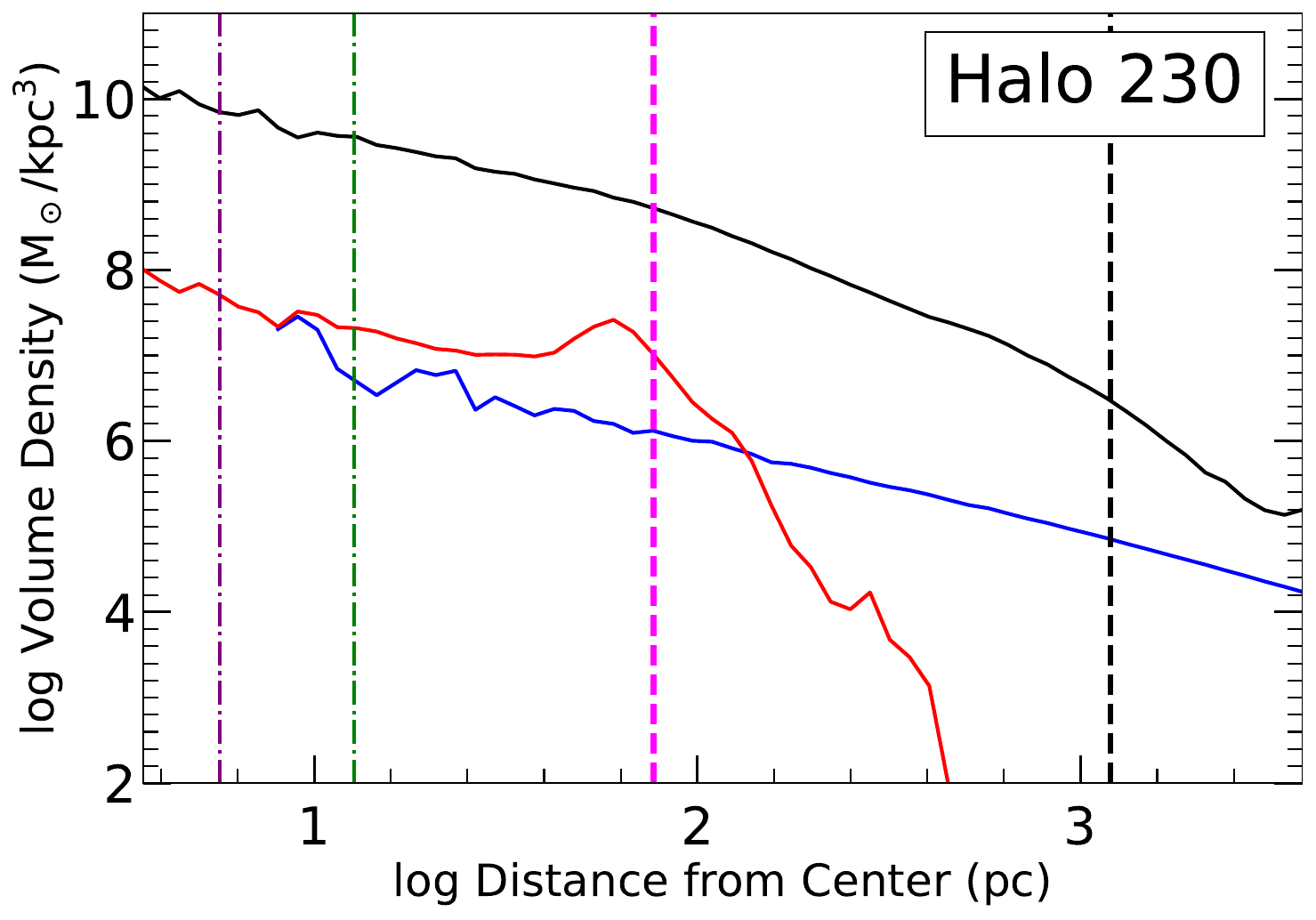}
      \put(60,52){\normalsize  \textbf{$t=1.2$ Gyr}}
    \end{overpic}\\
    \begin{overpic}[width=0.4\textwidth]{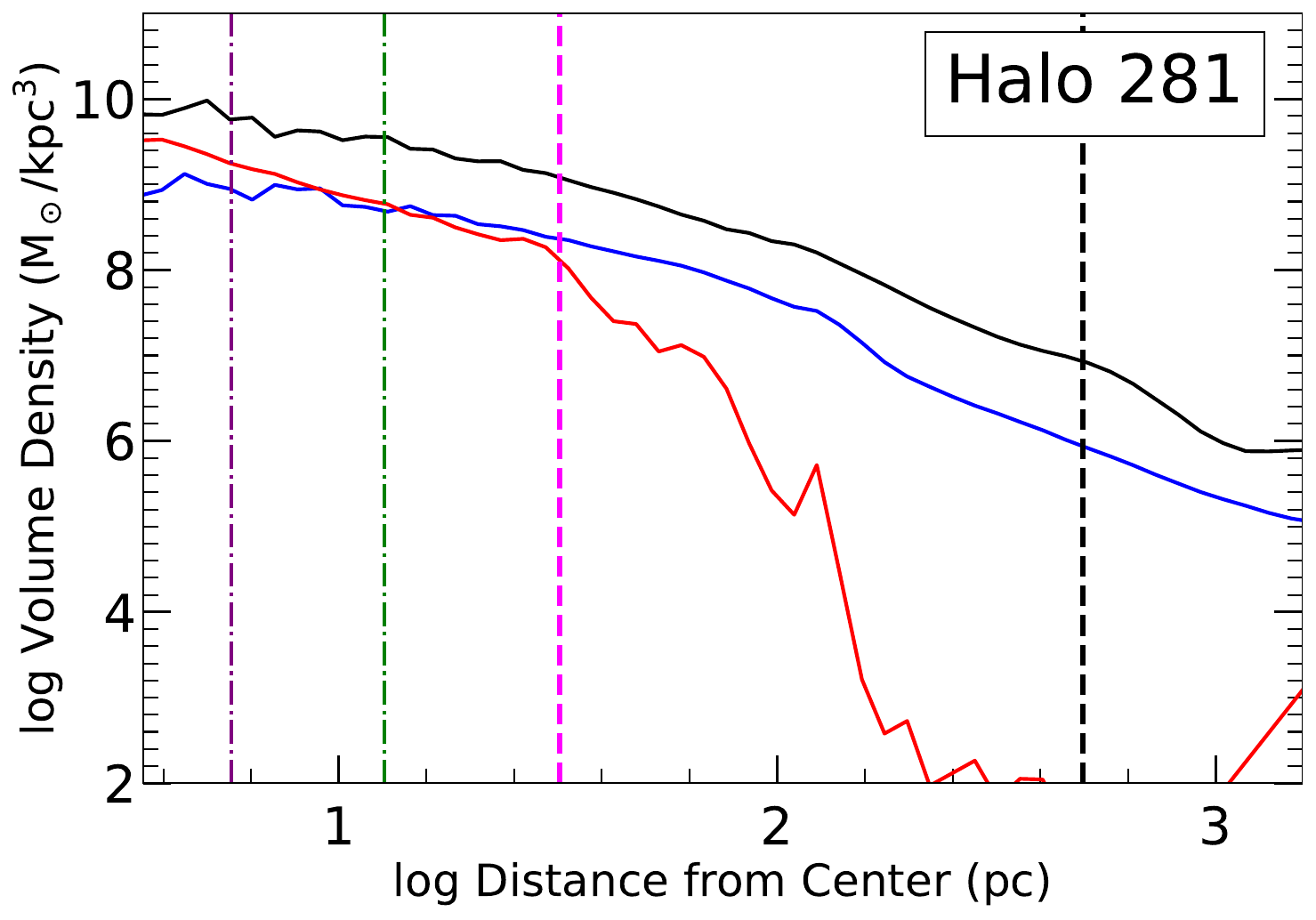}
      \put(65,52){\normalsize  \textbf{EoR}}
    \end{overpic}
    \begin{overpic}[width=0.4\textwidth]{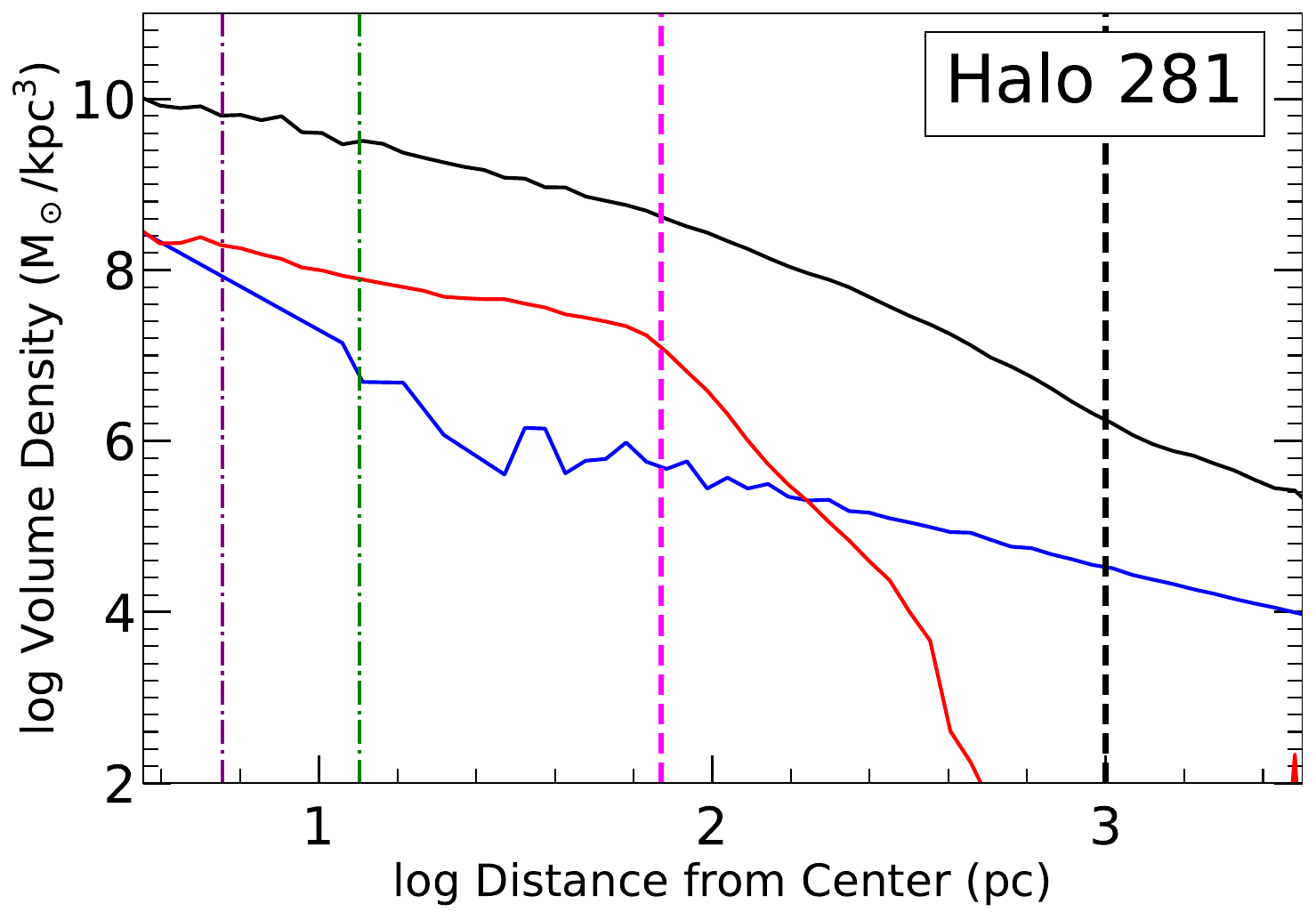}
      \put(60,52){\normalsize  \textbf{$t=1.2$ Gyr}}
    \end{overpic}\\
  \end{center}
  \caption{Same as Fig.~\ref{fig:ReRadial_high} but for the low-mass halos (Halos 215, 219, 230, and 281). {Alt text: Radial density profiles (DM in black, stars in red, gas in blue) for halos H215, H219, H230, H281 at EoR (left) and t=1.2 Gyr (right); vertical lines mark half‑mass, virial, and softening radii.}}
  \label{fig:ReRadial_low}
\end{figure*}

In Figs.~\ref{fig:ReRadial_high} and \ref{fig:ReRadial_low}, we present the density profiles at the EoR and the end of simulations. As shown in the figures, the stellar densities of the UCD-like ones, i.e., Halos 198 and 284, are as high as those of their dark matter density. In addition, their central dark matter density increased significantly from the EoR to the end of the simulations, whereas it did not in the other halos. In previous subsections, we categorized Halo 299 as a UCD-like galaxy. However, the central stellar density is lower than that of dark matter compared to Halos 198 and 284. Thus, Halo 299 can be a marginal object; indeed, it exceeded the SFE$=0.1$ criterion by only a small amount (see Fig.~\ref{fig:equ}). Although the central density of Halo 299 is lower than that of Halos 198 and 284, the density profile is smooth, unlike that of Halo 236, which is undergoing an ongoing merger at the end of the simulation. 

From the density profiles, we did not find a dark matter density criterion for the formation of UCD-like compact galaxies. At the EoR, the dark matter density of Halo 198 is relatively high compared to the others. However, the central dark matter density is similar to the others, although those of Halos 230 and 281 are even lower. Therefore, we consider that the dark matter halo density is not a necessary condition for forming a compact stellar distribution; rather, as a result of the formation of a compact stellar system, the dark matter density profile deepens. Such a phenomenon has also observed in previous simulations \citep{2025MNRAS.536..314M,2026arXiv260113765K}.

Furthermore, one may notice that the stellar density peak in Halo 281 decreases from $\sim 10^{10}\ \mathrm{M}_\odot\,\mathrm{kpc}^{-3}$ to $\sim 10^{8}\ \mathrm{M}_\odot\,\mathrm{kpc}^{-3}$. 
We confirmed that a merger involving dark matter occurred during this period, resulting in a more diffuse stellar system. This is consistent with previous simulations showing that mergers can increase the half-mass radius of dwarf galaxies \citep{revaz2023compactness}.

\section{Discussion: Relation between large-scale structures and star formation efficiency}

\begin{figure*}
    \begin{center}
        \includegraphics[width=0.33\textwidth]{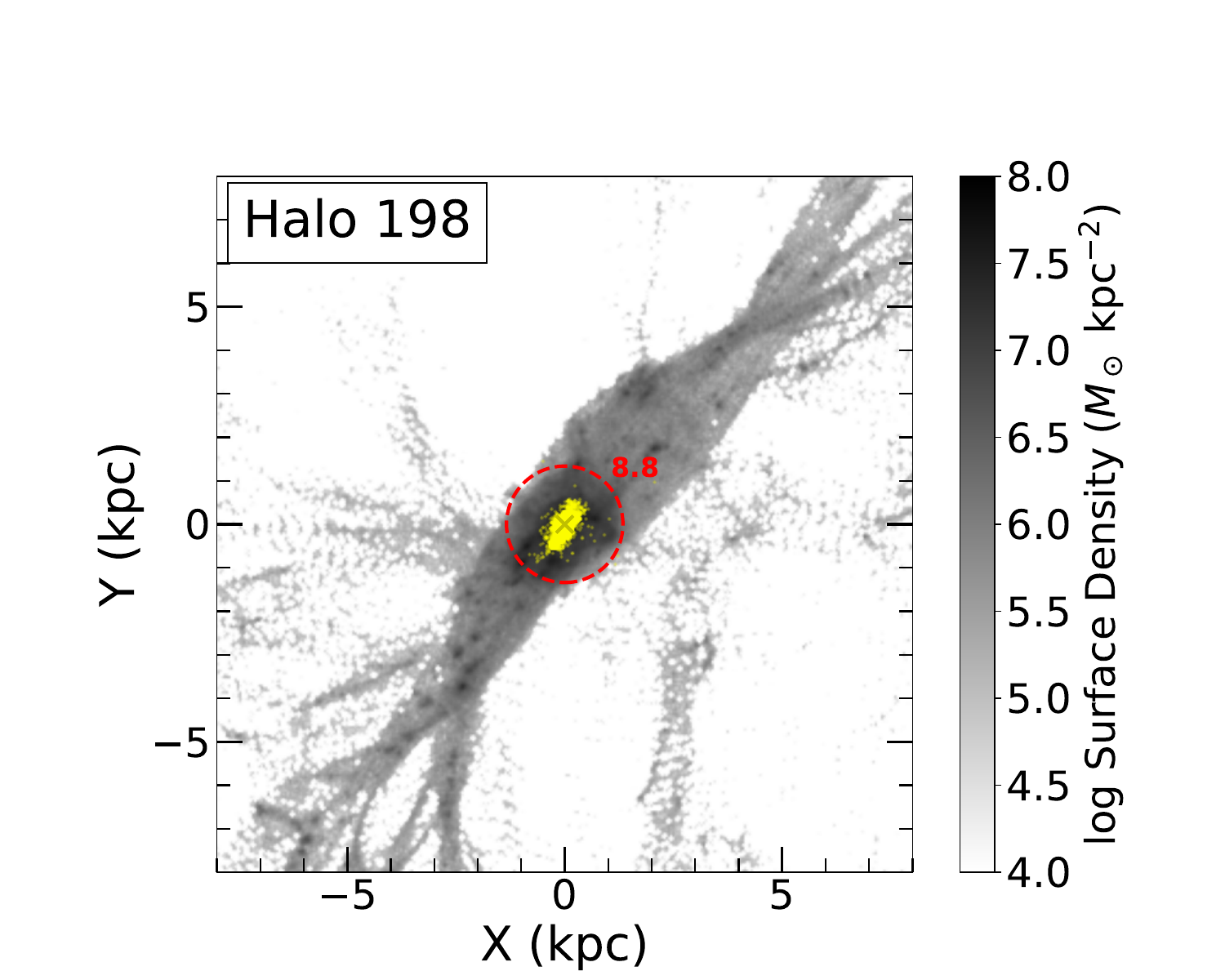}
        \includegraphics[width=0.33\textwidth]{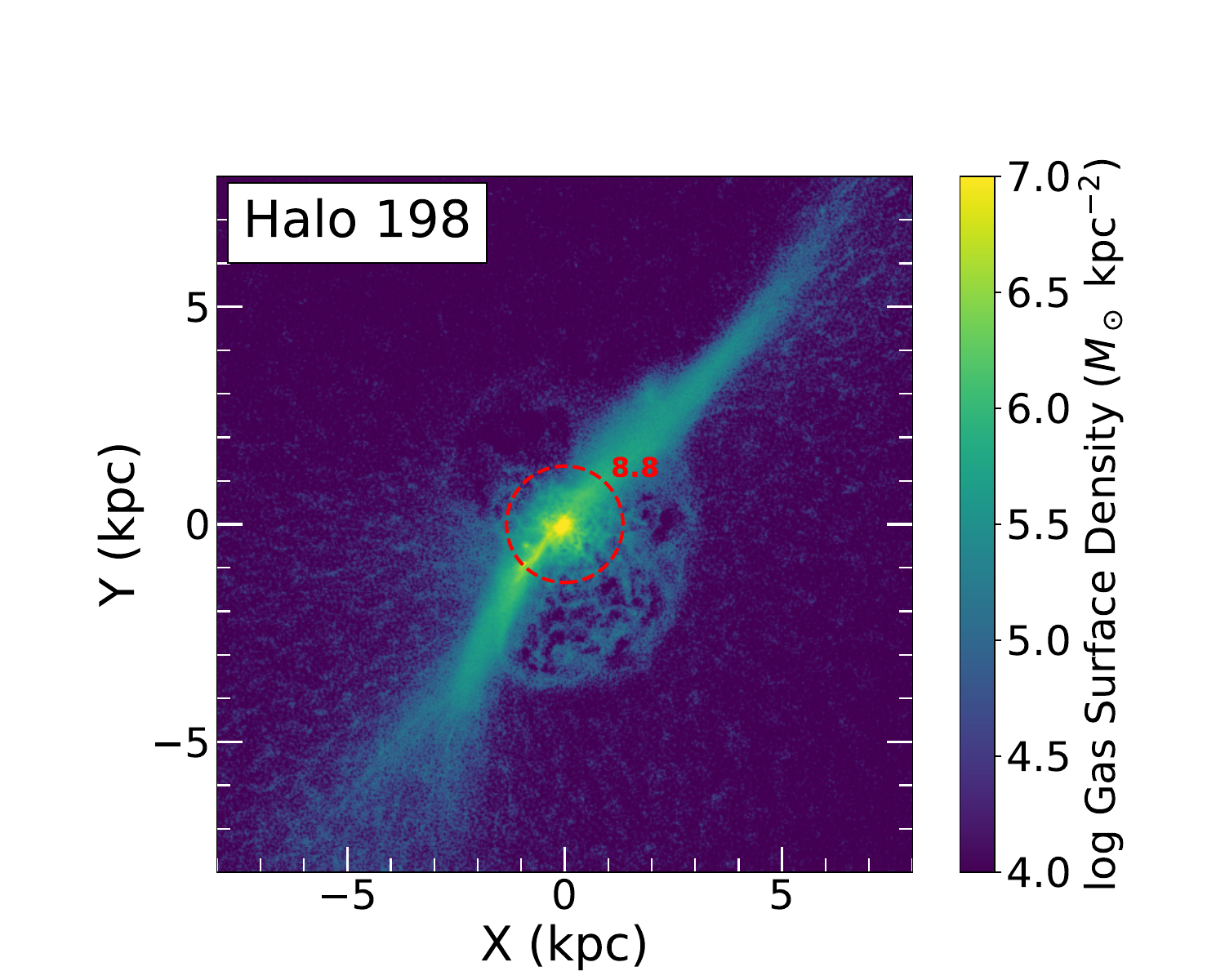} 
        \includegraphics[width=0.33\textwidth]{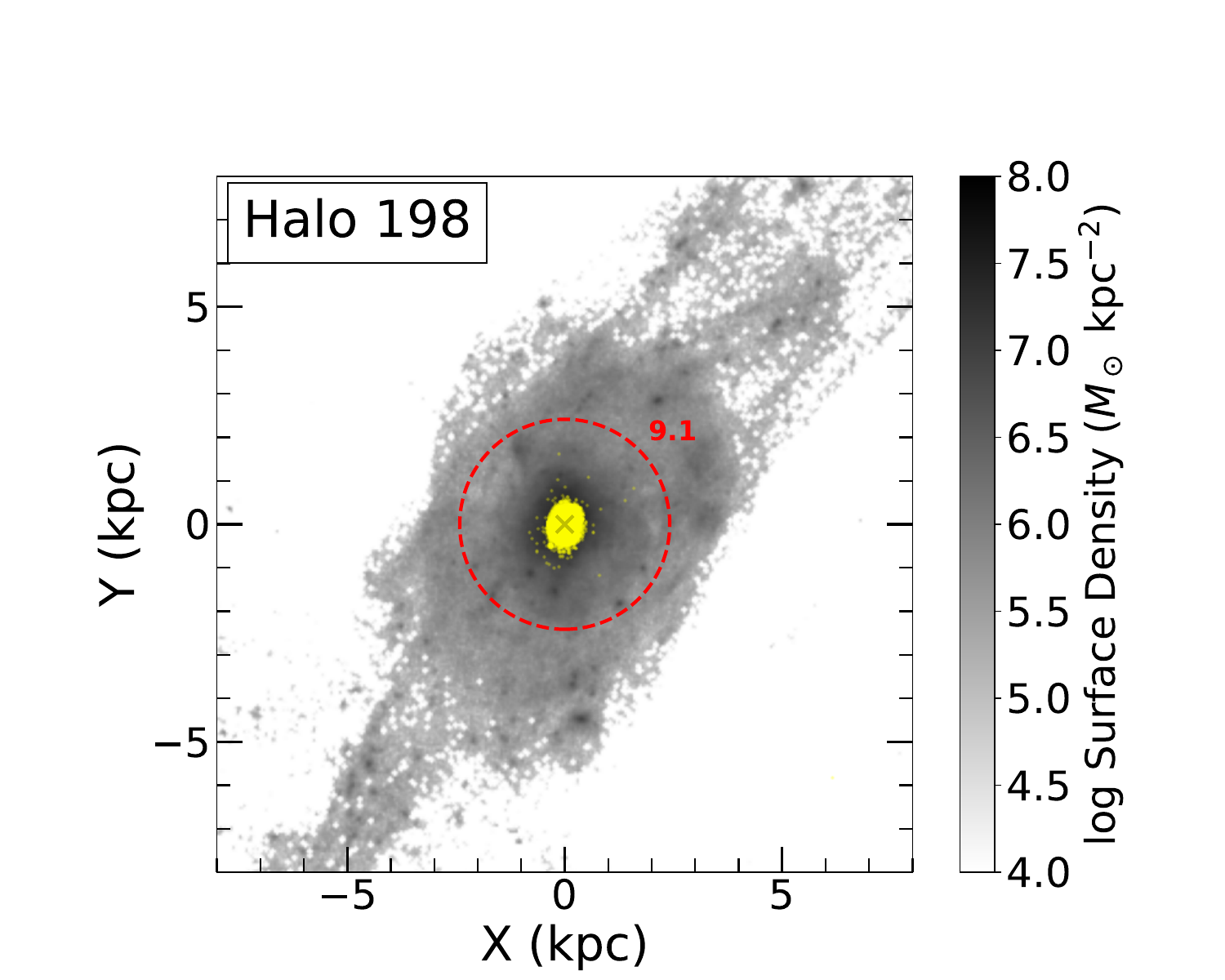}\\
        \includegraphics[width=0.33\textwidth]{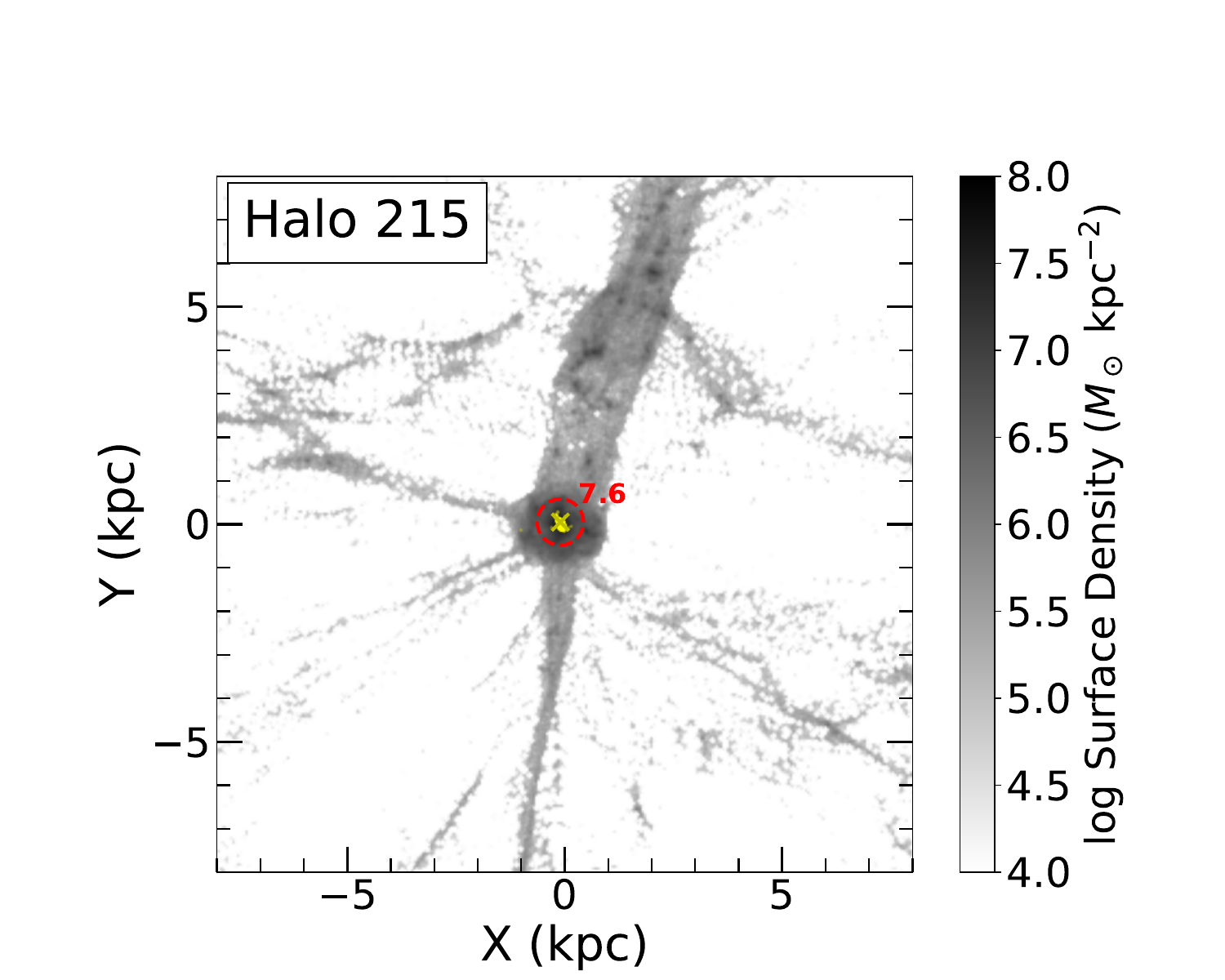}
        \includegraphics[width=0.33\textwidth]{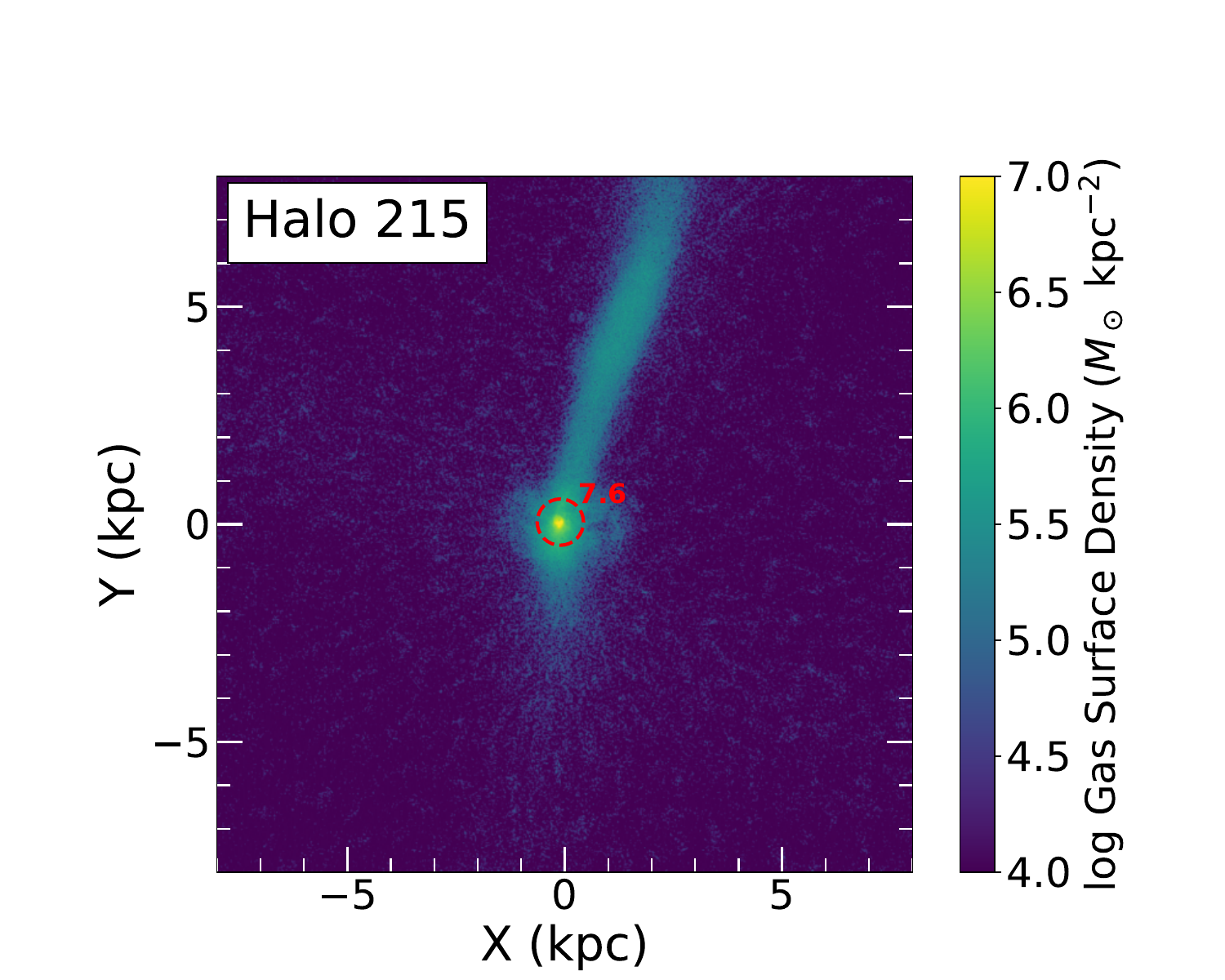}
        \includegraphics[width=0.33\textwidth]{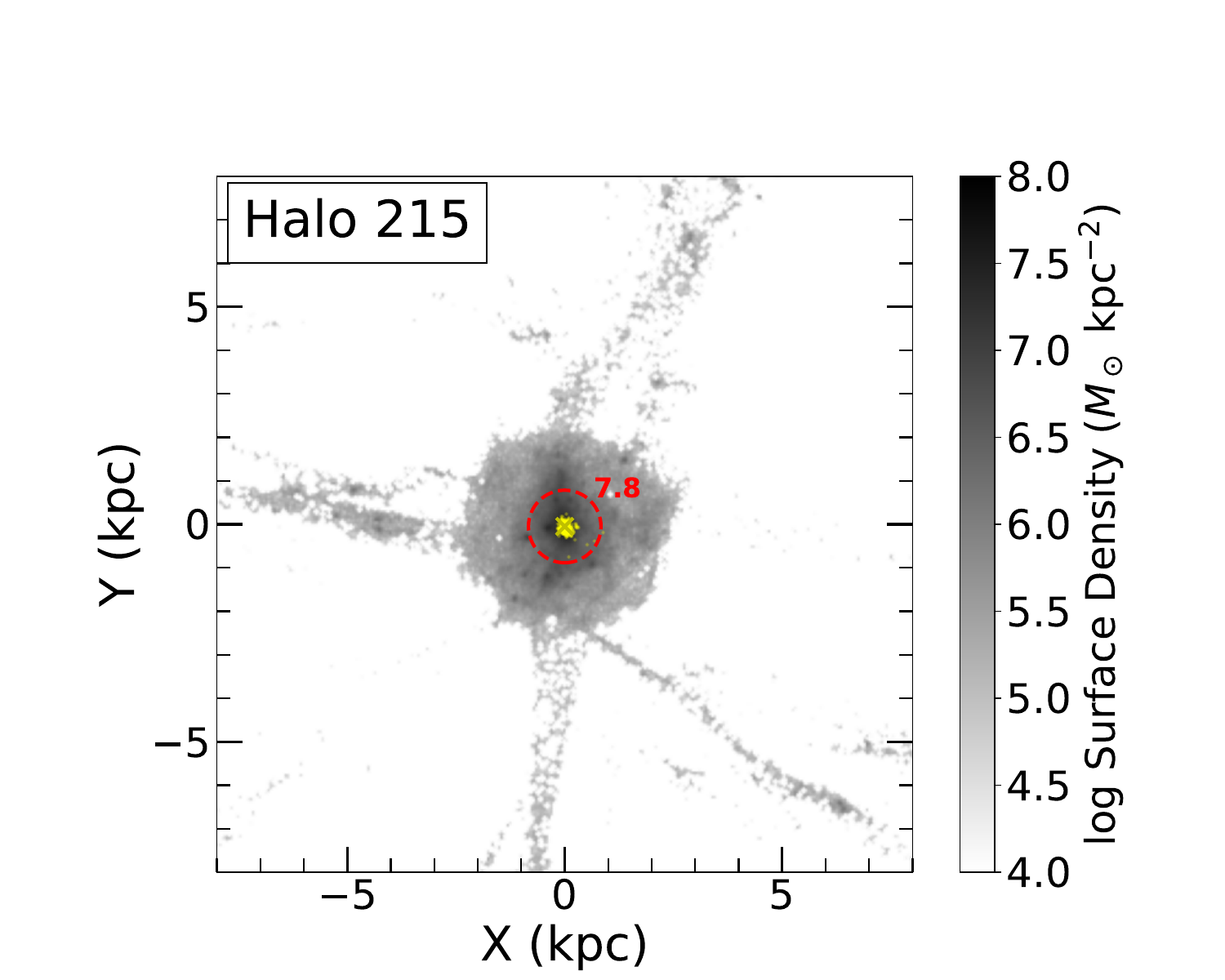}\\
        \includegraphics[width=0.33\textwidth]{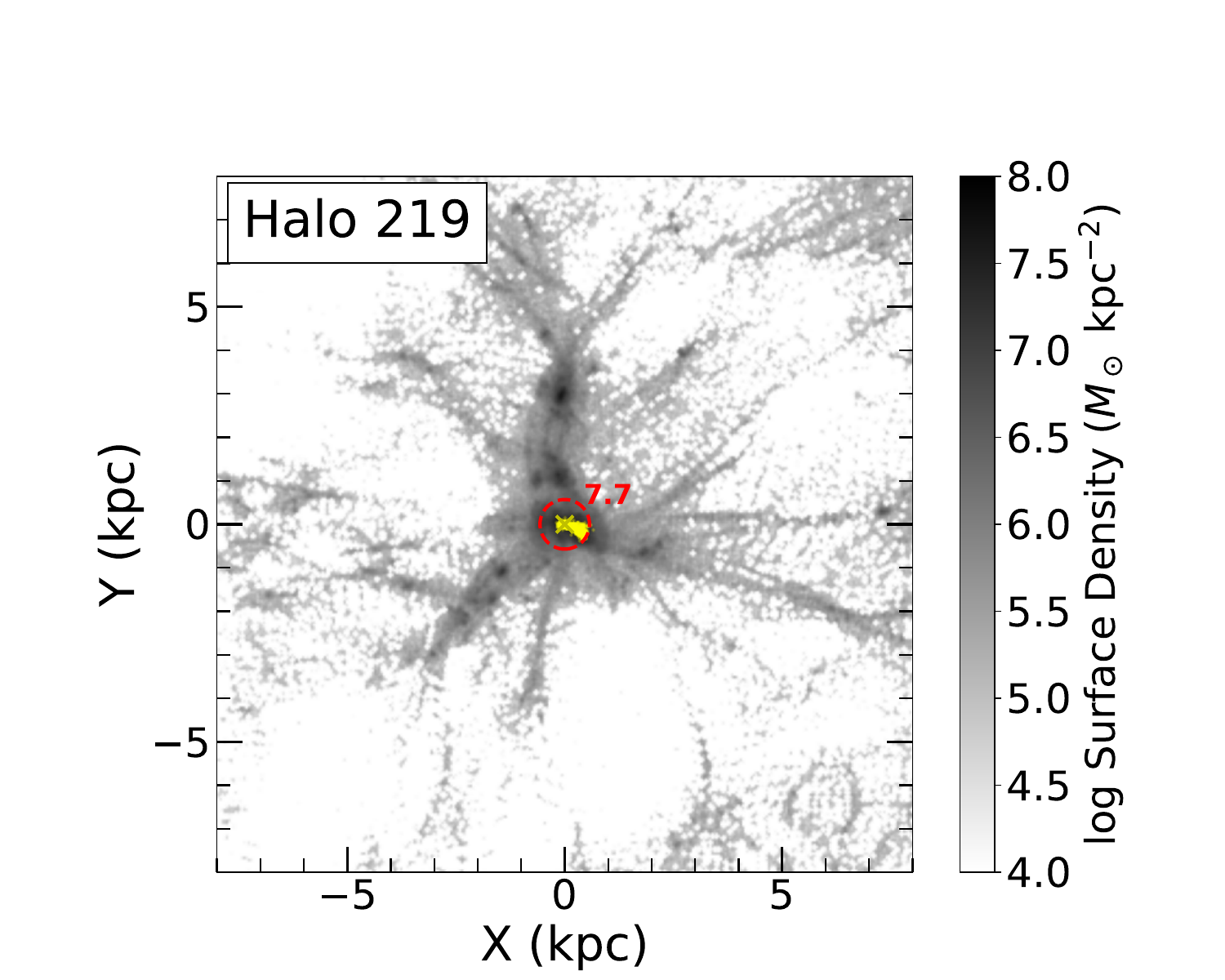}
        \includegraphics[width=0.33\textwidth]{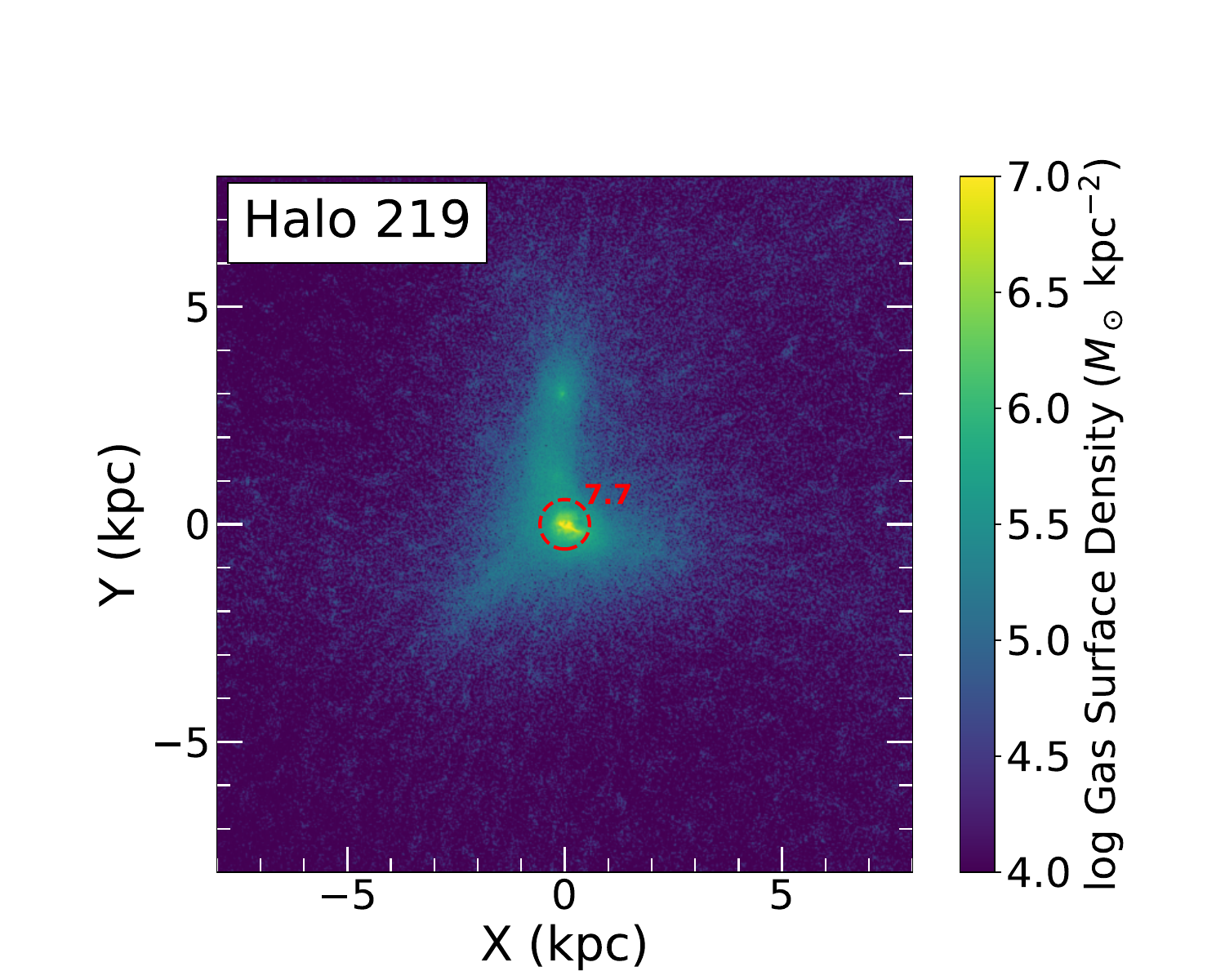}
        \includegraphics[width=0.33\textwidth]{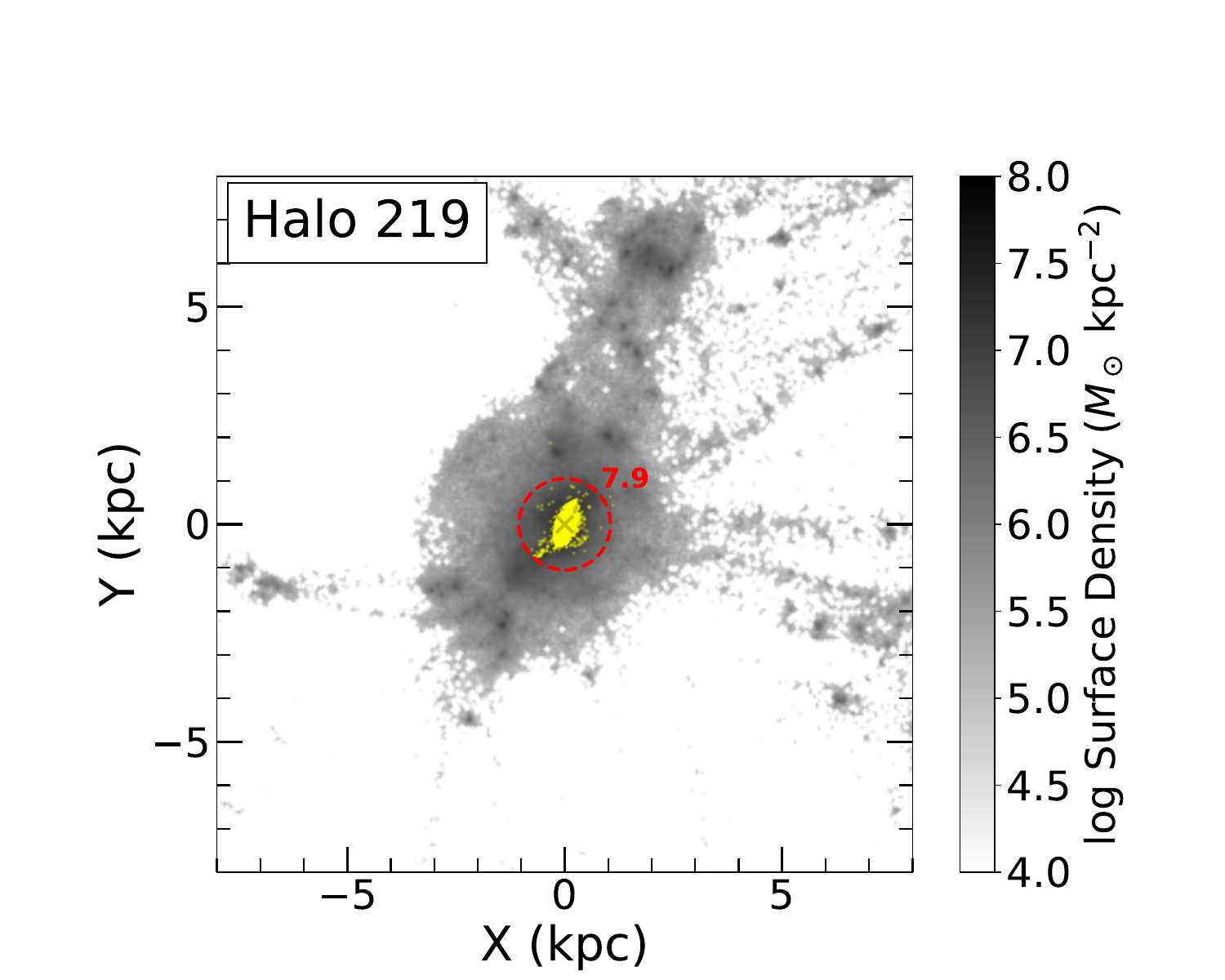}\\
        \includegraphics[width=0.33\textwidth]{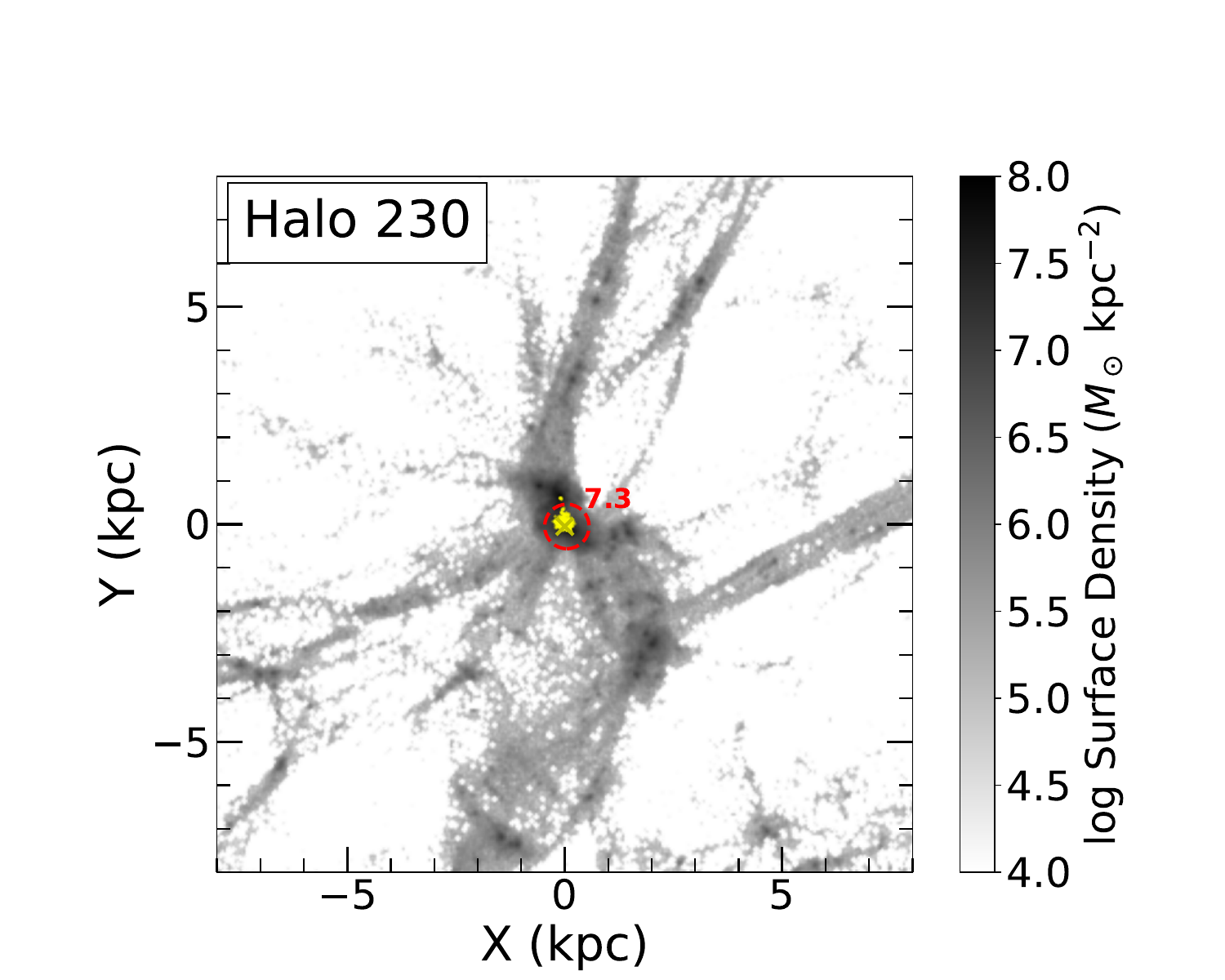}
        \includegraphics[width=0.33\textwidth]{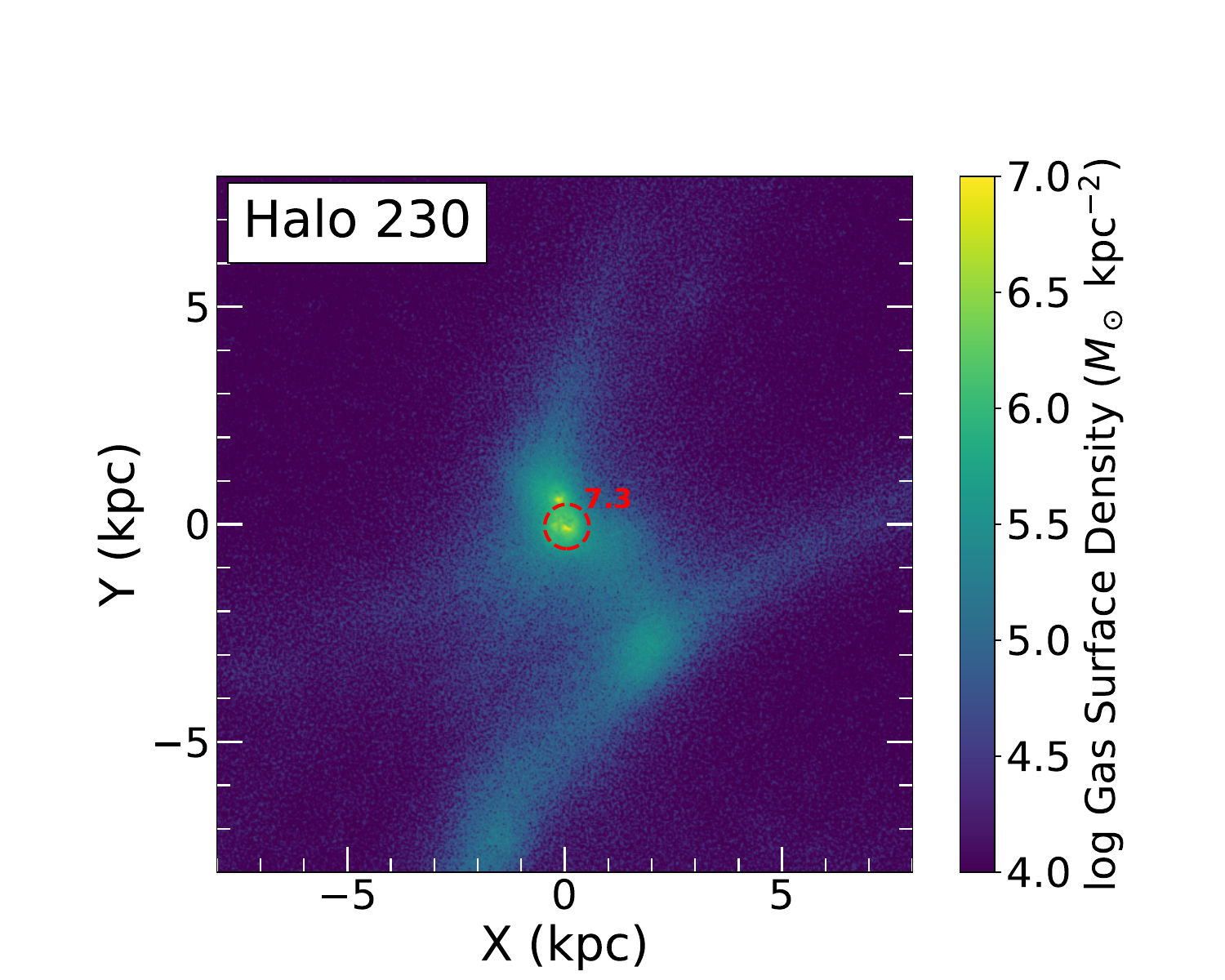}
        \includegraphics[width=0.33\textwidth]{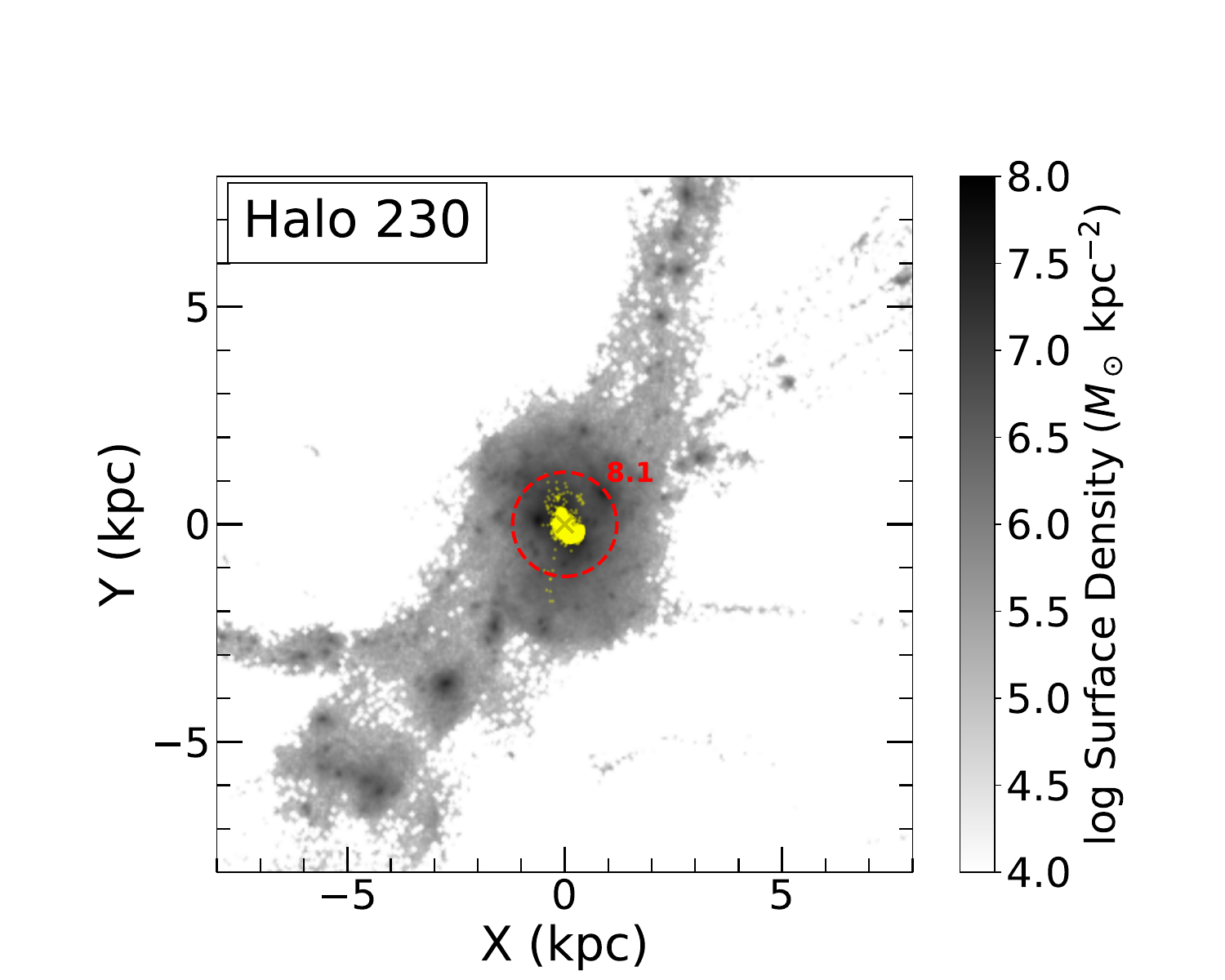}\\
    \end{center}
    \caption{Same as \rfig{fig:snapshots1} but for the large scale with 16 kpc $\times$ 16 kpc region. {Alt text: Snapshots of Halos 198, 215, 219, and 230 (rows top to bottom) in a 16 $\times$ 16 kpc region: left panels show dark-matter surface density (grayscale) with stellar positions as yellow dots at the epoch of reionization; middle panels show cold gas surface density (T<1000 K); right panels show dark-matter and stellar distribution at t = 1.2 Gyr; red circles and labels indicate virial radius and mass, yellow crosses mark halo centers.}}
    \label{fig:snapshots_large1}
\end{figure*}

\begin{figure*}
    \begin{center}
        \includegraphics[width=0.33\textwidth]{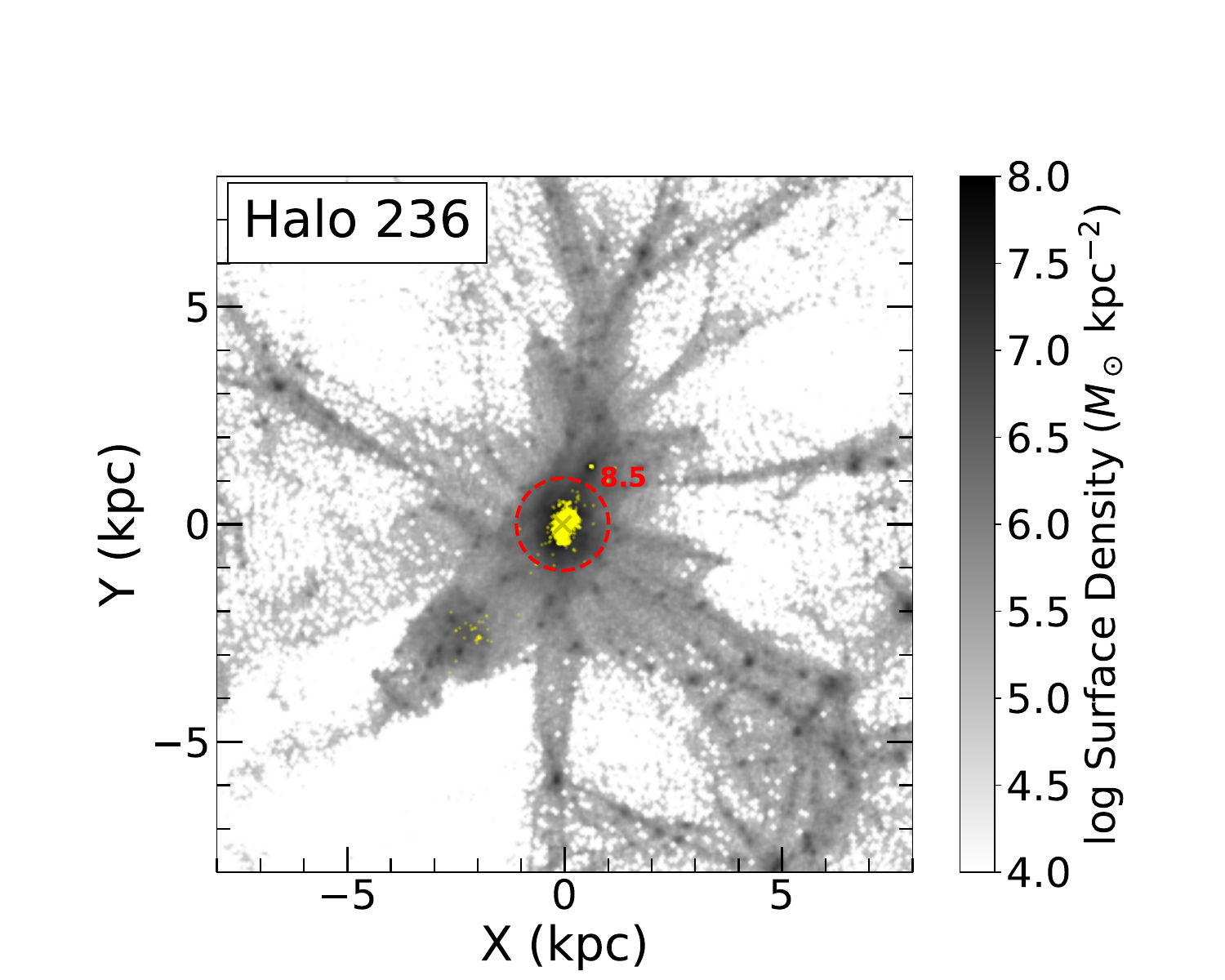}
        \includegraphics[width=0.33\textwidth]{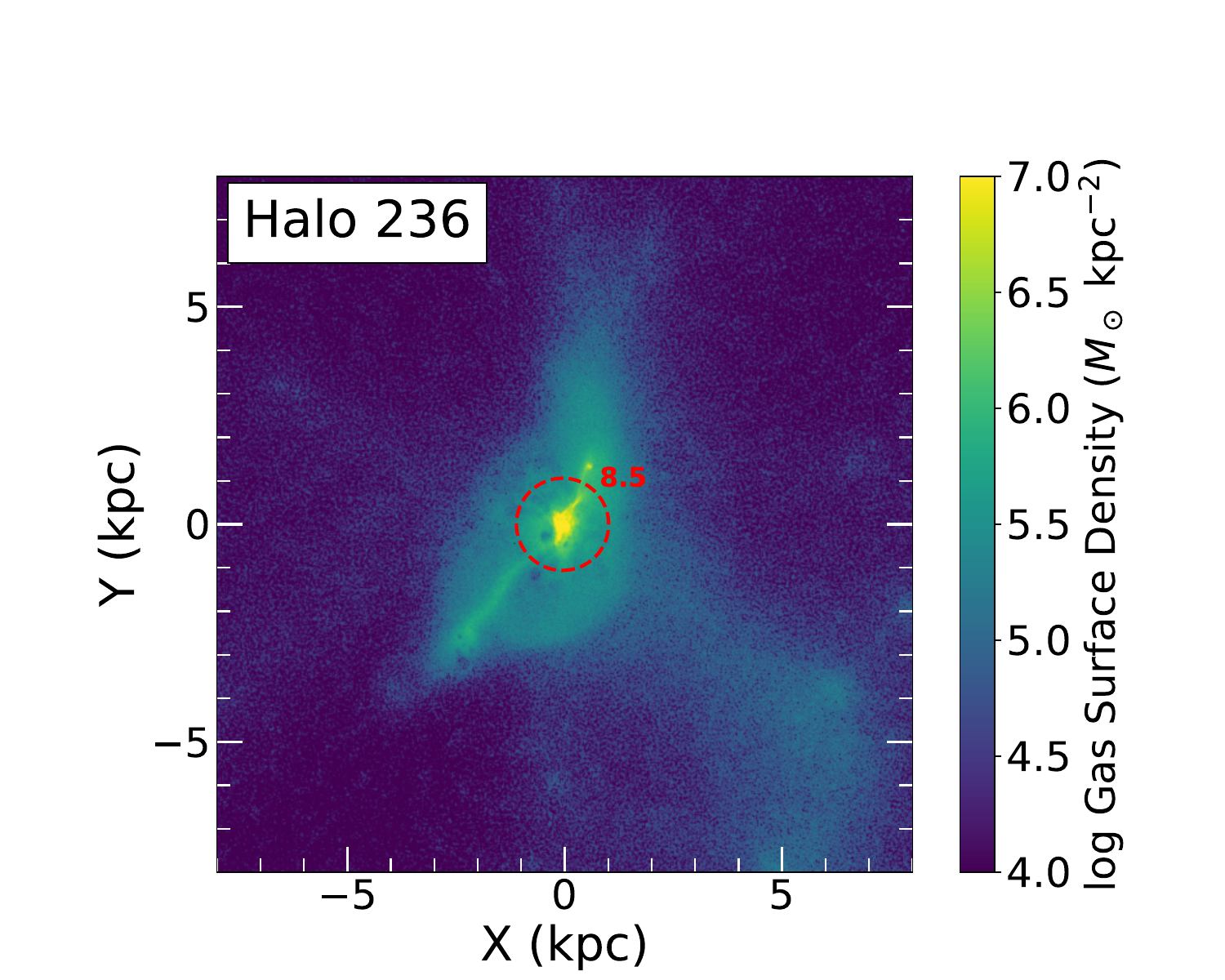}
        \includegraphics[width=0.33\textwidth]{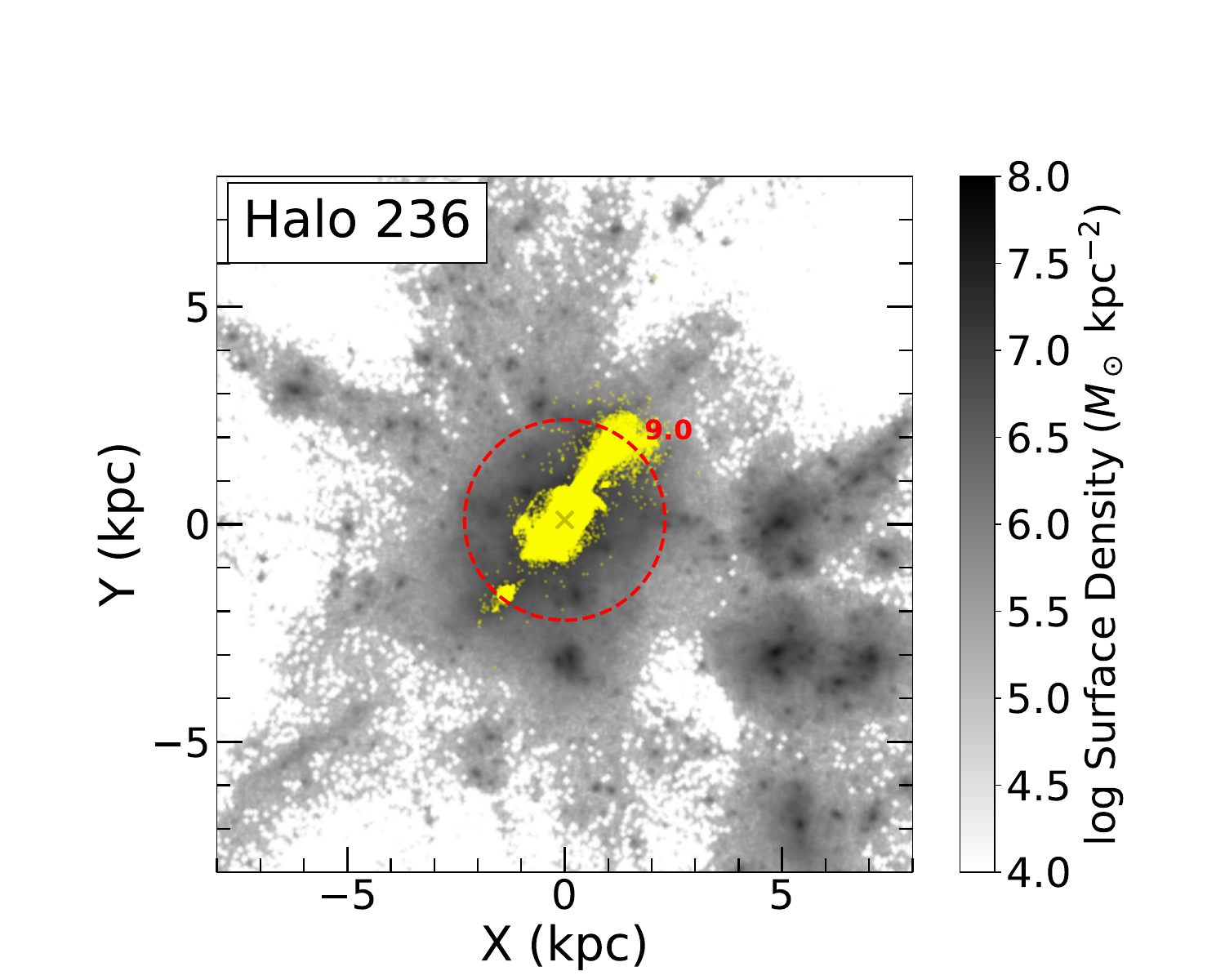}\\ 
        \includegraphics[width=0.33\textwidth]{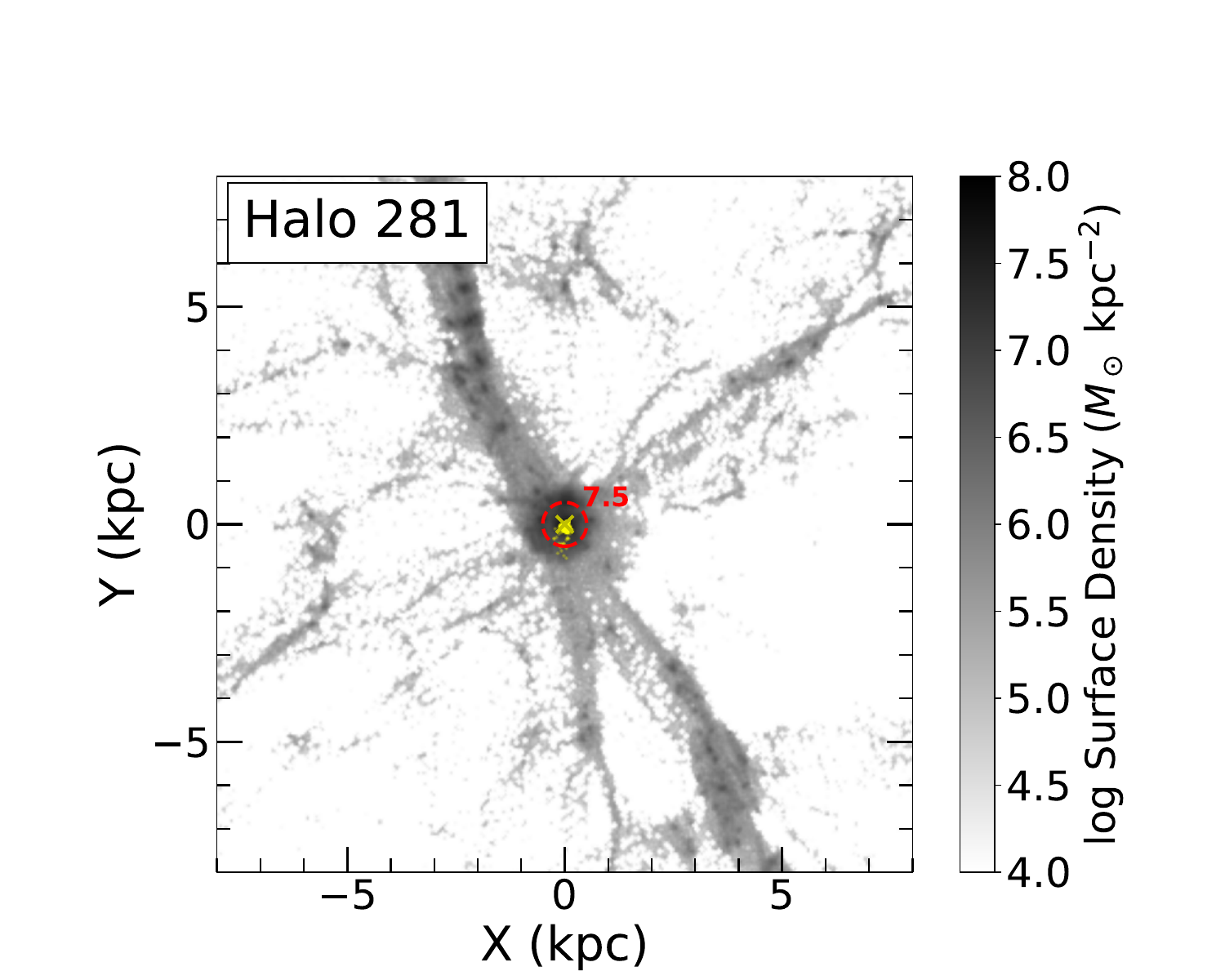}
        \includegraphics[width=0.33\textwidth]{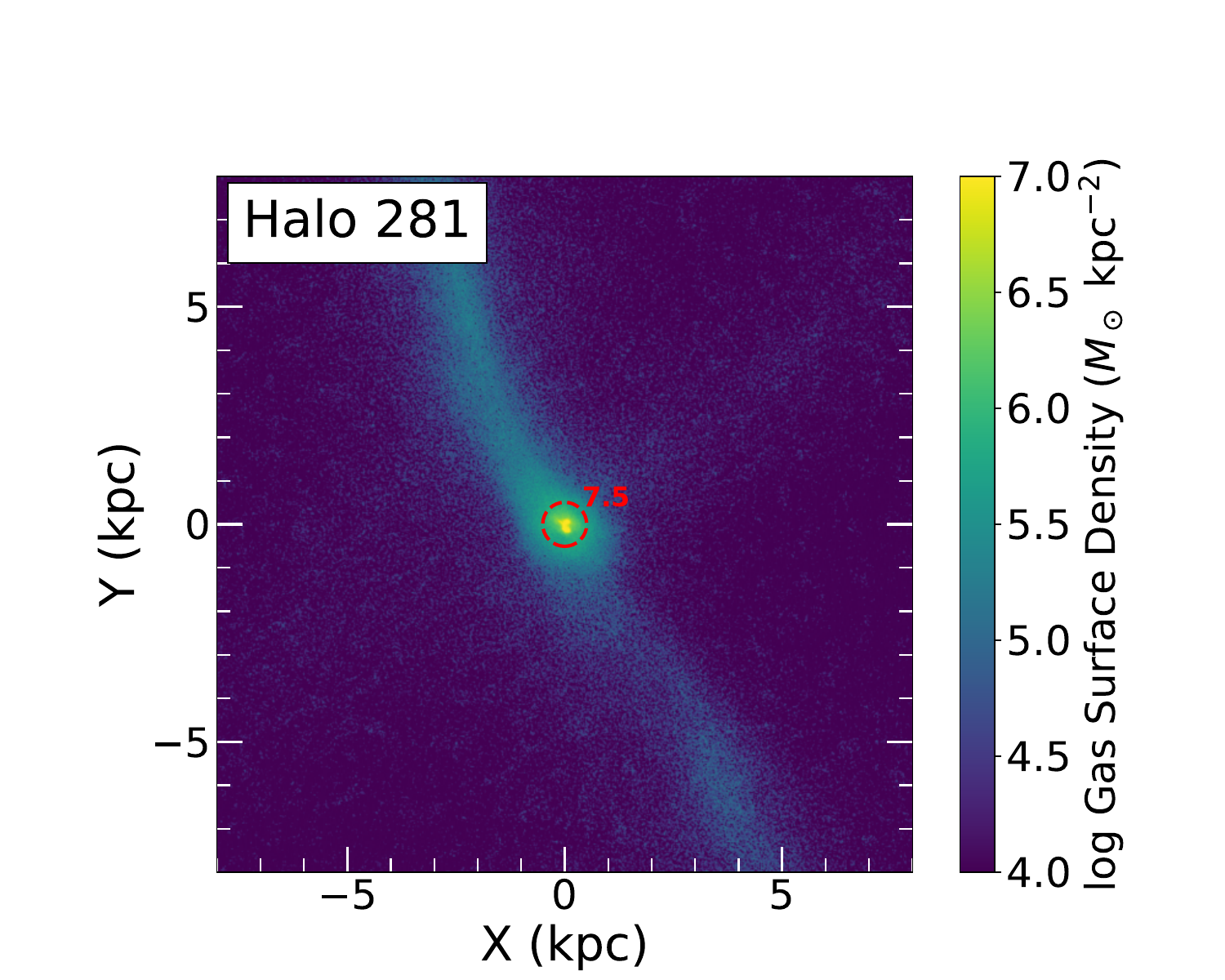}
        \includegraphics[width=0.33\textwidth]{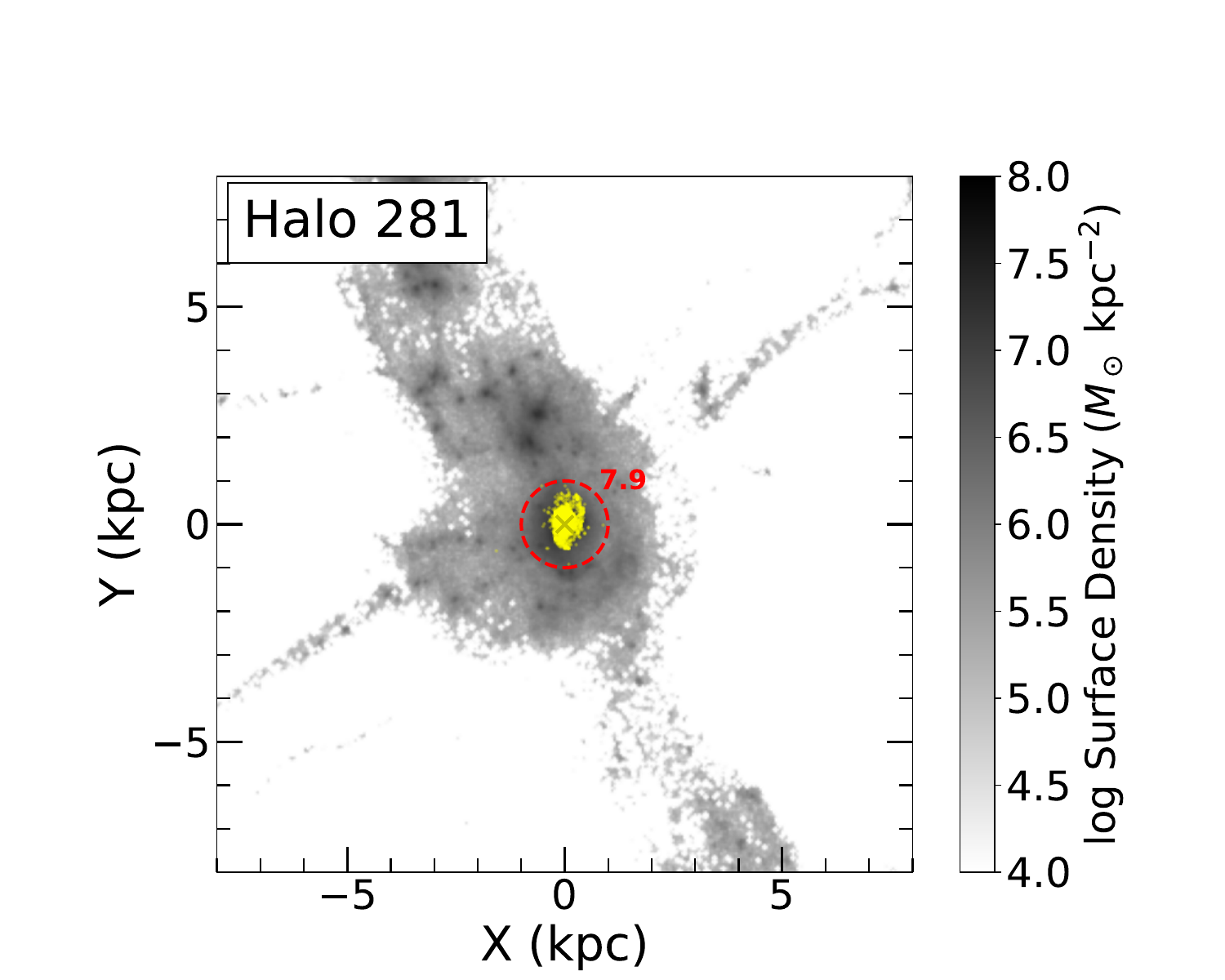}\\
        \includegraphics[width=0.33\textwidth]{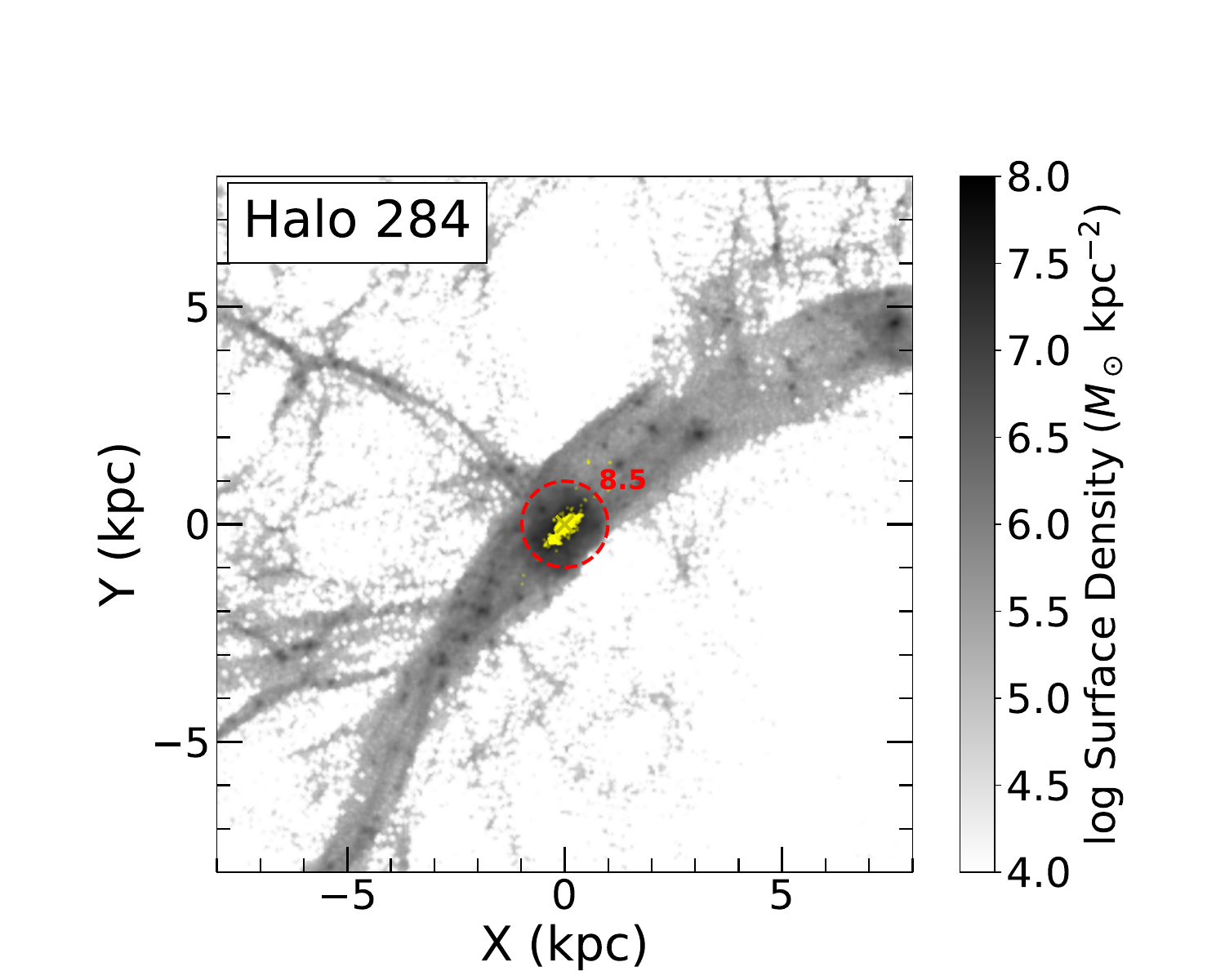}
        \includegraphics[width=0.33\textwidth]{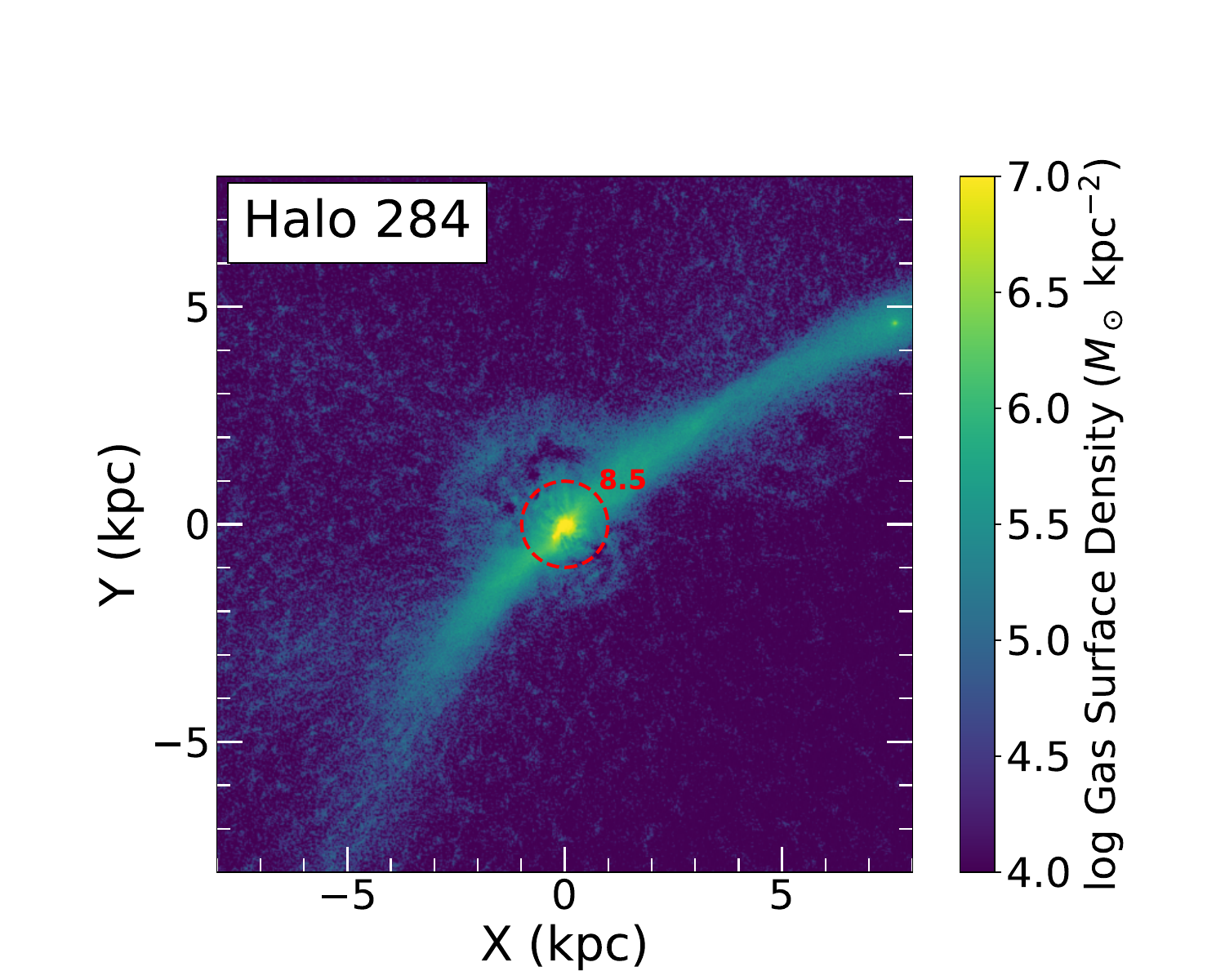}
        \includegraphics[width=0.33\textwidth]{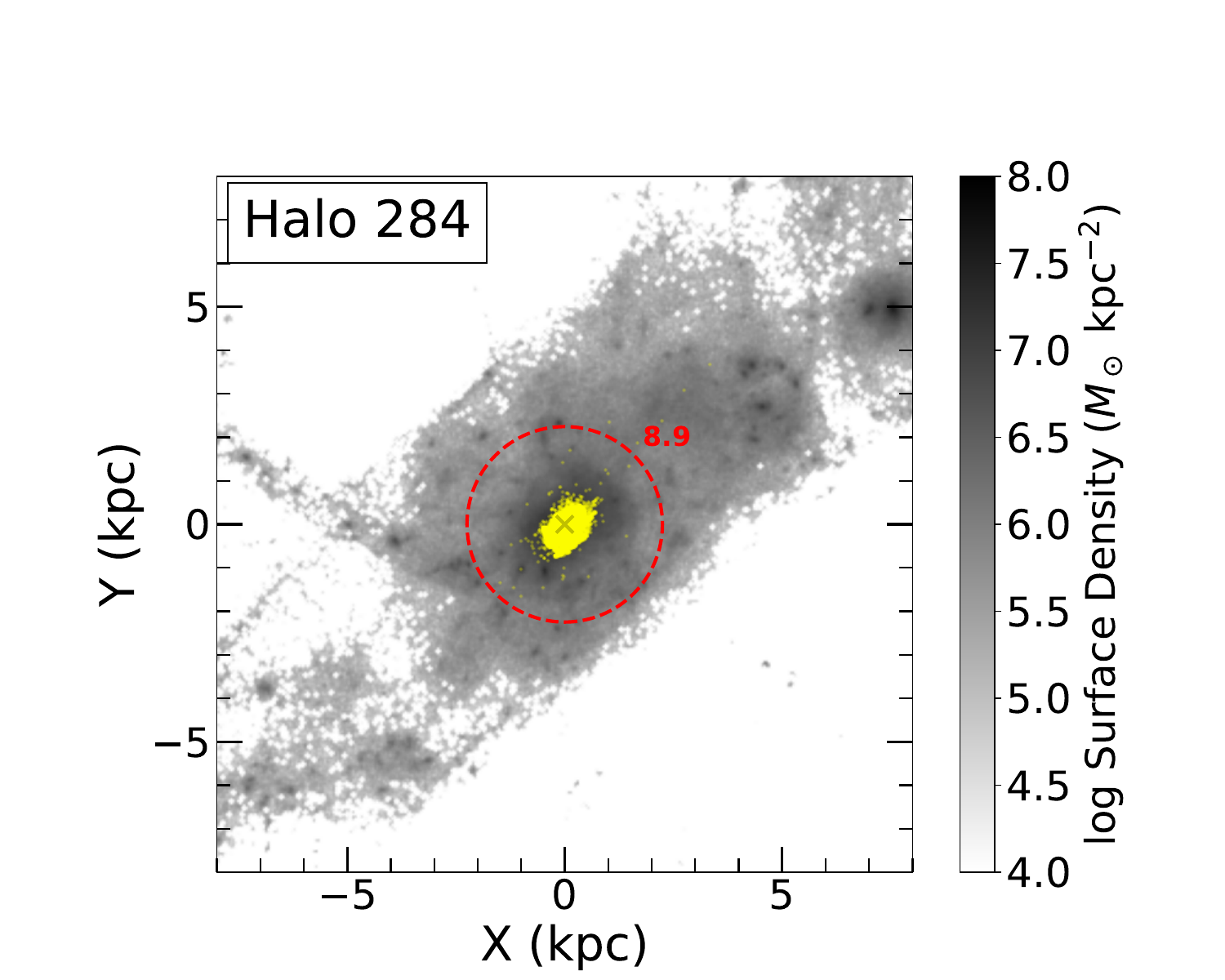}\\
        \includegraphics[width=0.33\textwidth]{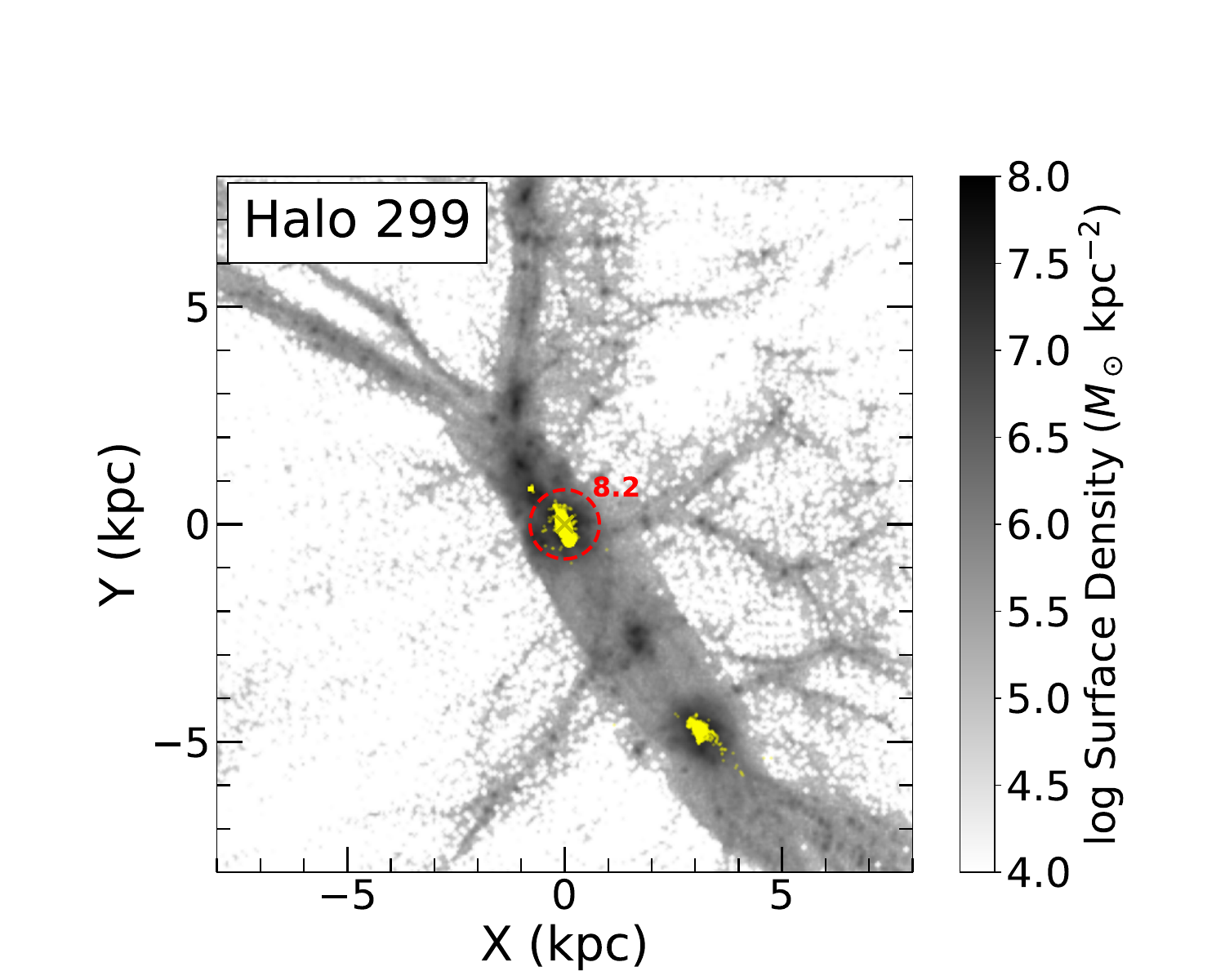}
        \includegraphics[width=0.33\textwidth]{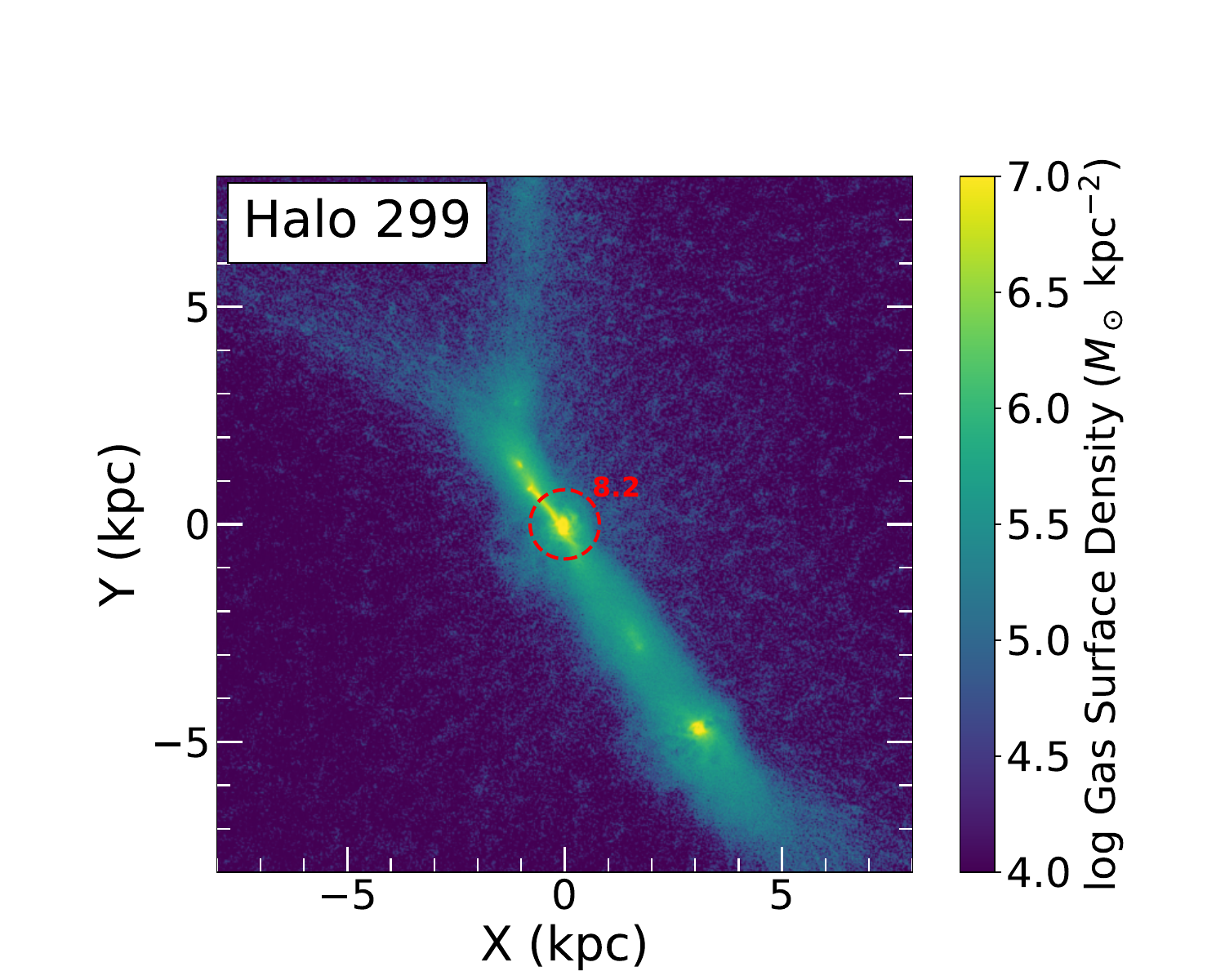}
        \includegraphics[width=0.33\textwidth]{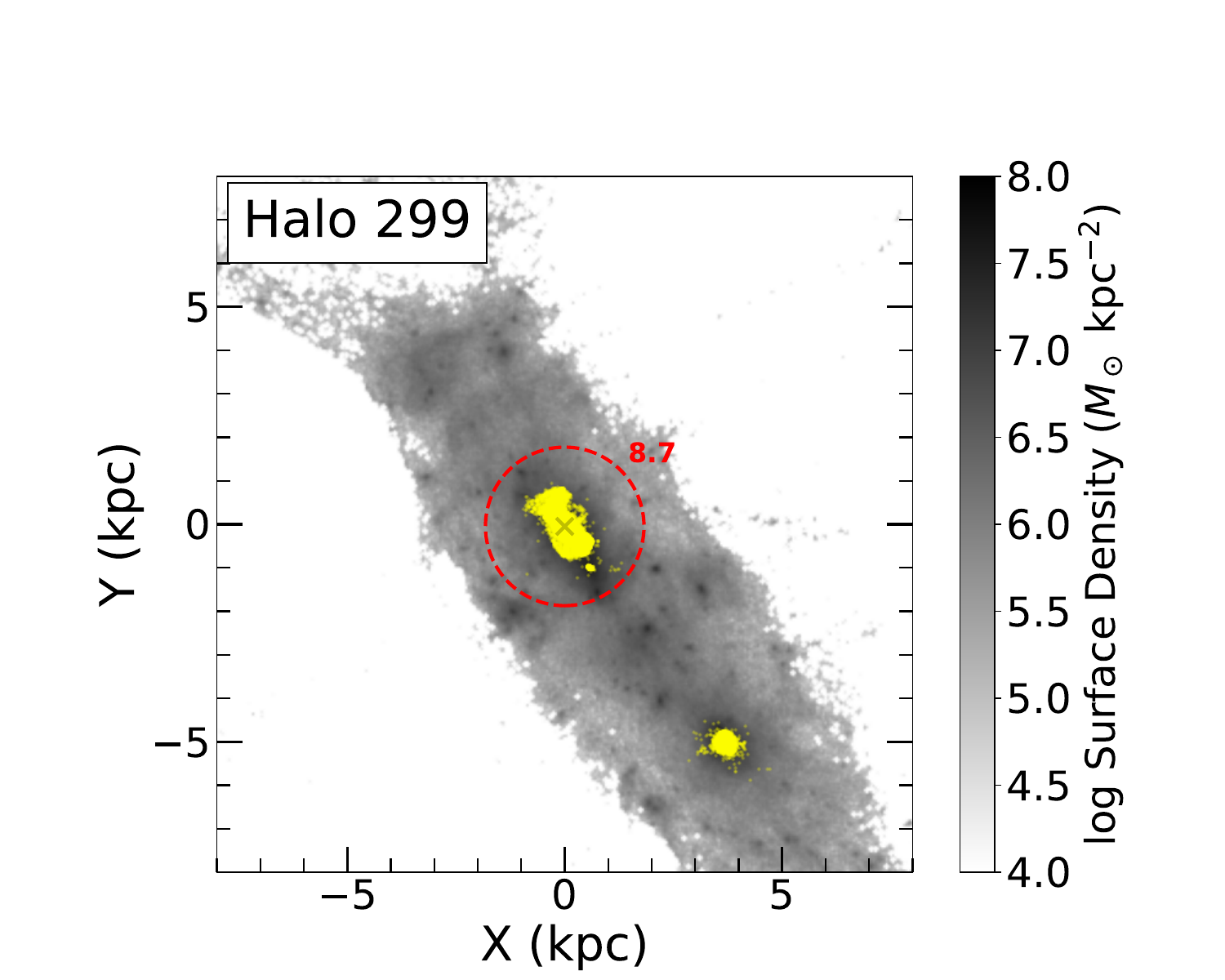}\\
    \end{center}
    \caption{Same as Fig.~\ref{fig:snapshots_large1} but for Halos 236, 281, 284, and 299. {Alt text: Snapshots of Halos 236, 281, 284, and 299 (rows top to bottom) in a 16 $\times$ 16 kpc region: left panels show dark-matter surface density (grayscale) with stellar positions as yellow dots at the epoch of reionization; middle panels show cold gas surface density (T<1000 K); right panels show dark-matter and stellar distribution at t = 1.2 Gyr; red circles and labels indicate virial radius and mass, yellow crosses mark halo centers.}}
    \label{fig:snapshots_large2}
\end{figure*}

We have shown that the gas density is essential for the formation of compact dwarf galaxies. However, the conditions required to achieve such a high density remain unclear. 
In Figs.~\ref{fig:snapshots_large1} and \ref{fig:snapshots_large2}, we present the snapshots of 16\,kpc$\times$16\,kpc. 
As shown in these plots, the dark matter distributions show filamentary structures, and the gas follows them. The halos that formed UCD-like compact stellar distributions, such as Halo 198, 284, and 299, are embedded in the middle of an arch-like, long, and thick filament extending over more than $10\,\mathrm{kpc}$. Gas appears to have accreted onto the halo center along these filaments. 
In contrast, the other galaxies are embedded in a spherical distribution of DM. We observe filamentary structures around these halos, but they are thinner than those observed in the halos that formed UCD-like galaxies. The gas distribution is also spherical and does not appear to be supplied by large-scale structures.

Halo 236 exhibits a high star formation rate, and therefore, we classified it as a High-Mass Halo. However, the DM density of Halo 236 is less dense compared to the others (Halo 198, 284, and 299) but more closely resembles that of the Low-Mass Halos. Indeed, its DM and gas distributions exhibit spherical rather than filamentary structures. More direct measurements of accreting cold gas may be able to provide a condition required for the formation of UCD-like galaxies, but it is beyond the scope of this paper. They will be investigated in a forthcoming paper.

\section{Summary}

To understand the formation processes of dwarf galaxies and their structural diversity, we performed a series of cosmological zoom-in hydrodynamical simulations. We followed the evolution of eight halos with $z=0$ masses of $M_{\rm vir}\sim 10^{9} \rm\ M_\odot$ up to t = 1.2 Gyr, by which time all halos had exhausted their cold, dense gas and ceased star formation.

We found variety in halo assembly histories, which directly translated into diverse stellar masses ($10^{4}$--$10^{6}\ \msun$). The final stellar mass correlates with the halo mass at the epoch of reionization (EoR). Our samples can be divided into two groups based on their halo mass at EoR: High-Mass Halos ($\gtrsim10^8\ \msun$ at EoR) exhibited star formation rates an order of magnitude higher than those of Low-Mass Halos and sustained star formation longer after reionization. In contrast, Low-Mass Halos rapidly lost their gas and quenched within $\sim 200$\,Myr after the reionization. Assuming the halo masses at $z=0$ from dark-matter-only simulations, the resulting stellar-to-halo mass ratios span a wide range ($5\times10^{-5}$ to $2\times 10^{-3}$), which are typical for ultra-faint dwarfs (UFDs) to classical dwarfs, although the halo masses are all $\sim 10^9M_{\odot}$.

We observed a bifurcation in the compactness of the stellar distributions for galaxies with $>10^6M_{\odot}$ in stellar mass. Three halos (Halos 198, 284, and 299) formed massive, highly compact stellar systems resembling Ultra-Compact Dwarfs (UCDs): Halo 198 ($r_h=28.9$\,pc, $M_\star=3.2\times10^{6}\ \msun$), Halo 284 ($r_h=26.3$\,pc, $M_\star=1.8\times10^{6}\ \msun$), and Halo 299 ($r_h=56.1$\,pc, $M_\star=1.04\times10^{6}\ \msun$). One (Halo 236) also has a mass of $>10^6M_{\odot}$, but the half-mass radius is $\sim 100$\,pc. These values are typical for classical dwarf galaxies. 
The remaining four halos also evolve into diffuse systems and follow the dwarf galaxy sequence observed in the Milky Way, similar to Halo 236. 

To identify the physical driver of this compactness, we adapted the theoretical framework of star cluster formation from \citet{fukushima2021radiation} and \citet{fukushima2022far}, but we extended it to include dark matter gravity and supernova feedback. In our simulations, gravity generally dominates over feedback, allowing star formation to proceed in all halos. 
We found that the formation of UCD-like structures requires a specific condition: $\Sigma_{\rm gas} \gtrsim 30\ \msun\,\mathrm{pc}^{-2}$, which is the condition for the formation of compact star clusters with a high star formation efficiency (SFE) of $>0.1$. In \citet{fukushima2022far}, the star formation rapidly proceeded once the star formation efficiency exceeded 0.1, and the gas cloud successfully formed compact stellar systems. We observed a similar process in our simulations. 
Only the three UCD-like halos (Halos 198, 284, and 299) satisfied this criterion. On the other hand, Halo 236, which is also in the High-Mass Halo group but had an extended morphology, did not reach this critical surface density. 
This reinforces that high halo mass alone is insufficient for UCD formation; high central gas density is the precondition.
We also investigated the density profiles of the simulated galaxies. The stellar density of the UCD-like halos (Halos 198 and 294) was as high as that of the dark matter. The dark matter density itself was not a criterion for the formation of UCD-like galaxies. 

The conditions under which dwarf galaxy progenitors achieve such high central gas densities remain an open question. One possible condition is their formation environment. As cold gas accretes into the halo center, it fuels star formation there. The snapshots showed that Halos 198, 284, and 299 reside in elongated, well-connected cosmic dark matter filaments that efficiently may channel gas to their centers, whereas the others are in rounder, clumpier environments with weaker gas replenishment. 
The relation between the dark matter structure and gas supply to the halo center will be investigated in a forthcoming paper. 
If we could connect the dark-matter structure to the formation of UCD-like galaxies, we would be able to evaluate the fraction of UCDs that formed via this evolutionary path from dark-matter-only simulations, which are less computationally expensive compared to full hydrodynamical simulations.



\begin{ack}

Simulations in this study were performed on Yukawa-2 at Kyoto University, and the analyses were conducted on analysis servers at the Center for Computational Astrophysics, National Astronomical Observatory of Japan.

\end{ack}

\section*{Funding}
This research was supported by KAKENHI Grant Numbers 23K22530, 21K03614, 21K03633, 22KJ0157, 22K03688, 24K07095, 25K01046, and 25H00664.
 
\section*{Data availability} 
The simulation data will be provided upon request.

\appendix 

\bibliographystyle{aasjournal}
\bibliography{bibtex}

\end{document}